\documentclass[11pt]{article}

\usepackage[utf8x]{inputenc}
\usepackage{cite}
\usepackage{graphicx}
\usepackage{amsmath}
\usepackage{amssymb}
\usepackage{color}
\usepackage{makeidx}
\usepackage{multicol}
\usepackage{url}
\usepackage[linktocpage]{hyperref}
\usepackage{placeins}
\usepackage{array}
\usepackage[font={small}]{caption}
\usepackage{pdfpages}
\usepackage[titletoc,title]{appendix}
\usepackage{tabularx}
\usepackage{ltablex}
\usepackage{lscape}
\usepackage{multirow}
\usepackage{xcolor}

\usepackage[textwidth=6.8in,textheight=9.0in,columnsep=16pt]{geometry} 

\allowdisplaybreaks[2]

\makeatletter

\newenvironment{figurehere}
{\def\@captype{figure}}
{}
\makeatother

\newcommand{\gev}{\operatorname{GeV}}

\newcommand{\sqrtsNN}{\mbox{$\sqrt{s_{_{\mathrm{NN}}}}$}}
\newcommand{\sqrts}{\mbox{$\sqrt{s}$}}

\newcommand{\ep}{\textit{e}+\textit{p}}
\newcommand{\eA}{\mbox{\textit{e}+A}}

\newcommand{\pPb}{\textit{p}+Pb}
\newcommand{\pA}{\textit{p}+A}

\newcommand{\eAu}{\mbox{\textit{e}+Au}}

\newcommand{\AuAu}{Au+Au}

\newcommand{\pp}{\mbox{\textit{p}+\textit{p}}}

\renewcommand{\AA}{A+A}

\newcommand{\lumi}{\mbox{$\mathrm{cm}^{-2}\ \mathrm{sec}^{-1}$}}

\begin{document}

\clearpage
\thispagestyle{empty}
\includepdf[width=8.51in]{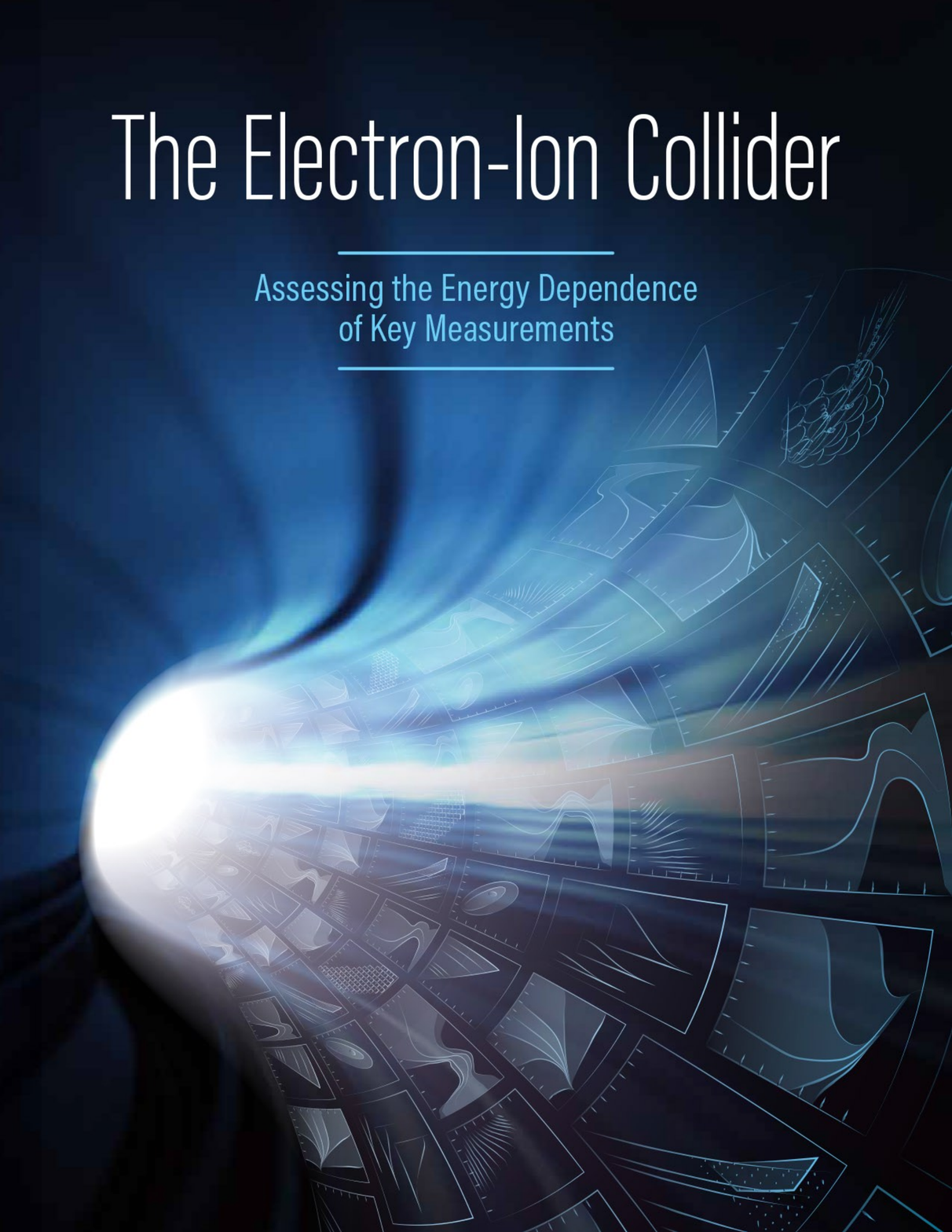}
\newpage
\thispagestyle{empty}
\mbox{}


\newpage
\thispagestyle{empty}
\hfill 
\begin{minipage}[t]{0.25\textwidth}
BNL Formal Report\\
BNL-114111-2017\\
\today\\
\end{minipage}\\

\vspace{2cm}
\begin{center}
\begin{minipage}[t]{\linewidth}
\bf\Huge The Electron-Ion Collider:\\
Assessing the Energy Dependence\\
of Key Measurements 
\end{minipage}
\vfill
Brookhaven National Laboratory\\
P.O. Box 5000\\
Upton, NY 11973-5000\\
www.bnl.gov\\
\end{center}

\newpage
\thispagestyle{empty}
\begin{center}
\begin{minipage}[b]{0.75\textwidth}
\centerline{\Large\bf Disclaimer}
\vspace{3mm}
This report was prepared as an account of work sponsored by an agency of the
United States Government. Neither the United States Government nor any
agency thereof, nor any of their employees, nor any of their contractors,
subcontractors, or their employees, makes any warranty, express or implied, or
assumes any legal liability or responsibility for the accuracy, completeness, or any
third party’s use or the results of such use of any information, apparatus, product,
or process disclosed, or represents that its use would not infringe privately owned
rights. Reference herein to any specific commercial product, process, or service
by trade name, trademark, manufacturer, or otherwise, does not necessarily
constitute or imply its endorsement, recommendation, or favoring by the United
States Government or any agency thereof or its contractors or subcontractors.
The views and opinions of authors expressed herein do not necessarily state or
reflect those of the United States Government or any agency thereof.
\end{minipage}
\end{center}


\label{authors}

\newpage
\renewcommand{\thefootnote}{\fnsymbol{footnote}}

\centerline{\Large\bf Authors}
\vskip 0.3in

\begin{center}
	E.C.~Aschenauer,
	S.~Fazio,
	J.H.~Lee,
	H.~M\"antysaari,
	B.~S.~Page,
	B.~Schenke,
	T.~Ullrich\footnote{Editors},
	R.~Venugopalan$^{*}$,
	P.~Zurita\\[1cm] 

	\noindent
	{\it  Brookhaven National Laboratory, USA}
\end{center}

\vfill

\noindent
{\Large\bf Acknowledgments}\\

We thank the following colleagues who made valuable contributions  to this document and the studies it is based on:
M.~Diehl (DESY, Germany), 
A.~Dumitru (Baruch), 
A.~Kiselev (BNL), 
H.~Paukkunen (Jyv\"askyl\"a, Finland),
R.~Sassot (Buenos Aires, Argentina), 
V.~Skokov (RBRC/BNL), 
M.~Stratmann (T\"ubingen, Germany),
 L.~Zheng (CCNU, China).

The authors are indebted to the following colleagues for critical and crucial comments in the preparation 
of this report: 
P.~Bond (BNL), 
A.~L.~Deshpande (BNL/Stony Brook),
M.~Diehl (DESY, Germany),  
B.~Jacak (LBNL),
D.~Kharzeev (BNL/Stony Brook),  
D.~Lissauer (BNL),  
B.~M\"uller (BNL), 
P.~Newman (Birmingham, UK),
G.~Sterman (Stony Brook), 
R.~Tribble (BNL), 
W.~Vogelsang (T\"ubingen, Gemany),
W.~A.~Zajc (Columbia).

We are grateful to T.~Bowman (BNL) for designing the cover for this report and D. Arkhipkin (BNL) and A. Kiselev (BNL) for generating the event display featured on the back cover. 

Notice: This manuscript has been authored by employees of Brookhaven Science Associates, LLC under Contract No. DE-AC02-98CH10886 and DE-SC0012704 with the U.S. Department of Energy. 
The United States Government retains a non-exclusive, paid-up, irrevocable, world-wide license to publish or reproduce the published form of this manuscript, or allow others
to do so, for United States Government purposes.

\setcounter{page}{1}
\pagenumbering{roman}
\newpage


\label{sec:abstract}

\newpage
\centerline{\Large\bf Abstract}
\vskip 0.3in
We provide an assessment of the energy dependence of key measurements within the scope of the machine parameters for a U.S. based Electron-Ion Collider (EIC) outlined in the EIC White Paper. We first examine the importance of the physics underlying these measurements in the context of the outstanding questions in nuclear science. We then demonstrate, through detailed simulations of the measurements, that the likelihood of transformational scientific insights is greatly enhanced by making the energy range and reach of the EIC as large as practically feasible. 

\newpage

\setcounter{tocdepth}{4}
\tableofcontents

\newpage
\thispagestyle{empty}
\mbox{}
\clearpage


%
%
\setcounter{page}{1}
\pagenumbering{arabic}

\section{Introduction}
\label{chapter:introduction}

An Electron-Ion Collider (EIC) is a key 
component of the future program of the nuclear physics community 
in the US. The EIC will be the world's first electron-nucleus collider and the world's first collider to scatter  polarized electrons off polarized protons. EIC-based science is extremely broad and diverse. It 
runs the gamut from detailed investigation of hadronic structure with unprecedented 
precision to explorations of new regimes of strongly interacting matter. This deep and varied study  will greatly expand our knowledge and understanding of Quantum Chromodynamics (QCD), the fundamental quantum theory of the quark and gluon fields making up nearly all the visible matter in the universe. EIC science can be characterized by a few distinguishing themes that reflect the major challenges facing modern science today, and that have deep links to cutting edge research in other sub-fields of physics.

The EIC White Paper \cite{Accardi:2012qut} released at the end of 2012, 
and updated in 2014, presents in detail the science case of an 
Electron-Ion Collider.  It lays out how the EIC's ability to collide 
high-energy electron beams with high-energy ion beams will 
provide access to kinematic regions in the nucleon and nuclei where 
their structure is dominated by gluons and how polarized beams 
will give unprecedented access to the spatial and spin structure 
of gluons in the proton. Thus, the machine design needs to aim at 
achieving highly polarized (70\%) electron and proton beams, ion 
beams from deuteron to the heaviest nuclei, high collision 
luminosity of $10^{33}-10^{34}$ \lumi, and center-of-mass
energies in a wide range up to 140 $\sqrt{Z/A}$ GeV. 
Several of the studies in the White Paper that demonstrate 
the physics reach and potential of an EIC  assumed
the higher energy range, and typical integrated luminosities of 10 fb$^{-1}$.

Now that the EIC has been embraced by the nuclear physics community 
in the 2015 Nuclear Science Advisory Committee (NSAC) Long Range Plan and the technology to build an EIC 
is becoming available, it is timely to review the arguments that led to the proposed center-of-mass energy range considered initially.  The purpose of this document is not to repeat arguments put forward in the EIC White Paper. We instead build upon these arguments and scrutinize more closely the energy dependence of key measurements that are essential to ensure a compelling EIC science program. To be specific, for \ep\ collisions we define the low energy range of center-of-mass energies to span 
 $\sqrt{s}=22$-$63$ GeV and the high energy range to span $\sqrt{s}=45$-$141$ GeV. The corresponding center-of-mass energies for \eA\ collisions off heavy nuclei are $\sqrt{s}=15$-$40$ GeV and $\sqrt{s}=32$-$90$ GeV. We add an important study omitted in the EIC White Paper, that of jets. Jets provide a highly precise characterization of the final state in deep inelastic scattering (DIS) that complements the precision provided by the electron probe.

To motivate a deeper appreciation of why a high energy EIC may be needed, we present in Section 2,  a  
``big picture" case for EIC science and how the proposed studies open new opportunities, generate excitement, and pose challenges. We note the lessons provided by past DIS experiments and  those learned from discoveries at other colliders. In Section 3,  we  examine in detail the energy dependence of several key measurements in order to guide discussions on the required energy reach of an EIC.  Most of the simulations presented were carried out with the same or improved procedures and programs that were used in the EIC White Paper. We conclude this document with a brief summary of our principal observations.

%
%
\section{The Big Picture}
		To further advance the compelling scientific case for a large scale EIC project, we will begin by placing its energy requirements in the broader context of the role of QCD and the strong interactions within the Standard Model of physics and by  articulating why QCD matters in the big picture. We will point to outstanding challenges in our understanding of how the many-body dynamics of quark and gluon fields give rise to the confined structure and dynamics of the strongly interacting matter that constitutes nearly all the mass of the visible universe. More specifically, we will sketch out the landscape of the QCD dynamics inside hadrons and nuclei that is probed by varying the energy and resolution of the electron probe. While a discovery can rarely be predicted, lessons from the past are instructive. We will examine lessons provided by completed DIS experiments and the discovery of the quark-gluon plasma to guide our study of the discovery potential of an EIC within the scope of machine parameters discussed in the EIC White Paper.

	\subsection{Why QCD Matters}
		\label{sec:bigPicture}
\begin{multicols}{2}
Quantum Chromodynamics (QCD) represents the apogee of a quantum field theory. It is a nearly self-contained fundamental theory of quark and gluon fields that is rich in symmetries; the only external parameters in QCD are the quark masses generated by the Higgs mechanism in the Standard Model~\cite{Wilczek:1999id}. Strongly interacting phenomena emerge from the interactions generated by the symmetries of QCD and from the breaking of these symmetries by the QCD vacuum and by the quark masses. The generation of the mass of strongly interacting matter is a striking example of an emergent phenomenon. Gluons carry no mass and the light quarks carry masses roughly a hundredth that of the proton. Despite this lightness of being, quarks and gluons, through their interactions with each other and the QCD vacuum, generate the mass of protons and neutrons, and other strongly interacting particles. The dynamics of quark and gluon fields are therefore responsible for nearly all the mass of visible matter in the universe.

Another striking emergent phenomenon is that colored quarks and gluons are permanently confined within hadrons on a length scale on the order of a Fermi. This emergent scale, nowhere evident in the QCD Lagrangian, dictates the maximal distance out to which the chromo-electric and chromo-magnetic fields of quarks and gluons can spread. At such separations, the forces between quarks and gluons are so large that it is energetically favorable for the QCD vacuum to spontaneously create quark-antiquark pairs, and gluons, which assist fundamentally in assembling colored quarks and gluons into colorless hadrons. How this phenomenon called ``confinement" occurs is still not understood and is one of the outstanding mysteries of physics.

The essence of all of the remarkable and confounding phenomena of QCD is its non-Abelian nature whereby, unlike photons, gluon fields can self-interact. QCD is therefore an intrinsically nonlinear theory, and much of the complexity of its dynamics can be traced to this feature of the theory. This nonlinear essence is also what makes the extraction of physics fiendishly difficult. In the 40 odd years since the discovery of QCD, powerful techniques have been developed to elucidate phenomena of nature's strong interaction. One of these is perturbative QCD (pQCD), which exploits yet another emergent phenomenon of the theory: asymptotic freedom. Quarks and gluons interact weakly at small separations, or equivalently, large momentum transfers. In this asymptotic limit, the QCD coupling constant is weak and manifestly nonlinear interactions are suppressed; as a consequence, computations can be undertaken that are highly precise and carry predictive power. In the opposite strongly interacting regime of QCD, effective field theories (EFT) ingeniously take advantage of symmetries and the separation of soft and hard emergent scales to extract physical information. A noteworthy example is chiral perturbation theory, that exploits the chiral symmetry of the QCD Lagrangian, and its breaking by the vacuum, for systematic computations at energies below the emergent QCD scale. 

The most powerful method of all is lattice gauge theory which discretizes and solves QCD in all its complexity  on the fastest computers available. Lattice computations have by now accurately reproduced {\it ab initio} the mass spectrum of hadrons, as well as other important quantities. Most importantly, lattice QCD has established beyond reasonable doubt that QCD is the correct fundamental theory of the strong interactions. Furthermore, lattice investigations of QCD symmetries, and their breaking, illustrate beautifully how complex phenomena such as the mass, spin, and structure of the lightest hadrons to the heaviest nuclei emerge out of seemingly nothing in QCD. 

Experimental progress has been no less impressive. Since the discovery of quarks in DIS experiments at SLAC, and of gluons in electron-positron collisions at DESY, a large number of experiments, across a wide range of energies, have uncovered fundamental information about QCD. The running of the QCD coupling predicted by asymptotic freedom is an established fact, in good agreement with theory predictions. QCD has become precision physics at high momentum transfers;  deviations of theory from experiment in this regime would be a harbinger of physics beyond the Standard Model. QCD studies continually reveal new discoveries. Candidates for novel tetraquark and pentaquark states have been identified, and the possibility of exotic hybrid and glueball hadrons is under investigation. The spin of hadrons, now understood as an emergent many-body phenomenon, is being quantified in terms of the spins of the underlying quarks and gluons. The discovery of the quark-gluon plasma (QGP) highlights the role of QCD in cosmology, from the QGP of the early universe, to Big Bang nucleosynthesis, the synthesis of heavy elements, and the high baryon density regime of neutron stars. Significant insight into the latter will be provided by the upcoming experiments with rare isotope beams. Explorations of the QCD phase diagram in temperature and baryon density, to uncover the phases of strongly interacting matter, are the focus of extensive experimental effort.

As the advances in theory and experiment demonstrate, QCD is clearly a mature subject. It is therefore important to clearly articulate, in this context, the urgency of the construction of an Electron-Ion Collider.

\end{multicols}
\FloatBarrier

	\subsection{The Landscape of QCD }
		\label{section:landscape}

\begin{figure*}[t!]
	\centering \includegraphics[keepaspectratio=true,width=0.75\linewidth]{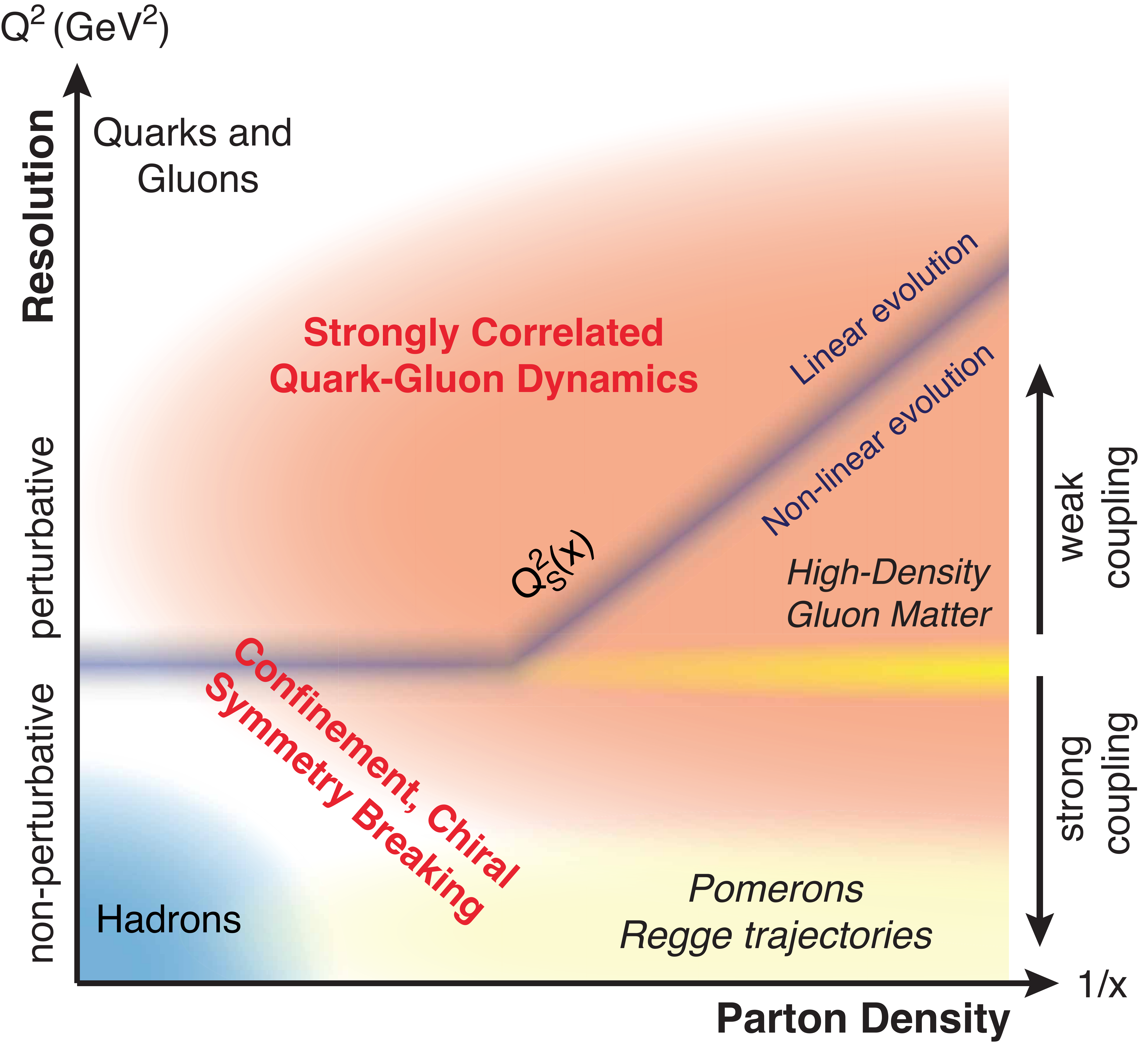}
	\caption{Landscape of QCD. The vertical axis represents varying resolution $Q^2$ and the horizontal axis varying parton density, an increasing function of $1/x$.}
	\label{fig:bigpicture}
\end{figure*}

\begin{multicols}{2}
The behavior of QCD at very short distances is well understood and can be used to predict the hard scattering of partons,  where ``partons" denote quarks, antiquarks, and gluons. However, the mechanisms by which QCD produces the bulk of the visible world remains mysterious and needs continuing and varied efforts to resolve.  A profound subset of outstanding questions concerns the internal structure of hadrons and nuclei at high energies. With increasing energy, gluons (and the accompanying ``sea" of quark-antiquark pairs) proliferate and the ``veil" that hides their dynamics at low energies is lifted. All emergent structure of hadrons such as mass and spin are a result of the confined many-body dynamics of partons. In Fig.~\ref{fig:bigpicture}, we show the landscape of hadron structure that opens up at high energies. The vertical axis represents varying resolution $Q^2$ and the horizontal axis represents varying parton density, an increasing function of $1/x$, where $x$ is the momentum fraction of the proton carried by a parton. When the parton density is fixed to be small (large $x$), increasing the energy of the probe increases $Q^2$, allowing one to finely resolve the partons inside protons and nuclei.

At low $Q^2$ and large $x$, strong interaction physics is described by hadrons and their interactions. In this regime, the QCD coupling is large, the fields are nonlinear, and the physics is nonperturbative. Moving up the  vertical axis at large $x$, to very large $Q^2$, the coupling becomes weak due to asymptotic freedom, and perturbative QCD describes well the interactions of quarks and gluons. How the transition from low to high $Q^2$ occurs, and what the degrees of freedom describing this transition region are, is not understood. Confinement and chiral symmetry breaking play a key role but how they influence many-body dynamics in this regime is unclear. At large $Q^2$, as one moves  towards higher parton density, many-body correlations between quarks and gluons become increasingly important. When the coupling is weak, it is possible to quantify these, and describe their evolution using the linear evolution equations of perturbative QCD. How such many-body correlations are modified in nuclei, or when the proton is polarized, is not known. 

At very large parton densities, or small $x$, gluon degrees of freedom become dominant. Total cross-sections in high energy scattering are dominated by the physics of small $x$ and low $Q^2$.  Nonperturbative QCD dynamics in this kinematic regime has historically been parametrized in terms of effective color singlet degrees of freedom called ``Pomerons" and ``Reggeons" and their interactions. How they are constituted fundamentally in terms of the dynamics of the underlying quark and gluon degrees of freedom, and that of the vacuum, is {\it terra incognita}. Remarkably, this corner of the QCD landscape, which contributes the most to the high energy scattering of hadrons, is perhaps the least understood. The takeaway message here is that even though there are aspects of the regimes discussed in  Fig.~\ref{fig:bigpicture} that are known, essential elements in constructing dynamical pictures of hadrons and nuclei that are important for a first principles understanding of strong interaction phenomena are missing.

A novel regime of QCD may exist in the upper right corner of Fig.~\ref{fig:bigpicture}, where parton densities are high.  For high but fixed values of the resolution $Q^2$, the density of gluons saturates with decreasing $x$.  In this gluon saturation regime, the chromo-electric and chromo-magnetic fields are as strong as allowed in QCD--the strongest fields in nature-- {\it even though asymptotic freedom ensures that the QCD coupling is weak}. The feature of weak coupling is key because it allows, for the first time, systematic computations of the many-body dynamics of quarks and gluons in an intrinsically nonlinear regime of QCD. Note that lattice QCD primarily studies static quantities--dynamical quantities, especially at high parton densities, are mostly outside its purview. 

The structure of QCD, and indeed the requirement that strongly interacting matter be stable, suggests that these saturated gluons {\it do not} form a Bose-Einstein condensate. Instead, they generate a unique form of strongly interacting matter, called a Color Glass Condensate (CGC)~\cite{Gelis:2010nm}. The strong interplay of attractive and repulsive many-body forces amongst gluons ensures that their typical momenta are peaked not at zero momentum as in a Bose-Einstein condensate, but instead at an emergent saturation momentum scale $Q_s$. Further, the interactions amongst gluons carrying larger or smaller fractions of the proton's momentum are time dilated to much longer than strong interaction time scales. This is typical for glassy material, and indeed, the equations that describe the CGC are similar in this respect to those for glasses. A prediction of the CGC description is that the saturation scale grows with energy and nuclear size and, for large values of both, can be much larger than the intrinsic QCD scale. As indicated in Fig.~\ref{fig:bigpicture}, for resolutions $Q^2 \gg Q_s^2$, the nonlinear dynamics of gluons goes over smoothly into the linear evolution of quarks and gluons that is characteristic of pQCD. 

Properties of the CGC in the gluon saturation regime can be computed in a weak coupling effective theory. The QCD evolution equations in this CGC effective theory are intrinsically nonlinear, and generate a hierarchy of many-body correlation functions whose renormalization group evolution is analogous to that of other strongly correlated systems. In particular, because interactions are time dilated, and the coherence length of probes is large relative to hadron sizes, the longitudinal and transverse dynamics of gluons nearly decouples  in the upper right corner of Fig.~\ref{fig:bigpicture}.
\end{multicols}

\FloatBarrier


	\subsection{Emergence and Universality in the Landscape}
		\label{sec:emergence}
\begin{multicols}{2}
The many-body physics of the QCD landscape provides a robust non-Abelian counterpart to those of other strongly correlated systems, where one observes remarkable emergent phenomena with universal features. Indeed, condensed matter and cold-atom physicists are exploring ways to mimic the strongly interacting dynamics of gauge theories  \cite{Wiese:2013}. The structure of QCD suggests strong discovery potential both by deep investigation of many-body dynamics in QCD and simultaneous attempts to reproduce it in other strongly correlated systems. 

An outstanding question is whether there exists matter inside protons and complex nuclei that is universal or whether the ``doping"  caused by quantum features that are unique to each hadron (such as their spin, flavor and baryon number) makes it impossible to isolate universal features of their dynamics. The conventional wisdom is that such quantum numbers are carried mostly by partons that possess large fractions of the hadron's momentum. In this picture, partons with smaller and smaller momentum fractions $x$ are increasingly like the spinless, flavorless and baryon-free QCD vacuum. Alternately, it is conceivable that the long arm of confinement ensures ``memory" of vacuum doping down to the smallest values of $x$. 

Models of high parton density matter predict that the strong chromo-fields of gluon saturation are achieved precociously in large nuclei at larger values of $x$ than in the proton. In particular, the saturation scale has a nuclear enhancement: $Q_s^2\sim A^{1/3}$. However, at {\it very} small $x$, if gluon interactions were universal, the values of $Q_s^2(x)$ in the proton and in more complex nuclei would be identical. This is only likely to be achieved for energies much larger than those at an EIC. Nevertheless, the deviations of $Q_s^2(x)$ from asymptotic expectations may have a specific structure that is accessible for the EIC range of energies. The observation of such systematics would therefore provide indirect evidence for the universality of saturated gluon matter. 

Besides this question of obvious interest, nuclei are an important ingredient towards a deeper understanding of our big picture landscape. An apparently elementary question of how the gluon distribution in the nucleus deviates from a simple convolution of nucleon and gluon distributions is unresolved. More detailed questions about how confinement operates in nuclei versus nucleons -- an example being the role of gluons and quarks in the short range nuclear forces that bind nuclei-have no clear answer at present.

Another fascinating window into the dynamics of confinement is the dynamical conversion of quarks and gluons into hadrons. Nuclei are a QCD laboratory to examine this hadronization question with ``fresh eyes". While QCD in nuclei has been explored for nearly 40 years, it has never been studied previously with deep inelastic scattering processes in collider mode. Energy reach is crucial in this regard because it allows i) wide exploration of the nuclear landscape of Fig.~\ref{fig:bigpicture} ii) novel tools such as heavy quark and jet observables, that were previously out of reach, and iii) access to a wider phase space to study the QCD dynamics of hadronization. 

How the spin of the proton emerges from the many-body dynamics of quarks and gluons presents several puzzles that high energies at an EIC can help resolve. The first of these is to understand how the relative contributions of the parton spins to the spin sum rule (see Eq.~\ref{Eq.sumrule}) hold across the QCD landscape despite the wide change in energy and resolution. Since the dynamics of spin also involves the orbital motion of quarks and gluons, this motivates developing tools that  allow one to image the transverse spatial and momentum structure of the proton. The imaging of the 3-D spin structure of the proton in the gluon dominated regime will be an important first for the EIC; this topic has been extensively discussed in the EIC White Paper. As noted there, uncertainties in the contributions to the quark and gluon spins at small $x$ are significant. Reducing these uncertainties, while maintaining the large $Q^2$ values essential for controlled computations, requires high energies: this is because  $Q^2 \sim x s$, where $s$ is the squared center-of-mass energy.

An important question, related to the previous discussion on the universality of the vacuum, is to understand how spin is transmitted to small $x$. So-called quantum anomalies, which cause symmetries satisfied by the QCD equations of motion to not be satisfied by the QCD amplitudes, may play an important role at small $x$ that is not transparent in the $x$-integrated quantities that contribute to the sum rule. The flavor structure of the QCD vacuum in polarized protons extracted from semi-inclusive DIS is an important ingredient in building a more complete picture. The high energies at an EIC will also enable precision DIS studies with charged current probes that can further help resolve the flavor structure of the proton. 

The precision study of the QCD landscape enabled by exploring, for the first time in DIS, nuclei and polarized protons in collider mode at high energies and luminosities will generate novel results. One such example is diffractive scattering off nuclei, where one can probe, with fine resolution, Pomeron color singlet exchanges that carry the quantum numbers of the vacuum. The wide range of such innovative measurements, a few of which we will focus on in this report, promise fundamental insight into the many-body nature of the outstanding mystery of confinement. Because several of these are first measurements in uncharted regions of the QCD landscape, discoveries that resolve fundamental questions are highly likely. 
\end{multicols}
\FloatBarrier


	\subsection{Discovery Science: Lessons from DIS Experiments}
		\label{sec:lessonsDIS}

\begin{multicols}{2}

\begin{figure*}[t!]
	\centering \includegraphics[width=0.95\linewidth]{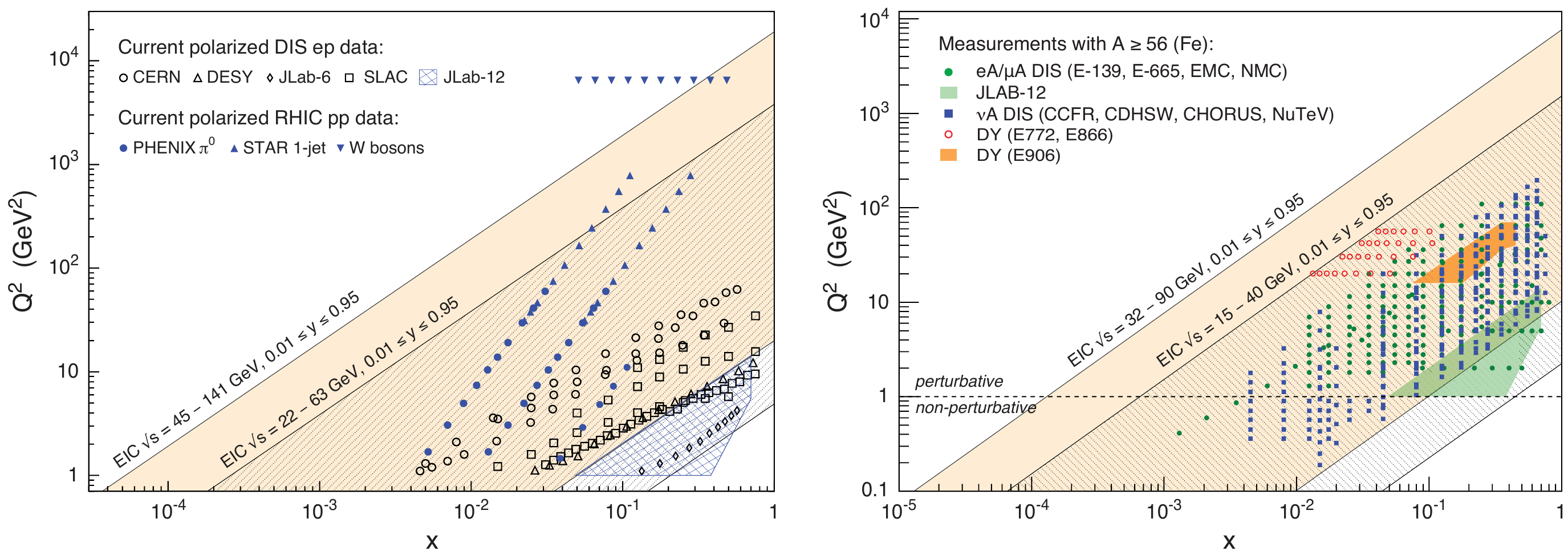}
	\caption{{\it Left:} The range in $x$ vs. $Q^2$, accessible with an EIC in polarized \ep\ collisions compared to past (CERN, DESY, SLAC) and existing (JLAB) facilities as
                                well as to polarized \pp\ collisions at RHIC.  Two different energy ranges from 22--63
                               GeV (hatched) and from 45--141 GeV (beige) are indicated.
                 {\it Right:} The kinematic acceptance in $x$ vs. $Q^2$ of completed lepton-nucleus%
                                 (DIS) and Drell-Yan (DY) experiments, as well as JLAB-12 (all fixed target) compared to the EIC acceptance 
                                 in two energy ranges, 15--40  GeV (hatched) and from 32--90 GeV (beige).}
\label{fig:x-q2-epeA_combo}
\end{figure*} 

The discovery of rapidly growing gluon distributions in DIS experiments at the HERA collider gave a clear experimental  indication that the proton at high energies is a complex many-body system where gluon degrees of freedom are dominant. This discovery gave impetus to the idea that there exists a novel saturation regime in QCD where many-body dynamics is intrinsically nonlinear. Precision measurements of quark and gluon distributions at HERA played a major role in the discovery of the Higgs boson and the characterization of its properties, and are a key tool in ongoing searches at the LHC for physics beyond the Standard Model~\cite{Anastasiou:2016cez}.

At the EIC, the focus is on parton distributions in nuclei and on polarized parton distributions in spin polarized protons.  The kinematic reach in Bjorken $x$ and the momentum resolution $Q^2$ for DIS for a range of EIC energies in \ep\ collisions (with and without polarized protons) is shown in Fig.~\ref{fig:x-q2-epeA_combo} (left). The kinematic reach in \eA\ collisions is shown in Fig.~\ref{fig:x-q2-epeA_combo} (right). For \ep\, the two energy ranges depicted are, i) a high energy range of center-of-mass range of  $\sqrt{s}=45$-$141$ GeV, and ii) a lower energy range of $\sqrt{s}=22$-$63$ GeV. In \eA\ collisions off heavy nuclei, the corresponding low energy center-of-mass range is $\sqrt{s}=15$-$40$ GeV and the higher energy range is $\sqrt{s}=32$-$90$ GeV. Diagonal lines on the plot represent lines of constant ``inelasticity" $y$. In the rest frame of the proton (or nucleus), the inelasticity is the ratio of the energy carried by the virtual photon divided by the energy of the incoming electron. Figure~\ref{fig:x-q2-epeA_combo} (left) also shows the $x$-$Q^2$ values for which data are available from fixed target DIS polarized \ep\ experiments as well as from polarized \pp\ collisions at RHIC. Correspondingly, Fig.~\ref{fig:x-q2-epeA_combo} (right) shows the $x$-$Q^2$ values for which data are available from fixed target \eA\ collisions. In both cases, for $Q^2 > 1$ GeV$^2$, there are no data below $x\sim 5\cdot 10^{-3}$. Alternately, for $Q^2 =1$ GeV$^2$, the kinematic reach of the EIC would exceed extant world data by nearly two orders of magnitude for polarized \ep\ scattering and a factor of 50 for \eA\ collisions. Thus, a region that is currently {\it terra incognita} for the extraction of gluon distributions and for the study of gluon saturation will become available for precision measurements at the EIC. 

\begin{figurehere}
	\vspace{4mm}
	\centering \includegraphics[width=\columnwidth]{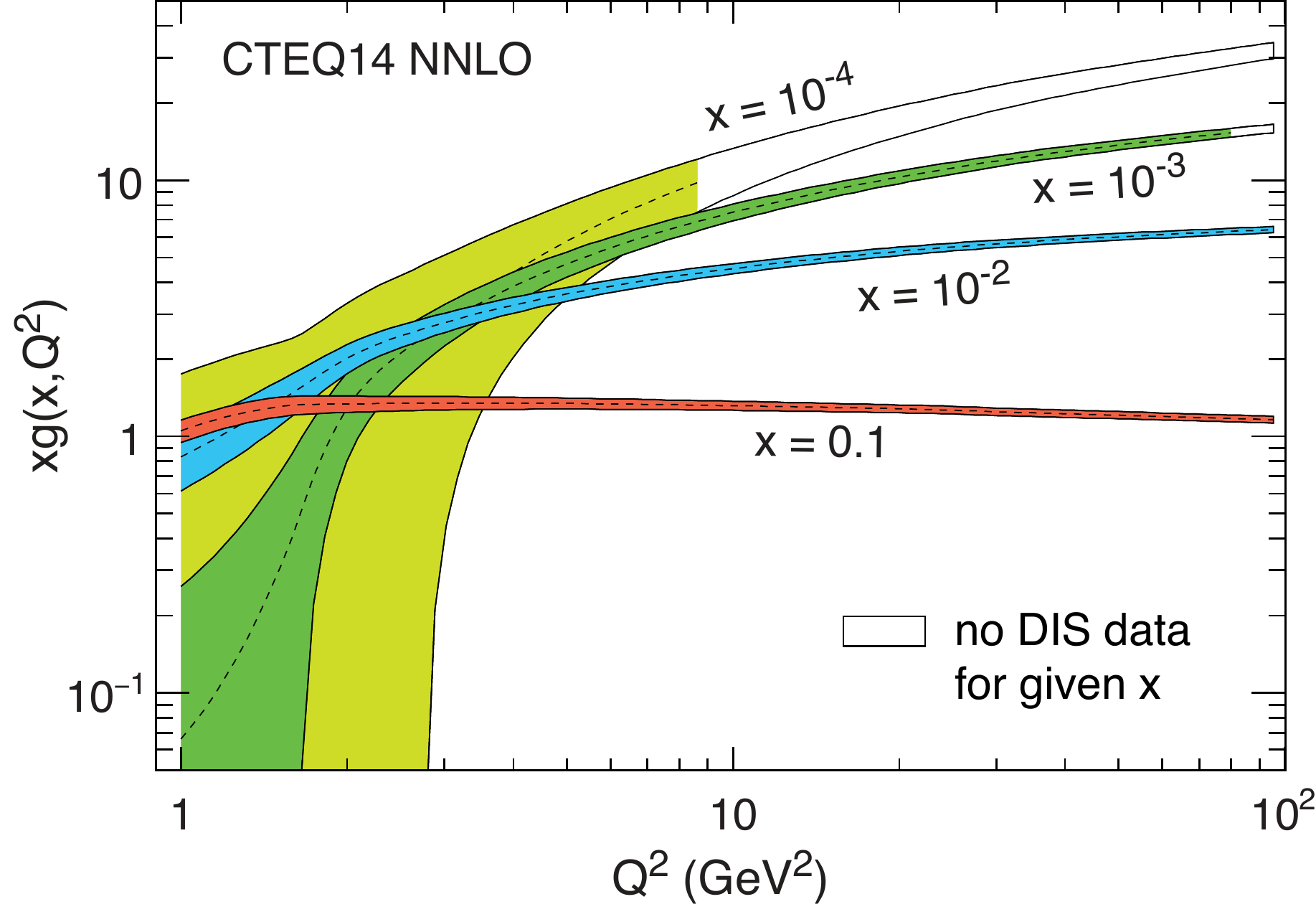}
	\vspace{-6mm}
	\caption{Proton PDFs of gluons as functions of $Q^2$ for various $x$ values
		as derived by the CTEQ  collaboration in NNLO  \cite{Dulat:2015mca}.The bands indicate the uncertainties in our knowledge of gluon PDFs. They are colored in the range where the relevant DIS data (HERA) is available.}
	\label{fig:CTEQ14-Q2dep_ep_combo}
	\vspace{4mm}
\end{figurehere} 

Even though gluons, unlike quarks, do not couple directly to electromagnetic probes, we can learn about their properties from ``scaling violations".  These in particular describe changes in quark distributions with $Q^2$ and Bjorken $x$. 
\begin{figure*}[t!]
	\centering \includegraphics[width=\linewidth]{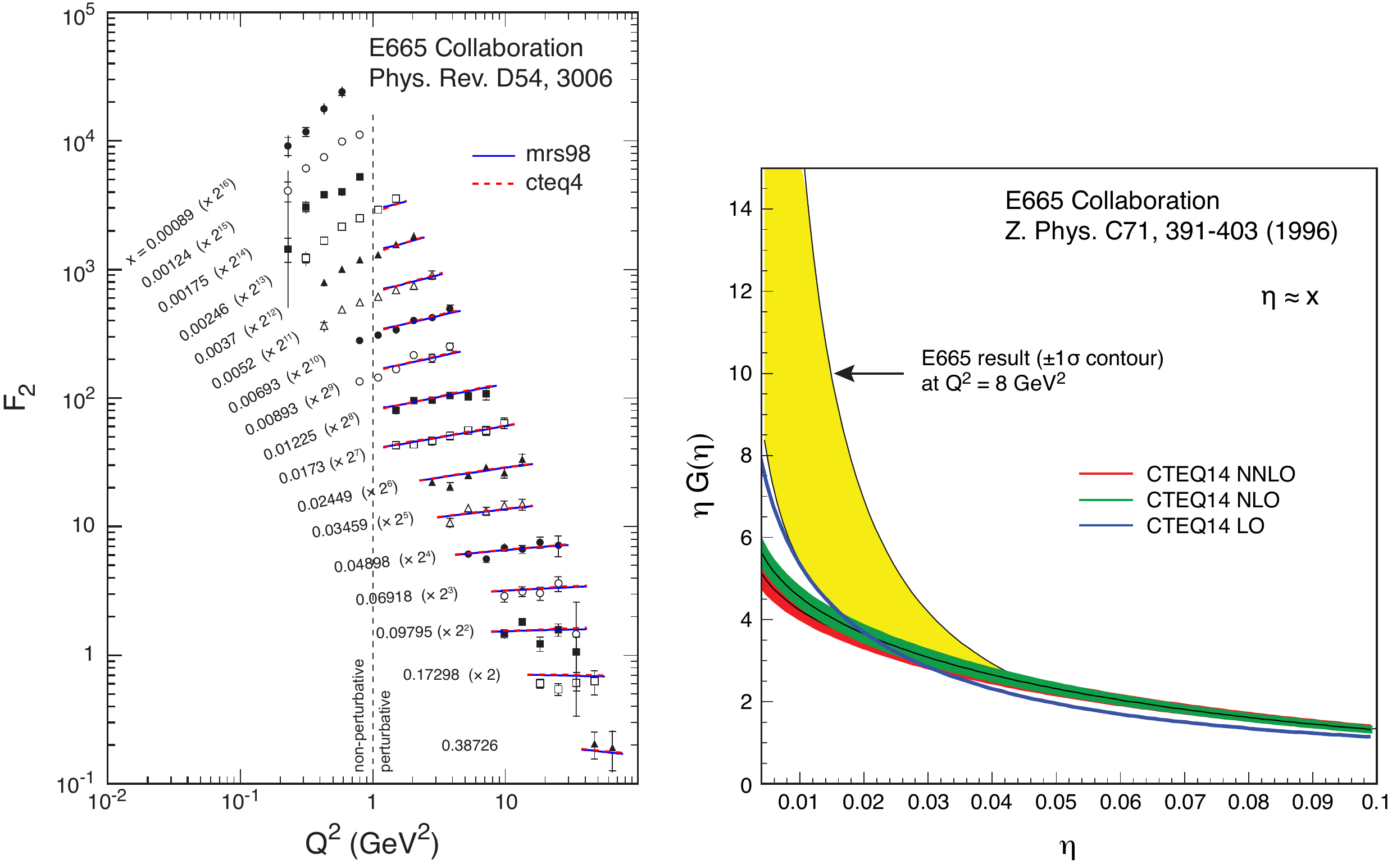}
	\caption{{\it Left:} Structure function $F_2$ of the proton as a function of $Q^2$ for various $x$ values measured by the E665 experiment at $\sqrt{s}=31$ GeV.
		{\it Right:} Gluon distribution derived from the E665 $F_2$ data (depicted on the left) as a function of $x$ for 
		$Q^2 = 8$ GeV$^2$ compared to results from state-of-the-art gluon PDFs  \cite{Dulat:2015mca}. }
	\label{fig:E665_F2_gluon-combo}
\end{figure*} 
The  evolution of  gluon distributions with $Q^2$ extracted from these scaling violations is described by the DGLAP renormalization group equations (RGE)~\cite{Gribov:1972ri,Altarelli:1977zs,Dokshitzer:1977sg} of perturbative QCD (pQCD). The renormalization group flow of information is towards smaller $x$ and larger $Q^2$.  
A wide lever arm in $Q^2$ is essential for the extraction of parton distributions while a wide coverage in $x$
is mandatory to access a broad dynamical regime. 

We see from  Fig.~\ref{fig:x-q2-epeA_combo} that the difference between the high and low energy ranges shown corresponds to a  factor of 5 increase in $x$ reach for a fixed $Q^2$, and likewise, a factor of 5 increase in $Q^2$ reach for fixed $x$. DIS measurements with data collected in this additional area can further constrain significantly nuclear gluon distributions and their extrapolations (via DGLAP evolution) to small $x$.

An example of lessons from electron-proton collisions at HERA for the EIC is depicted in Fig.~\ref{fig:CTEQ14-Q2dep_ep_combo}. The gluon distribution is parametrized at a low momentum resolution scale using the HERA electron-proton inclusive reduced cross-section (see Eq.~\ref{Eq:StructFunc}) data and evolved, using the DGLAP RGE, from this low $Q^2$ scale towards higher $Q^2$ at fixed $x$. We see that the next-to-next-to leading order (NNLO) DGLAP evolution performed by the  CTEQ collaboration \cite{Dulat:2015mca} generates gluon distributions to good accuracy for $x=0.1$ and $x=0.01$ in the entire $Q^2$ range plotted. However, at the smaller $x=10^{-3}$ and   $x=10^{-4}$ values, the  gluon distribution shows larger uncertainties, {\it especially  at the small $Q^2$ where  high quality data exist}. The precision of low $Q^2$ data in this case is ineffectual due to the lack of data at the larger $Q^2$ where the DGLAP RGE is initialized and evolved from. In contrast, the gluon distribution at  larger $x$ values is well constrained over the range shown by virtue of the larger $Q^2$ lever arm.  This example illustrates why a greater EIC energy will not only improve our knowledge of the gluon distribution over a wider $Q^2$ range but also more precisely in the range that is already accessible at lower energies.

The lesson drawn, of the importance of expanded reach in $x$ and $Q^2$, is starker and more pertinent for \eA\  collisions at the EIC.  In this case, the parametrization of the data at the initial scale will not have the $x$-$Q^2$ reach of \ep\ collisions at HERA. To illustrate this, Fig.~\ref{fig:E665_F2_gluon-combo} (left) shows the structure function $F_2$ extracted for the $x$-$Q^2$ reach of the fixed target E665 data at $\sqrt{s}=31$ GeV. Though it is for \ep\ data, it holds an important lesson for the lower  \eA\ center-of-mass energy of 40 GeV. Scaling violations, the variation of $F_2^p(x,Q^2)$ with $Q^2$, are clearly visible only for $x\leq 0.01$. The small $x$ and large $Q^2$ region therefore governs the precision with which gluon distributions can be extracted from these scaling violations.

The extracted proton gluon distribution is shown in Fig.~\ref{fig:E665_F2_gluon-combo} (right). 
It  has large uncertainties at $Q^2=8$ GeV$^2$ and $x=0.01$.  These are significantly larger than the CTEQ evolved gluon distribution shown previously that utilizes HERA data, which had a much wider coverage in $x$ and $Q^2$. For nuclear gluon distributions, the  companion plot to Fig.~\ref{fig:CTEQ14-Q2dep_ep_combo}, demonstrating the precision to which these can be extracted at the two representative center-of-mass energies, is shown and discussed in Sec.~\ref{section:nuclearStructureFunctions}. 

\end{multicols}

\FloatBarrier

		
	\subsection{Discovery Science: Lessons from Heavy-Ion Collisions}
		\label{sec:lessonsRHIC}
\begin{multicols}{2}
The energy, luminosity and kinematic reach needed for ground breaking discoveries are in advance uncertain by their very nature. In the previous section, we discussed some lessons from past DIS experiments. Experience drawn from QCD discoveries in related sub-fields can also provide important guidance. We will discuss here the experience gained from heavy-ion collisions, in particular the discovery of the Quark Gluon Plasma.  

\begin{figurehere}
	\begin{center}
	\includegraphics[width=\columnwidth]{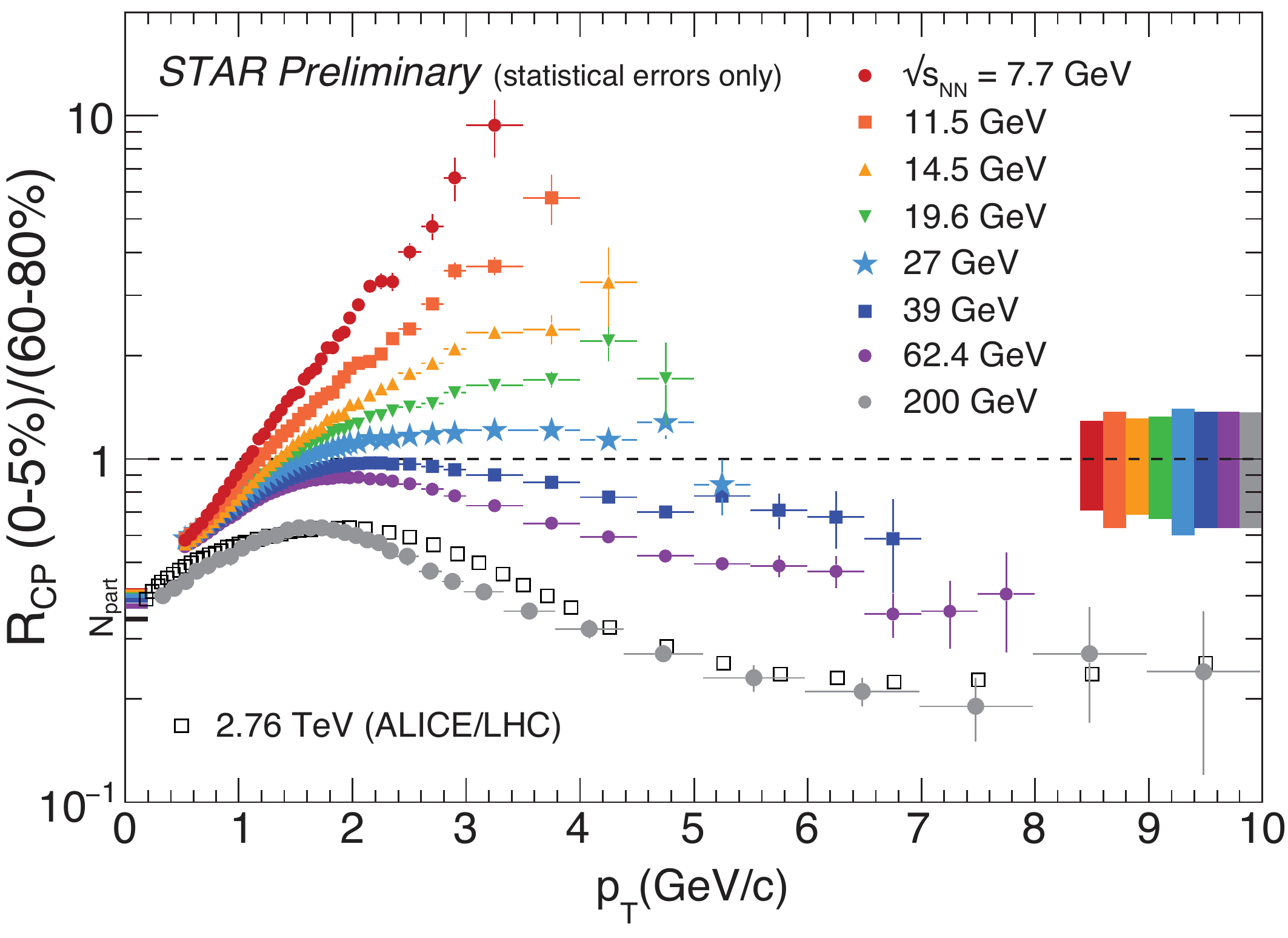}
	\end{center}
	\vspace{-6mm} 
	\caption{Charged hadron suppression, expressed by $R_{CP}$,  measured at RHIC energies, \sqrtsNN\ = 11.5--200 GeV and at the LHC at \sqrtsNN\ = 2.76 TeV. The error bands at unity on the right side of the plot correspond to the $p_T$ independent systematic uncertainties.}
	\label{fig:Rcp_RHIC_LHC}
\end{figurehere}
\vspace{3mm} 

Compelling evidence that there was striking new physics occurring in heavy-ion collisions at RHIC was provided by the discovery of ``jet quenching"~\cite{Adcox:2001jp,Adler:2002xw,Adler:2002tq}. The observable was the ratio $R_{\rm CP}$ of the inclusive hadron spectrum in central \AuAu\ relative to that in peripheral \AuAu\ collisions, each of them normalized by the respective number of collisions. The absence of nuclear effects corresponds to $R_{\rm CP}=1$.  Figure~\ref{fig:Rcp_RHIC_LHC} plots $R_{\rm CP}$ for a wide range of center-of-mass energies per nucleon at RHIC, from \sqrtsNN\ = 11.5 GeV to \sqrtsNN\ = 200 GeV. Also shown in the plot is $R_{\rm CP}$ from the LHC at \sqrtsNN\ = 2.76 TeV.

At CERN's SPS, the highest energy heavy-ion experiments before RHIC, heavy-ion beams on fixed targets had a maximal per nucleon energy of 158 GeV, corresponding to \sqrtsNN\ = 17.2 GeV. This is close to the  \sqrtsNN\ = 19.6 GeV RHIC curve in Fig.~\ref{fig:Rcp_RHIC_LHC} \cite{Horvat:2016vxp,Aamodt:2010jd,Adams:2003kv}. At these energies, one observes a depletion at small $p_T$ and an enhancement at large $p_T$, a phenomenon often called the ``Cronin Effect". An explanation of this effect is that the multiple scattering of partons depletes the lower $p_T$ part of the spectrum and shifts their average momenta to higher $p_T$. At RHIC, for \sqrtsNN\ = 200 GeV, $R_{\rm CP}\sim 0.2$ --a decrease of a factor of 5 from unity. The observation of this large suppression indicated that high $p_T$ partons were suffering significant energy loss in their interactions with the medium and thereby provided strong evidence that a QGP had been created. More differential measurements, fortified by more refined theory computations of jet quenching, have subsequently reinforced the early role played by $R_{\rm CP}$ as a key measurement in the discovery of the strongly interacting QGP.

While jet quenching is unambiguous at \sqrtsNN\ = 200 GeV, it is also noteworthy in retrospect that quenching is seen marginally at \sqrtsNN\ = 39 GeV and more clearly at \sqrtsNN\ = 62.4 GeV. However, the evidence for this result requires significant $p_T$ reach; quenching is seen clearly only above $p_T=3$ GeV. Since physics at higher $p_T$ is luminosity hungry, it took over a decade of further running and a beam energy scan to obtain the beautiful systematic behavior established by Fig.~\ref{fig:Rcp_RHIC_LHC}. 

Increasing the energy of the ion beams beyond the original design values is prohibitively expensive. Consequently, it was important that RHIC's design energy enabled a  ``day one" discovery. Note that the suppression of $R_{\rm CP}$ shown in  Fig.~\ref{fig:Rcp_RHIC_LHC} is maximal at the highest RHIC energy and is not exceeded by data from the LHC, as also shown in the figure. The fact that the highest RHIC energy saturates jet quenching is remarkable and was not anticipated in early models of jet quenching. 

Continual luminosity upgrades, resulting from experience with the RHIC accelerator, have led over time to more than a forty-fold increase in luminosity over its design value. This experience appears to be a generic feature of high energy colliders. In addition, the RHIC collider has shown tremendous versatility in running at a variety of energies, employing ion species from the lightest ions to Uranium. This versatility, and the implementation of electron cooling of heavy-ion beams to further increase the luminosity in RHIC's BES II phase, is key to sustaining its future discovery potential. 

RHIC was the world's first heavy-ion collider; the EIC will be the world's first electron-nucleus and polarized electron-polarized proton collider. A key lesson from RHIC's success and future potential in exploring the ``hot and dense QCD" phase diagram is that the discovery of similarly novel many-body dynamics in the QCD landscape sketched in Fig.~\ref{fig:bigpicture} requires a significant energy range and reach well beyond those of prior fixed-target DIS machines.

In a DIS collider, the kinematic equivalent of varying the center-of--mass energy in heavy-ion collisions is the expanded range in the Bjorken variable $x$ for fixed $Q^2$. In polarized electron-polarized proton collisions, for fixed $Q^2$, the reach in $x$ at the highest proposed EIC energy is two orders of magnitude greater than at fixed-target DIS experiments. For DIS collisions off heavy nuclei, at fixed $Q^2$, the range in $x$ is a factor of 50 greater than available at fixed-target machines. This extended reach and range is comparable to that of RHIC relative to prior fixed-target heavy-ion experiments. 

The success of RHIC suggests that the discovery potential of a high energy EIC is high and that energy reach and range plays a pivotal role. Very little is known about spin and angular momentum of quarks and gluons at small $x$, as well as their transverse spatial and momentum distributions. Nuclear gluon distributions are {\em terra incognita}, as is our understanding of the propagation of jets and heavy quarks in cold nuclear matter. We shall now demonstrate, with a simple case study, the utmost importance of energy range and reach for the physics of gluon saturation. 
\end{multicols}

\FloatBarrier


    \subsection{Case Study: Accessing the Saturation Regime}
		\label{sec:dipole}
\begin{multicols}{2}

\begin{figure*}[t!]
	\begin{center}
				\includegraphics[width=0.8\textwidth]{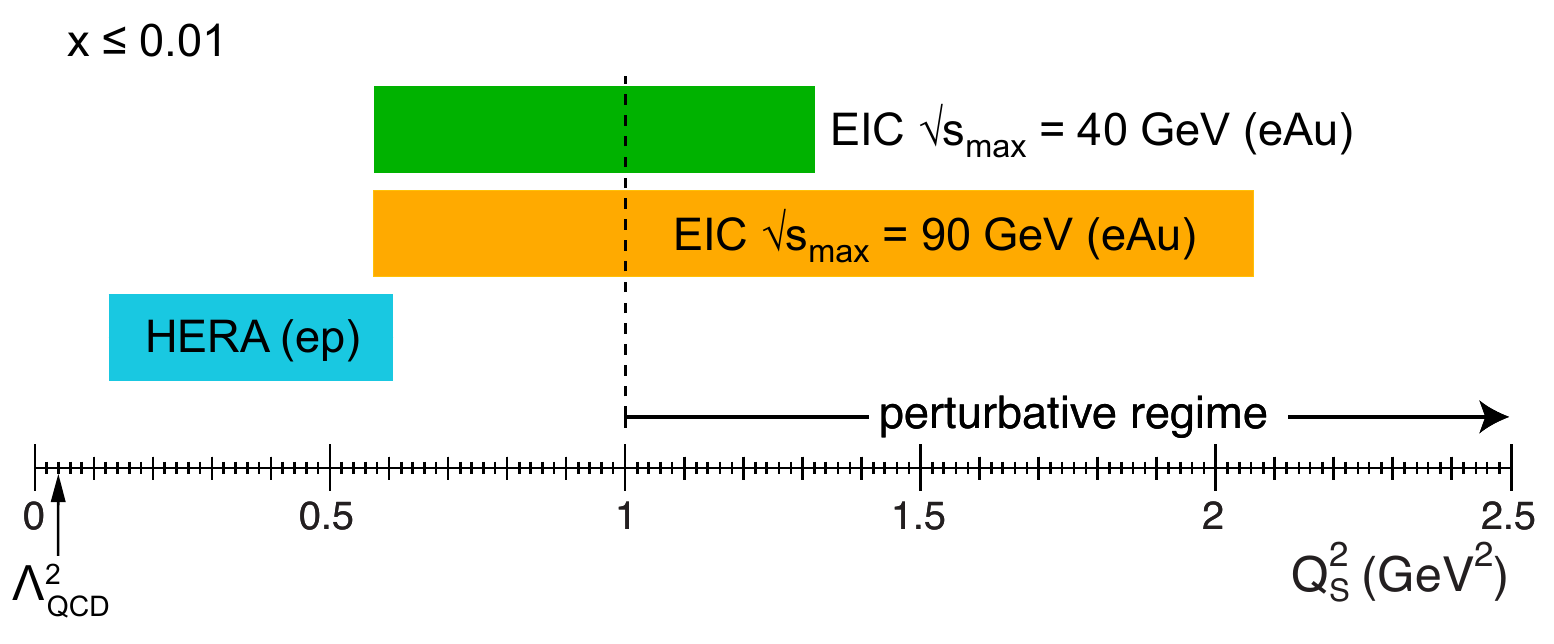}
	\end{center} 
	\vspace{-7mm}
	\caption{\label{fig:QsReach_eRHIC_JLEIC} Accessible values of the saturation scale $Q_s^2$ at an EIC in \eA\ collisions  assuming two different maximal center-of-mass energies. The reach in $Q_s^2$ for \ep\ collisions at HERA is shown for comparison.}
\end{figure*} 

In Fig.\,\ref{fig:QsReach_eRHIC_JLEIC}, the saturation scales $Q_s^2$ for \eA\ collisions at an EIC, with two different maximal $\sqrt{s}$, are compared to the values achievable in \ep\ collisions at HERA. The first point to note is that the projected saturation scales in \eA\ collisions at both EIC energies are significantly larger than those in \ep\ collisions, even though the HERA $\sqrt{s}$ value is approximately eight times greater than the lower EIC energy. This enhancement of $Q_s^2$ in nuclei is a striking consequence of the high energy DIS probe interacting simultaneously with partons in different nucleons along its path through the nucleus.

We also observe that the maximal $Q_s^2$ in \eA\ collisions at the EIC is approximately 50\% larger for the higher energy of $\sqrt{s}_{\rm max}=90\,{\rm GeV}$, compared to $\sqrt{s}_{\rm max}=40\,{\rm GeV}$. The  difference in $Q_s^2$ may appear relatively mild but we will demonstrate in the following that this difference is sufficient to generate a dramatic change in DIS observables with increased center-of-mass energy. This is analogous to the message from Fig.~\ref{fig:Rcp_RHIC_LHC} where we clearly observe the dramatic effect of jet quenching once \sqrtsNN\ is increased from 39 GeV to 62.4 GeV and beyond. 

To compute observables in DIS events at high energy, it is advantageous to study the scattering process in the rest frame of the target proton or nucleus. In this frame, the scattering process has two stages. The virtual photon first splits into a quark-antiquark pair (the color dipole), which subsequently interacts with the target. This is illustrated in Fig.~\ref{fig:DipoleScatAmp}. Another simplification in the high energy limit is that the dipole does not change its size $r_\perp$ (transverse distance between the quark and antiquark) over the course of the interaction with the target.

Multiple interactions of the dipole with the target become important when the dipole size is of the order $|\vec{r}_\perp|\sim 1/Q_s$.
In this regime, the imaginary part of the dipole forward scattering amplitude $N(\vec{r}_\perp, \vec{b}_\perp, x)$, where $\vec{b}_\perp$ is the impact parameter, takes on a characteristic exponentiated form~\cite{Mueller:1989st}:
\begin{equation}\label{eq:dipoleampl}
  N = 1 - \exp\left( -\frac{r_\perp^2 Q_s^2(x,\vec{b}_\perp)}{4} \ln \frac{1}{r_\perp \Lambda} \right),
\end{equation} 
where $\Lambda$ is a soft QCD scale.

At high energies, this dipole scattering amplitude enters all relevant observables such as the total and diffractive cross-sections. It is thus highly relevant how much it can vary given a certain collision energy. If a higher collision energy can provide access to a significantly wider range of values for the dipole amplitude, in particular at small $x$, it would allow for a more robust test of the saturation picture.

\begin{figurehere}
	\centering
	\includegraphics[width=0.5\columnwidth]{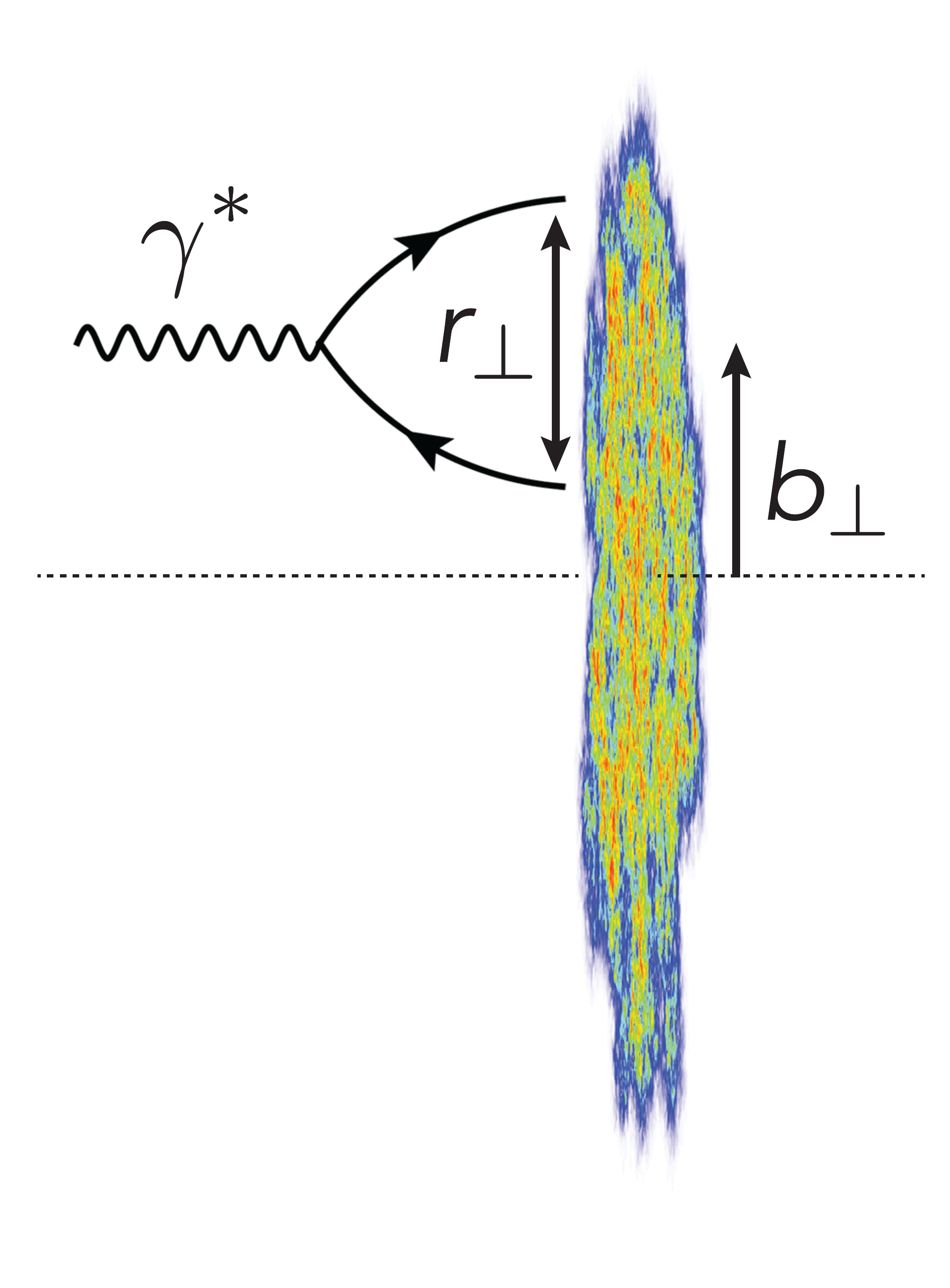}
	\vspace{-7mm}
	\caption{\label{fig:DipoleScatAmp} The forward scattering amplitude for DIS on a nuclear target. The virtual photon splits into a $q\bar{q}$ pair of fixed size $r_\perp$, which then interacts with the target at impact parameter $b_\perp$.}
	\vspace{5mm}
\end{figurehere}

To study the effect of a varying reach in $Q^2$, one may, to good approximation, replace $r_\perp$ in (\ref{eq:dipoleampl}) by the typical transverse resolution scale $2/Q$ to obtain the simpler expression $N \sim 1 - \exp\left\{-Q_s^2/Q^2 \right\}$. The appearance of both $Q_s^2$ and $Q^2$ in the exponential is crucial. Its effect is demonstrated in Fig.\,\ref{fig:q2qs_shaded}, where the dipole amplitude $N$ is plotted as a function of $Q^2$ for fixed $x=10^{-3}$. While the variation of this quantity within the $Q^2$ reach at  $\sqrt{s}=40$ GeV is only approximately 20\%, it reaches up to a factor of 6 within the $Q^2$ reach at $\sqrt{s}=90$ GeV. To further draw on the analogy we raised previously with RHIC, the 20\% suppression in $R_{\text CP}$ seen at $\sqrtsNN =39$ GeV is not robust given the systematic uncertainties shown in Fig.~\ref{fig:Rcp_RHIC_LHC}. In contrast, the factor 5 suppression seen at $\sqrtsNN =200$ GeV, is an unambiguous signature of discovery. 

As also shown in Fig.~\ref{fig:q2qs_shaded}, this effect is further enhanced for diffractive events. In that case, the {\it square} of the dipole amplitude enters the cross-section. This quantity changes by a factor 25 over the $Q^2$ range at $\sqrt{s}=90\,{\rm GeV}$, but only by 1.7 for the lower center-of-mass energy. See Section~\ref{subsection:diffraction} for further discussion.

In addition to the wider range in $Q^2$, the wider reach in $Q_s^2$ (see Fig.~\ref{fig:QsReach_eRHIC_JLEIC}) will allow for measurements with a similar lever arm in $Q^2$ also at lower values of $x$. Therefore in this simple dipole model case study, we have demonstrated that because the higher center-of-mass energies provide a significantly broader reach in $x$ and $Q^2$, saturation effects can be large in DIS even if the $Q_s^2$ values do not differ widely. This is crucial for probing the physics of gluon saturation and for testing as well as constraining existing models.

\begin{figurehere}
	\vspace{4mm}
	\centering
	\includegraphics[width=\columnwidth]{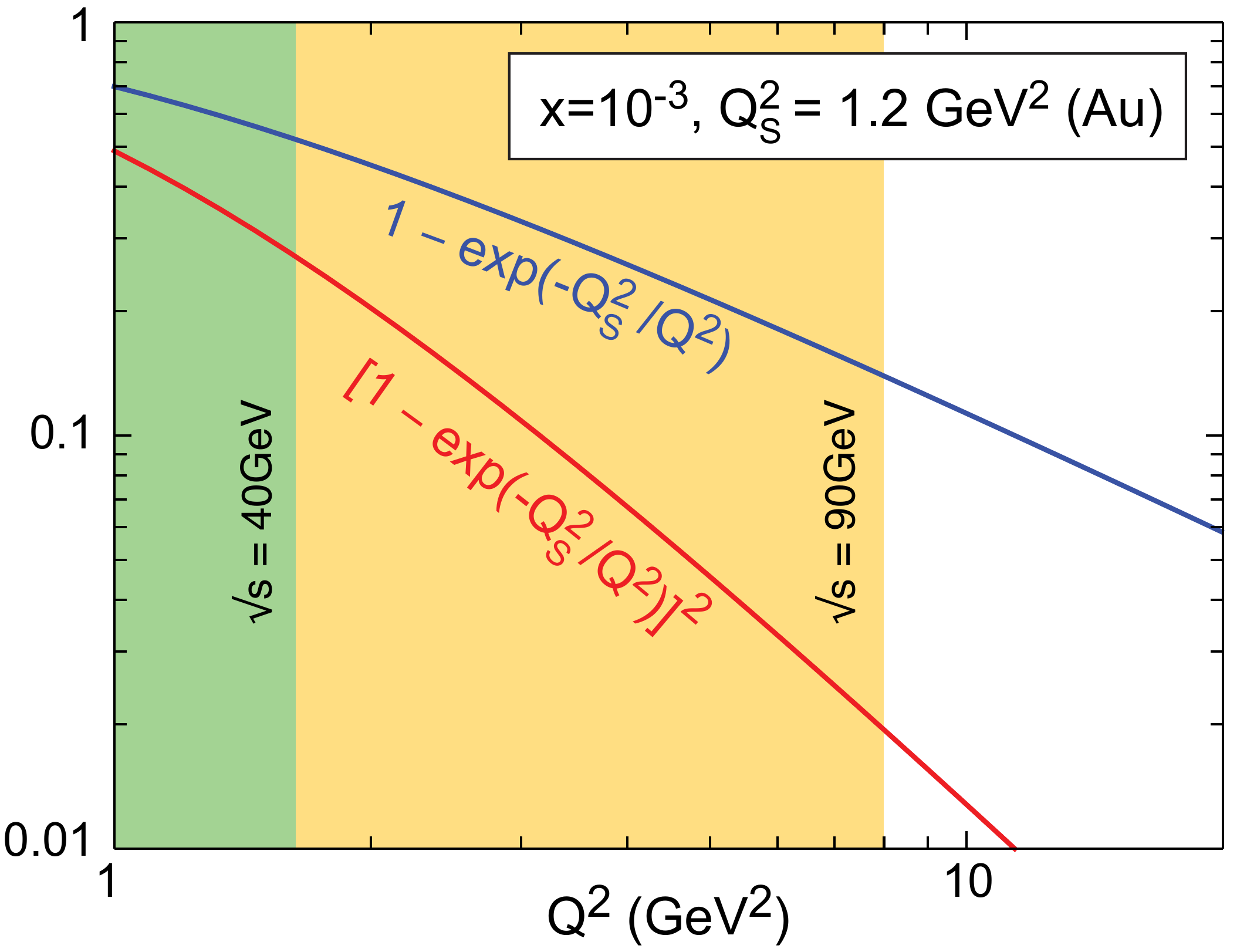}
	\caption{\label{fig:q2qs_shaded} The curve in blue depicts the variation in the dipole scattering amplitude as a function of $Q^2$. Inclusive DIS cross-sections are sensitive to this quantity. The shaded areas indicate the $Q^2$ reach at two different center-of-mass energies. The curve in red depicts the variation in the square of the dipole scattering amplitude with $Q^2$. This quantity enters diffractive cross-sections -- see Section~\ref{subsection:diffraction}.}
\end{figurehere}

\end{multicols}

\FloatBarrier


%
%
\section{Impact of Collision Energy on Key Measurements}
			\label{chapter:energyCase}
In the following, we will examine the energy requirements of key measurements within the range of energies, and assuming the acceptance of a model detector, as outlined in the EIC White Paper.  Subsections~\ref{section:spinSection} and~\ref{section:imaging} respectively present the case for precision studies of the spin of the proton and the three dimensional imaging of parton distributions within.  Subsections~\ref{section:nuclearStructureFunctions} and~\ref{section:chargeSection} focus on parton distributions in the nuclear medium. Measurements that are sensitive to gluon saturation are discussed in subsection~\ref{section:saturation}. We conclude in subsection~\ref{section:jets} by presenting a novel study of jets at an EIC.

	\subsection{Precision Measurement of the Proton Spin}
			 \label{section:spinSection}
\begin{multicols}{2}

\begin{figure*}[t!]
	\center
	\includegraphics[width=1.\linewidth]{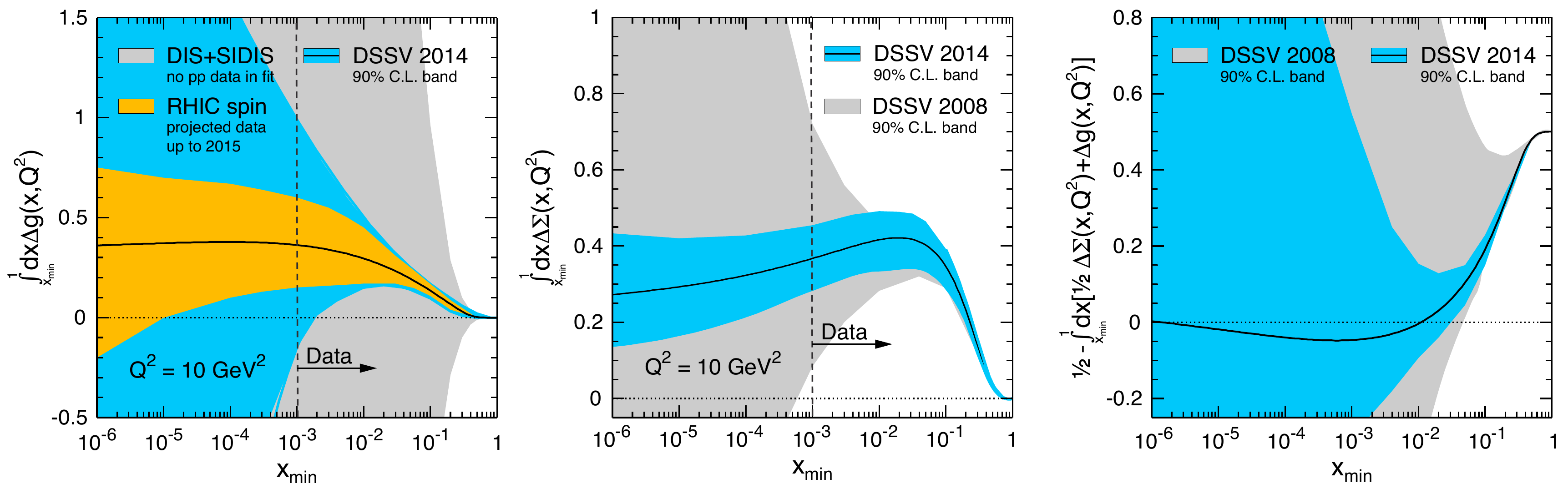}
	\vspace{-6mm}
	\caption{90\% C.L. uncertainty estimates for the running integrals of the gluon helicity ({\it left}), quark helicity ({\it middle}), and orbital angular momentum ({\it right})  distribution at $Q^2 = 10~\text{GeV}^2$ as a function of $x_\textrm{min}$. The gray-shaded band denotes the DSSV08~\cite{deFlorian:2009vb} fit which includes primarily DIS data. The blue-shaded band is based on the DSSV14 fit~\cite{deFlorian:2014yva}, which includes polarized \pp\ data from RHIC collected prior to 2012. The yellow-shaded band is a projection, which accounts for the most recent RHIC data~\cite{Aschenauer:2015ata}. The region constrained by current data lies to the right of the vertical dashed lines.}
	\label{Fig:Spin1}
\end{figure*}

Understanding how the spin of the proton emerges from the properties and dynamic interactions of its constituents is an outstanding puzzle in hadronic physics and a key motivation for the realization of a  polarized EIC. This topic is addressed in Sec.~2 of the EIC White Paper~\cite{Accardi:2012qut}. 
The extent to which quarks and gluons with a given momentum fraction $x$ have their spins aligned with the spin direction of a nucleon is encoded in the helicity dependent parton distribution functions.  Knowledge of these fundamental quantities, along with estimates of their uncertainties, is gathered from comprehensive QCD analyses~\cite{deFlorian:2008mr,deFlorian:2009vb,deFlorian:2014yva} of all available data taken in spin-dependent DIS and proton-proton collisions, with and without additional identified hadrons in the final state. 
By integrating these polarised parton distributions over the momentum fraction $x$ from 0 to 1 at a fixed $Q^2$, the spin of the proton can be written in terms of its constituents using the Jaffe--Manohar sum rule~\cite{Jaffe:1989jz}
\begin{eqnarray}
\frac{1}{2} &=& \frac{1}{2} \int_{0}^{1}\text{d}x\Delta\Sigma\left(x,Q^{2}\right) + \nonumber \\
&&\int_{0}^{1}\text{d}x\Delta g\left(x,Q^{2}\right) + {\cal{L}}(Q^2)\,, 
\label{Eq.sumrule}
\end{eqnarray}

\noindent
where $\frac{1}{2}\Delta\Sigma(x,Q^{2})$ represents the quark helicity contribution, and $\Delta g(x,Q^{2})$ represents the gluon helicity contribution to the total spin of the proton. The respective orbital angular momenta of quarks and gluons are represented by ${\cal{L}}(Q^2) = \sum_q\left[L_q(Q^2)+L_{\bar{q}}(Q^2)\right] + L_g(Q^2)$. 

Figure~\ref{Fig:Spin1} shows
an extraction of the integrals of the quark and gluon contributions in Eq.~\ref{Eq.sumrule}, running between $x=x_\textrm{min}$ and $x=1$ with their 90\% confidence level (C.L) uncertainties.
The gray-shaded band is the outcome of the DSSV08~\cite{deFlorian:2009vb} analysis, which is almost exclusively based on the existing DIS data. The blue-shaded band shows the result of the DSSV14~\cite{deFlorian:2014yva} fit, which includes polarized \pp\ data from RHIC. The yellow-shaded region shows the projected constraints on the parton distributions once all RHIC data collected through 2015 is included. In the plots, the region to the right of the dashed vertical line is constrained by current data. It is clear that precision data are needed to determine the parton contribution to the proton's spin, especially at low $x$. 

\begin{figurehere}
	\begin{center}
		\includegraphics[width=\columnwidth]{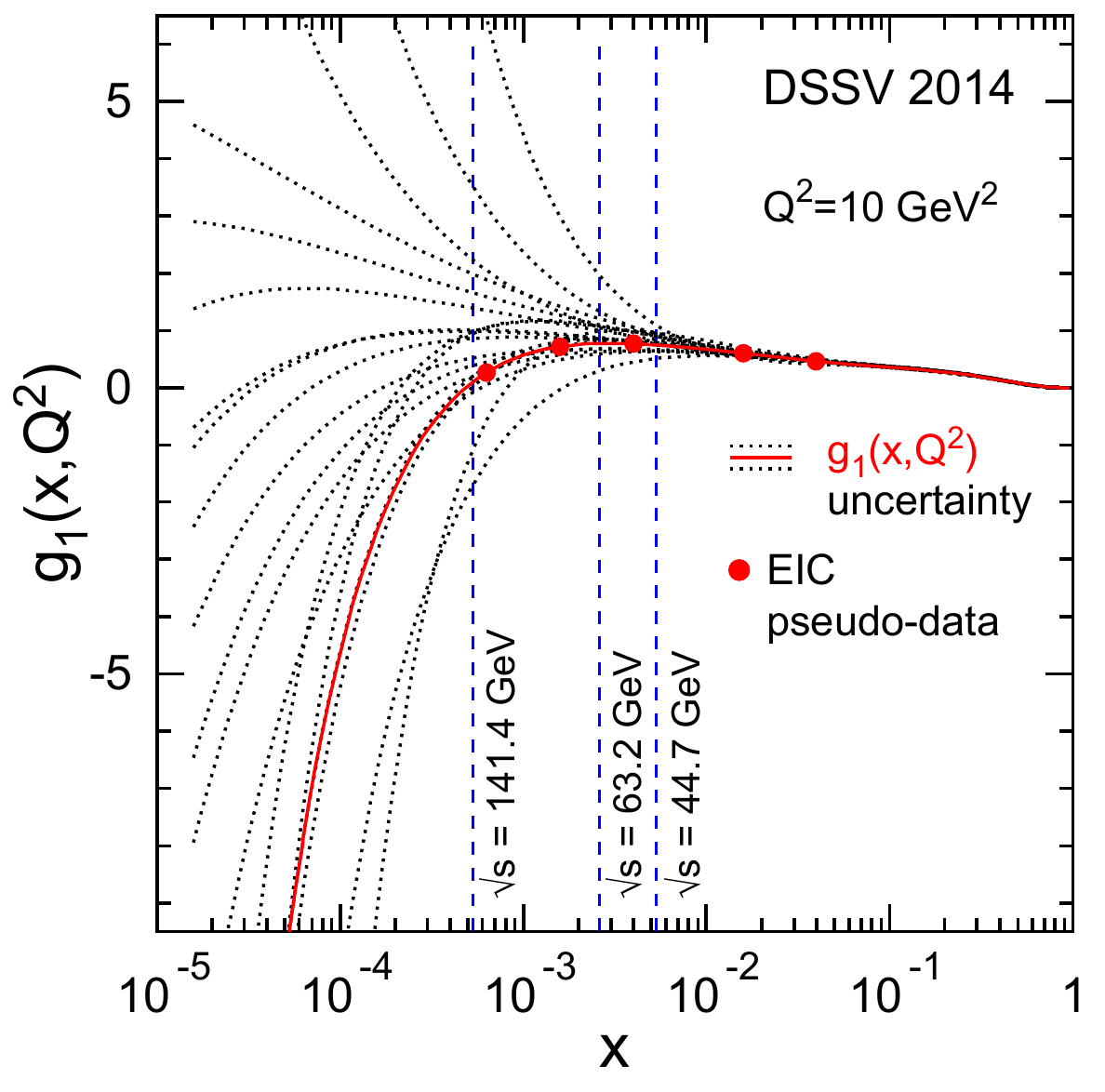} 
	\end{center}
	\vspace{-5mm}
	\caption{Present knowledge of the evolution in $x$ of the structure function $g_{1}$, based on the DSSV14 extraction~\cite{Aschenauer:2015ata}. The dotted lines show the results for alternative fits that are within the 90\% C.L. limit.}
	\label{Fig:g1-0} 
	\vspace{4mm}
\end{figurehere}

The fraction of the spin from angular momenta can be obtained by subtracting $\frac{1}{2}\Delta\Sigma(Q^{2})$ and $\Delta G(Q^{2})$ from the total spin of the proton, using the sum rule in Eq.~\ref{Eq.sumrule}. The right panel in Fig.~\ref{Fig:Spin1} shows how the angular momenta contribution is totally unconstrained at moderate to low $x$. A key observable in disentangling the various parton contributions to the proton spin is the polarized structure function $g_{1}(x,Q^{2})$.  It is proportional to the difference of the neutral current cross-sections of DIS events, with the beams polarized parallel and anti-parallel in the longitudinal direction,
 \begin{equation}
 \frac{1}{2} \left[ \frac{\mathrm{d}^{2}\sigma^{\leftrightarrows}}{\mathrm{d}x\mathrm{d}Q^{2}} - \frac{\mathrm{d}^{2}\sigma^{\rightrightarrows}}{\mathrm{d}x\mathrm{d}Q^{2}} \right] \simeq \frac{4\pi \alpha^{2}}{Q^{4}}y(2-y) g_{1}(x,Q^2).
 \end{equation}
\begin{figurehere}	
	\begin{center}
	        \includegraphics[width=\columnwidth]{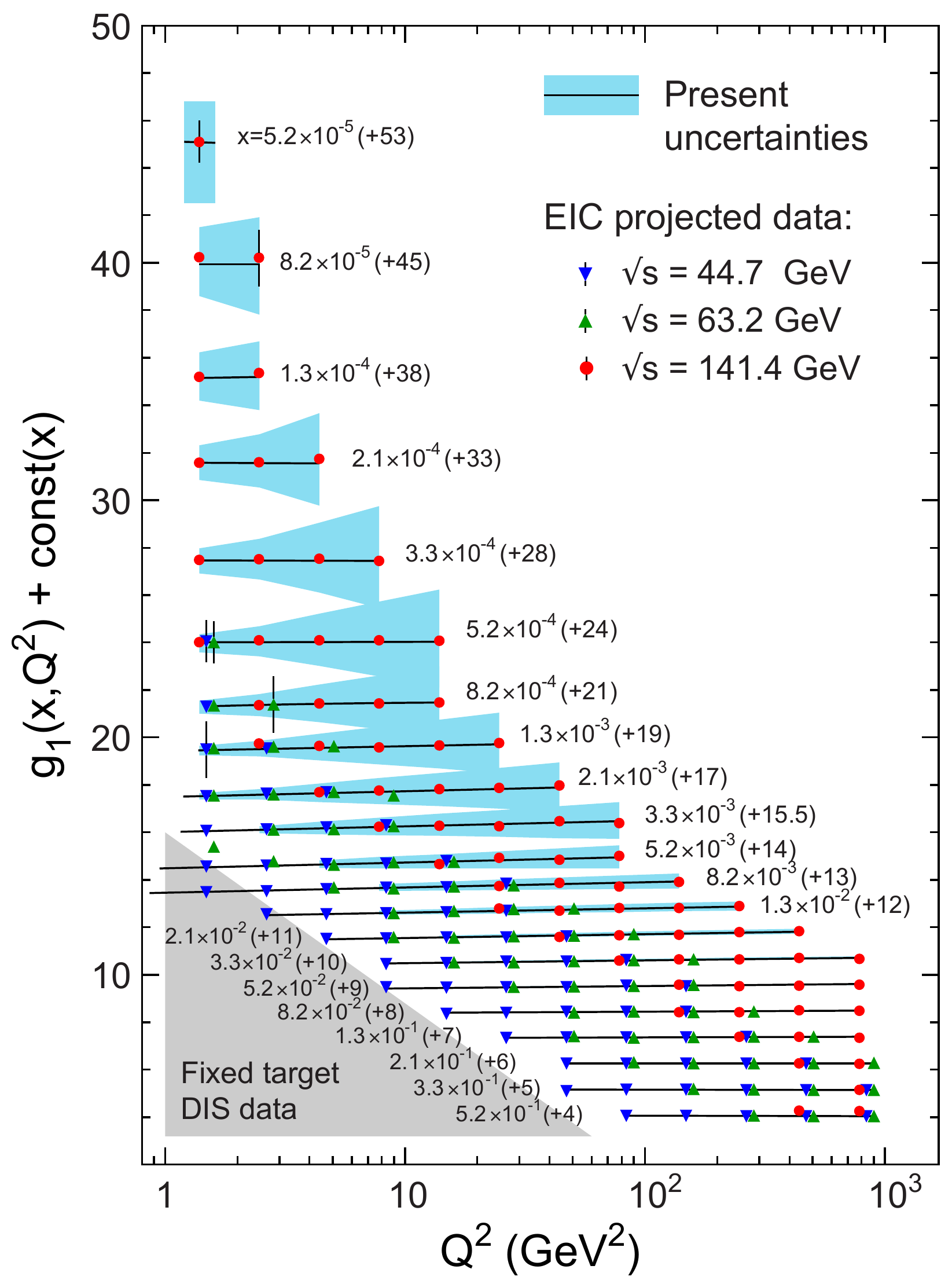}
	\end{center}
	\vspace{-4mm}
	\caption{Projections for the structure function $g_{1}$ at different $\sqrt{s}$, compared with a  model extrapolation and its uncertainties~\cite{deFlorian:2014yva}. The curves correspond to different values of $x$ that are specified next to each curve. For clarity, constants are added to $g_1$ to separate different $x$ bins; moreover, multiple data points in the same $x$-$Q^2$ bin are displaced horizontally. The gray area marks the phase space currently covered by fixed target experiments. See text for details.}
	\label{Fig:g1} 
\end{figurehere}
\vspace{5mm}
The integral of the structure function over $x$ is sensitive to the contribution from the quarks and the derivative versus $Q^2$ is sensitive to the gluon distribution. Therefore $\Delta g(x,Q^{2})$ can be accessed in DIS data via scaling violation fits $\sim \text{d}g_{1}\left(x,Q^{2}\right)/\text{d}\text{ln}Q^{2}$. However, a precise scaling violation fit requires, depending on the respective uncertainties, a sufficiently large lever arm in $Q^{2}$ at any given value of $x$.  
Figure~\ref{Fig:g1-0} shows how the present knowledge of the structure function $g_{1}$ rapidly deteriorates and uncertainties explode at low $x$.
The EIC pseudo-data are depicted by the red data points. The uncertainties
are smaller than the symbols illustrating the enormous constraining power an EIC will have on $g_{1}$.

\vspace{2mm}
\begin{figurehere}
	\begin{center}
		\includegraphics[width=\columnwidth]{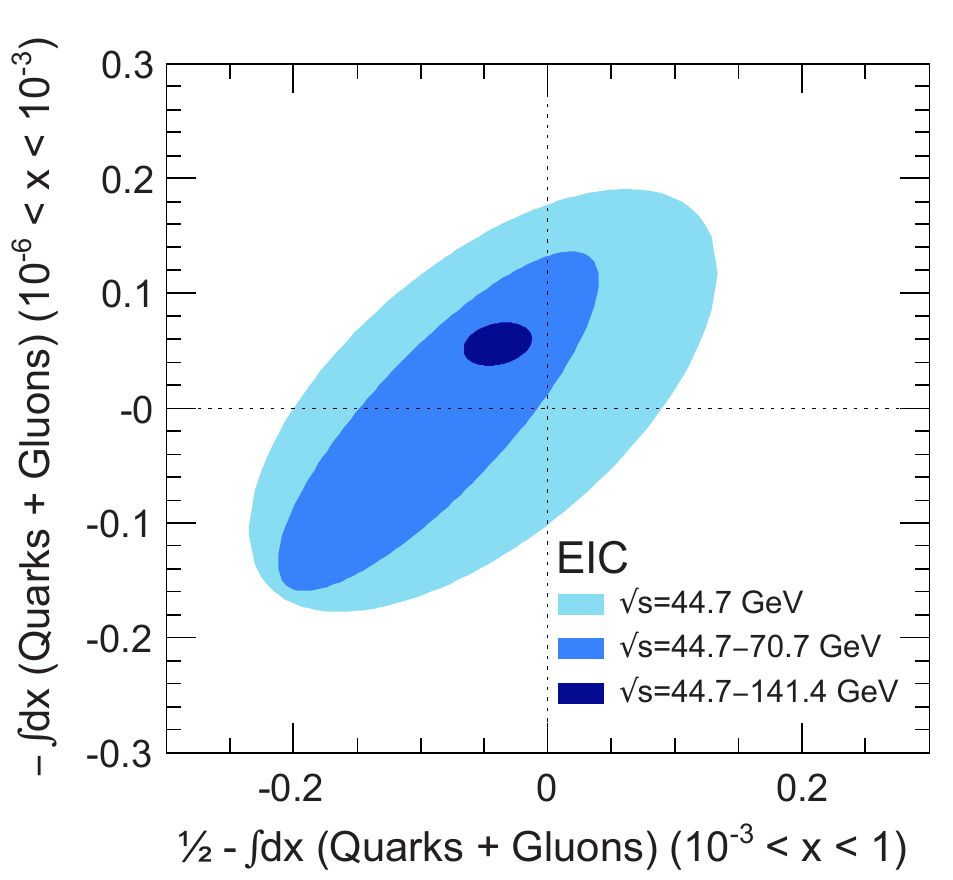}
	\end{center}
	\vspace{-3mm}
	\caption{The EIC's impact on the knowledge of the  integral of the quark and gluon spin contribution in the range $10^{-6}<x<10^{-3}$ ($y$-axis) versus the contribution from the orbital angular momentum  in the range $10^{-3}<x<1$ ($x$-axis).}
	\label{Fig:g1-2} 
\end{figurehere}
\vspace{5mm}

Figure~\ref{Fig:g1} shows the structure function $g_{1}(x,Q^{2})$ in \ep\ collisions at  $\sqrt{s}=44.7, 63.4, 141.4$~GeV from EIC pseudo-data, compared with the phase space currently reached by fixed target experiments. 
The error bars indicate only the statistical precision and correspond to a sampled luminosity of 10~fb$^{-1}$. The uncertainties of the  DSSV14 theoretical prediction~\cite{deFlorian:2014yva} are shown by the blue bands. It is clear that the assumed sampled luminosity is already enough to get really precise measurements, whereas the larger $\sqrt{s}$ extends greatly the reach to lower $x$ values where present uncertainties are large. Given the high statistical precision, it will be critical to constrain experimental systematic uncertainties to below a few percent~\cite{Aschenauer:2015ata}.

Figure~\ref{Fig:g1-2} uses simulated data to clearly demonstrate the EIC's impact on the knowledge of the integral of the proton's quark and gluon spin contributions for $10^{-6} < x < 10^{-3}$ versus the contribution to the orbital angular momentum for the range $10^{-3} < x < 1$.  A dramatic shrinkage of the uncertainties in the parton helicities is seen with the largest energy reach. The underlying reason for this rapid shrinkage can be traced to the very unstable behavior of $g_1(x,Q^2)$ due to the lack of data at small $x$ shown in Fig.~\ref{Fig:g1-0}. Data obtained in the small $x$ region constrain this behavior.

\end{multicols}

\FloatBarrier

	\subsection{Spatial Imaging of Quarks and Gluons}
	       	\label{section:imaging}
\begin{multicols}{2}

The parton structure of the proton changes significantly across the QCD landscape sketched in Fig.~\ref {fig:bigpicture} of Section~\ref{section:landscape}.  We illustrate schematically in Fig.~\ref{Fig:Evol-cartoon} how varying $x$ from high values ($x\sim1$) to low values ($x\sim10^{-4}$) at a given resolution scale $Q^2$ of a few GeV$^2$ reveals the complex many-body structure of quarks and gluons inside the proton. The structure revealed by dialing down in $x$  changes from the valence quark dominated regime, to a regime where the proton's constituents are gluons and sea quark-antiquark pairs generated through QCD radiation,  and finally at small $x$ to an intrinsically nonlinear regime where the gluon density is so large that the gluons radiate and recombine at the same rate.

\begin{figurehere}
	\vspace{3mm}	\begin{center}
		\includegraphics[width=\columnwidth]{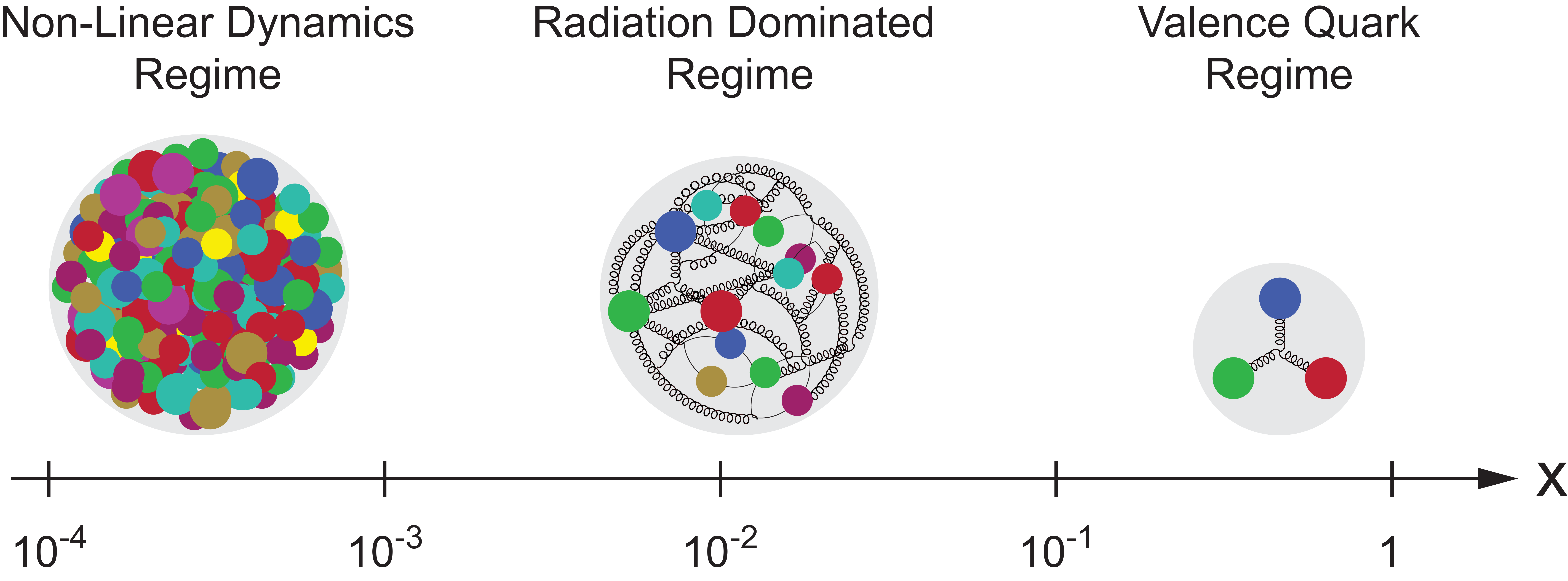} 
	\end{center}
	\vspace{-3mm}
	\caption{The development of the internal quark and gluon structure of the proton going from high to low $x$. Decreasing $x$ corresponds to increasing the center-of-mass energy.}
	\label{Fig:Evol-cartoon}
\end{figurehere}
\vspace{4mm}

High luminosities at the EIC, combined with a large kinematic reach, open up a unique opportunity to go far beyond our present largely one dimensional picture of the proton. It will enable  parton ``femtoscopy'' by correlating information on parton
contributions to the proton's spin with 
their transverse momentum and spatial distributions inside the proton.  Such three dimensional 
 images  have the potential to radically impact our understanding of the confining dynamics of quarks and gluons in QCD.  This is because one will be able to probe, with fine resolution $Q^2$, parton dynamics as a function of impact parameter in the proton, out to length scales where their interactions are no longer weakly coupled but become increasingly strongly coupled generating the phenomena of chiral symmetry breaking and confinement. 
 \begin{figure*}[t!]
 	\center
 	\includegraphics[width=0.9\textwidth]{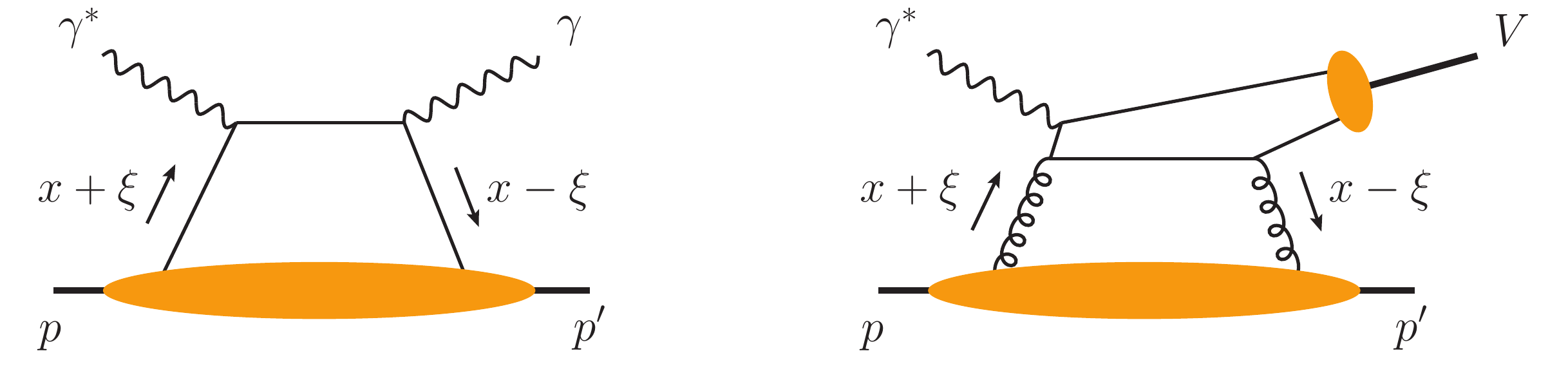} 
 	\caption{Diagrams depicting deeply virtual Compton scattering ({\it left}) and exclusive vector meson production ({\it right}) in terms of GPDs, represented by the yellow blobs. The upper filled oval in the right figure represents the meson wave function. The symbol $\xi$ reflects the asymmetry in the longitudinal momentum fraction of the struck parton in the initial and final state.}
 	\label{Fig:DVMP-diagram}
 \end{figure*}

The three dimensional parton structure of hadrons is uncovered in DIS by measurements of exclusive final states, wherein the proton remains intact after scattering off the lepton probe. The transverse position of the scattered quark or gluon  is obtained by performing a Fourier transform of the differential cross-section $\mathrm{d}\sigma/\mathrm{d}t$, where $t$ is the squared momentum transfer  between the incoming proton and the scattered proton. Examples of exclusive processes are deeply virtual Compton scattering (DVCS) and the exclusive production of vector mesons. These are illustrated in Fig.~\ref{Fig:DVMP-diagram}. 

The nonperturbative quantities that encode such spatial tomographic information are often referred to as Generalized Parton Distributions (GPDs) and are defined at a nonperturbative factorization scale that separates the nonperturbative information encoded from perturbative dynamics at short distances. Powerful renormalization group arguments, analogous to those of the DGLAP equations for the one dimension parton distributions, can be employed to understand how the three dimensional dynamics encoded in the GPDs changes as this factorization scale is varied~\cite{Diehl:2003ny,Belitsky:2005qn}. 

GPDs provide important insight into the three dimensional structure of polarized protons. Famously, the second moment of one set of quark GPDs gives the total quark angular momentum of the proton, and another set of  gluon GPDs can identically be related to the total gluon angular momentum. From the ``Ji sum rule"~\cite{Ji:1996ek}, the proton's spin can be expressed as the sum of these total angular momenta. 
In  Sec~\ref{section:spinSection}, we discussed the Jaffe-Manohar spin sum rule that decomposes the spin of the proton into the sum of the quark and gluon helicities, and their respective angular momenta. Therefore, in principle, GPD measurements can be combined with the direct measurements of quark and gluon spin helicities, to provide further insight into quark and gluon orbital momenta. However, there are a number of subtle issues that need to be resolved before this program can be realized fully~\cite{Wakamatsu:2014zza,Leader:2013jra}.

At present, our empirical knowledge about GPDs from DVCS data is mostly limited to the valence quark region, from the HERMES~\cite{Airapetian:2009cga,Airapetian:2011uq,Airapetian:2012mq,Airapetian:2012pg,Airapetian:2013gta} experiment at HERA,  the Jefferson Lab 6~GeV experiments~\cite{Defurne:2015kxq,Girod:2007aa,Jo:2015ema} and COMPASS~\cite{Joerg:2017ubq} at CERN. In the near future, one anticipates results from the Jefferson Lab 12~GeV experiments. There is also limited relatively low precision HERA data on sea quarks and gluons from the H1~\cite{Aktas:2005ty, Aaron:2009ac} and ZEUS~\cite{Chekanov:2008vy} experiments, and in the near future, a glimpse into sea quark distributions will be provided by COMPASS. A high energy, high luminosity EIC will extract sea quark and gluon GPDs with unprecedented reach and precision. Transverse spatial distribution of quarks and gluons, in both protons and complex nuclei, will be extracted through precise measurements of the $t$-dependence of  DVCS and exclusive cross-sections for production of $J/\psi$, $\phi$, $\pi$, $K$ and other mesons. For protons, the interval $0 \approx |t| \le 1.5~\text{GeV}^2$ will enable one to map out parton distributions down to an impact parameter of $\sim 0.1$ fm.

 \begin{figure*}[t!]
	\center
	\includegraphics[width=1.0\textwidth]{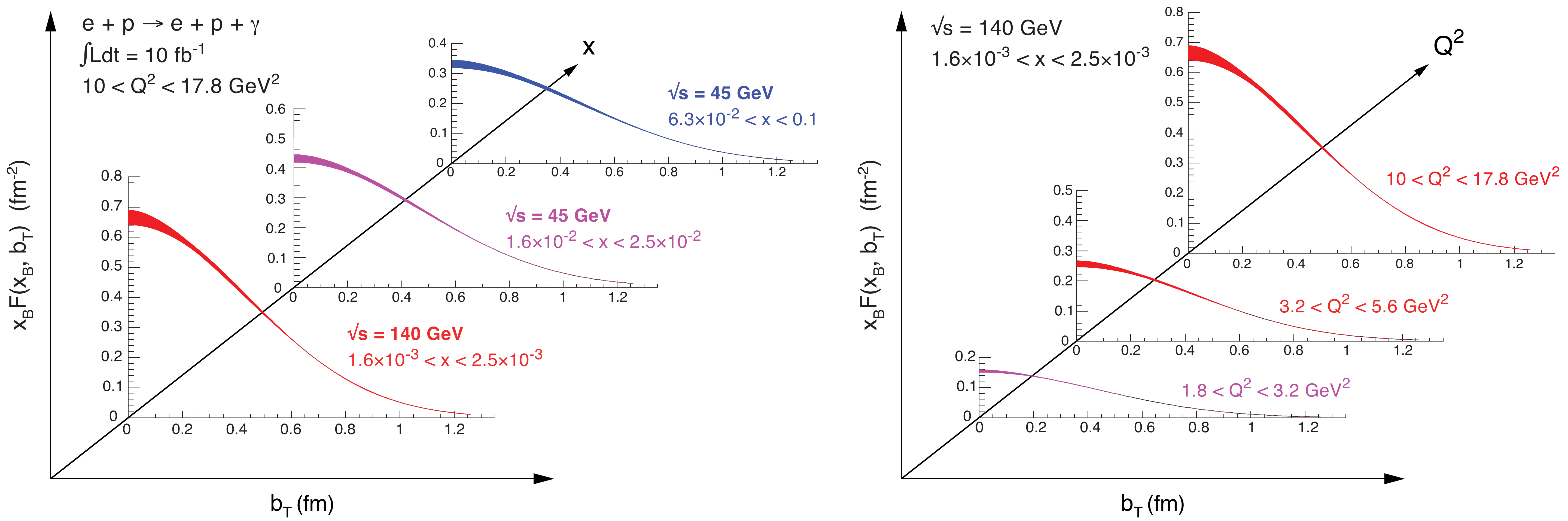} 
	\caption{The projected precision of the transverse spatial distribution of partons obtained from the Fourier transform of the measurement of the unpolarized DVCS cross-sections as a function of $|t|$ at an EIC for a targeted luminosity of 10~fb$^{-1}$ at each center-of-mass energy. $b_{T}$ is the distance from the center of the proton, known also as ``impact parameter''. 
		Left plots show the evolution in $x$ at a fixed $Q^{2}$ ($10 < Q^{2} < 17.8~\text{GeV}^{2}$). Right plot shows the evolution in $Q^{2}$ at a fixed $x$ ($1.6\times10^{-3} < x < 2.5\times10^{3}$). See text for more details.}
	\label{Fig:DVCS}
\end{figure*}

Figure~\ref{Fig:DVCS} shows the precision that an EIC can provide for imaging of quarks as obtained by a Fourier transform of the unpolarized DVCS cross-sections as a function of $t$. These simulated data are based on GPDs extracted from a fit to the world DVCS data. Bearing in mind the $x,Q^{2}$ kinematic coverage shown in Fig.~\ref{fig:x-q2-epeA_combo}, each bin can be accessed either only at lower center-of-mass energy (blue band) or at higher energy (red band). The purple band represents a region typically reachable at both low and high energies. The impact parameter dependent parton distribution functions obtained show clearly the growth of parton distributions at low $x$ and high $Q^{2}$, where sea quarks are important.  The evolution in $x$ and $Q^2$ can therefore teach us about the relative spatial distributions of valence quarks, the quark sea and gluons. The plot in Fig.~\ref{Fig:DVCS} demonstrates that a wide window in $Q^2$ resolution is available at the high center-of-mass energy for $x\sim 10^{-3}$. Such a $Q^2$ reach at fixed $x$ is important to extract the gluon spatial distribution through the scaling violation of the DVCS cross-section, just as is the case for the $g_1$ and $F_2$ structure functions.

From these data, one can also extract the mean squared radius $\langle b_T^2\rangle$ of partons in the proton as a function of Bjorken $x$. This is shown in Fig.~\ref{fig:fitB}. At small $x$, this dependence is closely related to  the QCD string tension in the Regge framework. In this framework, the transition from large to small $x$ contains important information that allows one to deduce how the dynamical degrees of freedom transition from Reggeon exchanges to so-called Pomeron exchanges, or -- in parton language -- from quark to gluon exchanges, where the latter carries the quantum numbers of the QCD vacuum.
The evolution over a large range in $Q^2$ can teach us how the
the string tension evolves from this nonperturbative stringy picture to that of QCD bremsstrahlung. 
One can thus study with unprecedented precision how the dynamics changes when going upwards from the lower right corner in Fig.~\ref{fig:bigpicture}.

In Fig.~\ref{fig:fitB}, an inelasticity of $y\le 0.6$ was chosen; this is important to ensure that the DVCS cross-section is not dominated by the Bethe-Heitler background; details of the analysis are given in Ref.~\cite{Aschenauer:2013hhw}. As a result, the values of $x$ do not go below $x=10^{-3}$.
The analysis of data with higher $y$ and lower $x$ is possible but more involved. 
These considerations are also valid at lower $\sqrt{s}$. Therefore, at lower energies there is limited reach beyond the Reggeon exchange dominated region. 

Another important exclusive channel is that of  $J/\psi$ production, which provides unique access to the unpolarized gluon GPD through the dominant photon-gluon fusion production mechanism; this mechanism is discussed further in Sec.~\ref{section:nuclearStructureFunctions} and illustrated in Fig.~\ref{fig:CharmProd}. Transverse spatial  images obtained from Fourier transforming the $t$-dependent  $\gamma^{*}p \rightarrow J/\psi + p^{\prime}$  $J/\psi$ cross-section for $\sqrt{s}=140$ GeV show that gluon distributions can be accessed across the entire transverse plane with fine resolution at small $x$.  

\begin{figurehere}
	\vspace{2mm}
	\begin{center}
		\includegraphics[width=1.0\columnwidth]{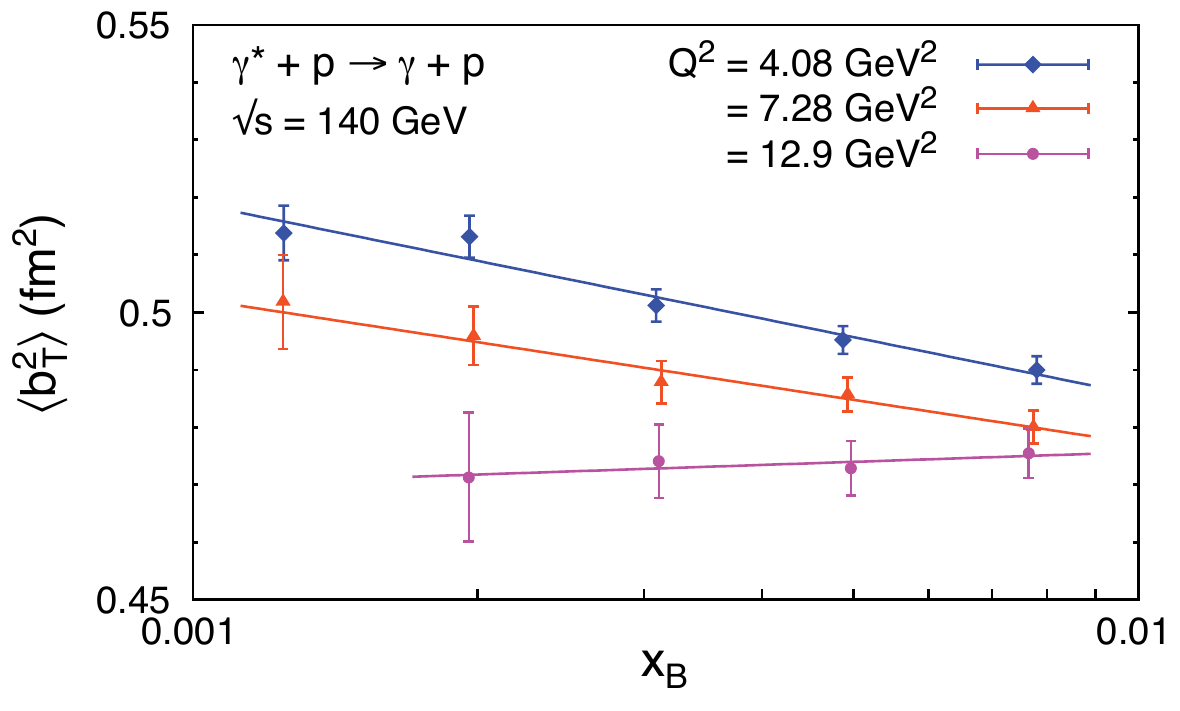} 
	\end{center}
	\vspace{-6mm}
	\caption{The average value of the mean squared parton radius of the proton, extracted from the DVCS cross-section, plotted as a function of 
		Bjorken $x$. Results are shown for three different values of $Q^2$. Plot from the EIC White Paper~\cite{Accardi:2012qut}.}
	\label{fig:fitB}
	\vspace{6mm}
\end{figurehere}

Incoherent exclusive scattering is characterized by the breakup of the proton. These processes are unique in that they are sensitive to  the color charge {\em fluctuations} in the proton. This is discussed later on page~\pageref{page:incoherentdiff}. A combined study of the coherent processes discussed here  (where the proton stays intact), with incoherent exclusive reactions, may allow one to reconstruct how gluon saturation sets in through the progressive clumping of gluons in the transverse plane. Conversely, one may be able to reconstruct where the transition to pion degrees of freedom sets in by quantifying the relative distribution in impact parameter of gluons and sea quarks.

\end{multicols}
\FloatBarrier

	\subsection{Opportunities with Charged Current Processes}
             \label{section:chargeSection}
	\begin{figure*}[b!]
		\centering \includegraphics[keepaspectratio=true,
		width=\linewidth]{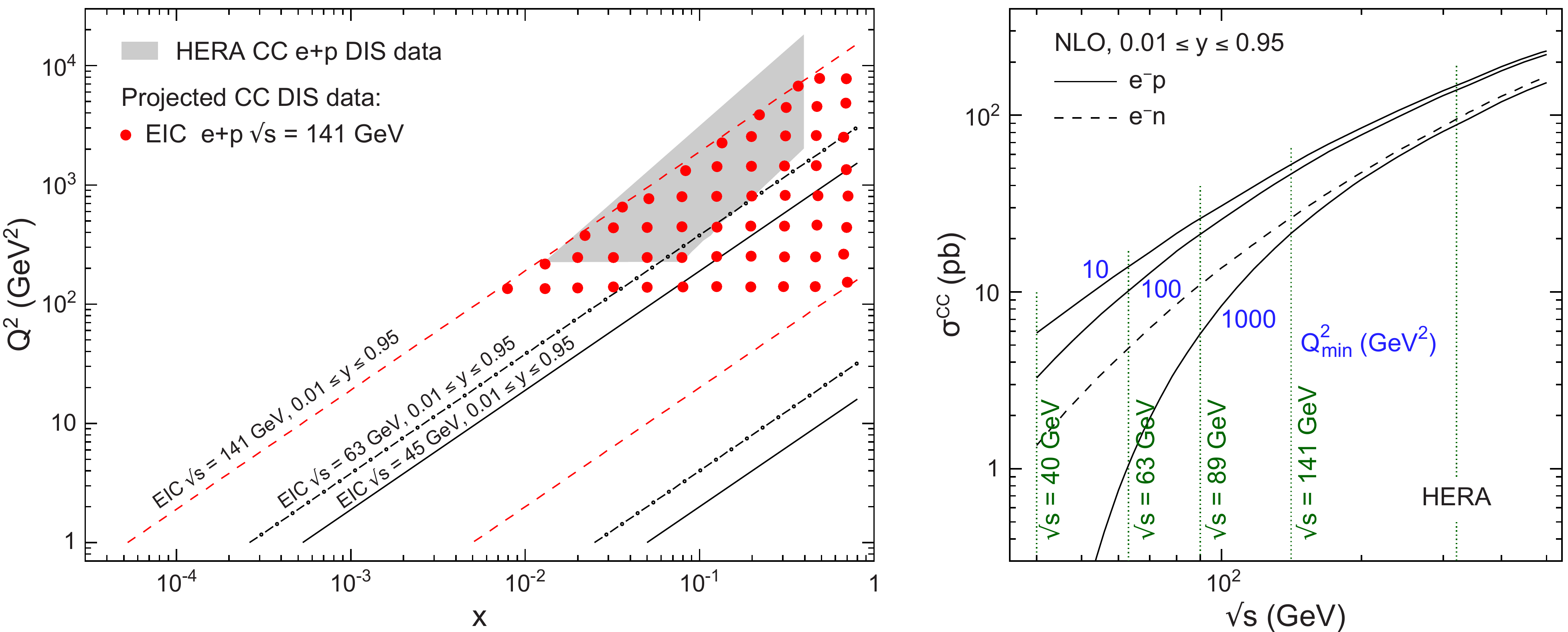}
		\caption[Charged current DIS cross-section]{{\it Left:} Kinematic reach in $x$ and
			$Q^2$ for three EIC center-of-mass energies with $0.01 \leq y \leq 0.95$. The 
			filled circles represent the projected CC DIS coverage for 
			$\sqrt{s} = 141$~GeV and the shaded region gives the extent of HERA CC data. 
			{\it Right:} Integrated unpolarized 
			CC DIS cross-section for electron-proton 
			and electron-neutron (dashed line) 
			scattering at NLO accuracy as a function of center-of-mass energy 
			for $Q^2 > Q^{2}_{\mathrm{min}}$ and $0.01 \leq y \leq 0.95$.}
		\label{fig:CCXSEC}
	\end{figure*} 
\begin{multicols}{2}	
At sufficiently large $Q^2$ values in DIS, the cross-section for the exchange of 
virtual $W^{\pm}$ bosons becomes comparable to that of the virtual photon case. These charged current (CC) events access different combinations of quark and anti-quark flavors than photon-mediated DIS. 
(For a recent discussion of the energy requirements of neutral current mediated measurements at the EIC, see Ref.~\cite{Zhao:2016rfu}.) 
As discussed in \cite{Aschenauer:2013iia,Boer:2011fh}, these events can be used to constrain flavor-dependent 
parton distribution functions. In this context, CC DIS measurements 
have an advantage over semi-inclusive DIS (SIDIS) studies. The latter can also 
constrain flavor-dependent PDFs, but only with the additional input of fragmentation functions relating the out-going hadron with the flavor of the progenitor quark. Charged current measurements provide access to  flavor-dependent PDFs without the uncertainties introduced by fragmentation functions. The theoretically clean CC channel will be complementary to SIDIS, providing stringent tests of SIDIS computations as well as addressing the universality of the PDFs. While CC DIS was studied extensively at HERA \cite{Aaron:2009aa}, 
the ability of an EIC to provide polarized proton beams will allow access to 
polarized flavor-dependent PDFs (via the single target spin asymmetry) which are 
important for solving the proton spin puzzle as well as investigating nonperturbative aspects of proton structure. Note that in this regard $W$ measurements in polarized \pp\ collisions at RHIC~\cite{Adamczyk:2014xyw,STAR:2011aa,Aggarwal:2010vc,Adare:2010xa} provide important complementary information. Models of PDF behavior at high $x$ based on helicity retention, which 
for example predict that $\Delta d/d \rightarrow 1$ as $x \rightarrow 1$ can also be tested \cite{Avakian:2007xa}. In addition, charm production in CC DIS will give access to the strange sea 
\cite{Boer:2011fh} and measurements on neutrons (via deuterium or $^3\mathrm{He}$ beams) 
can provide information on possible charge symmetry violation, which may be 
relevant for the EMC effect \cite{Cloet:2012td,Cloet:2009qs}. It should be noted that studies on parity violating single-spin observables which are sensitive 
to neutral current $\gamma - Z$ interference have also been performed \cite{Zhao:2016rfu}.

In order to maximize the impact of CC measurements at an EIC, it will be 
important to achieve the largest possible $\sqrt{s}$, as both the CC cross-section and available 
kinematic reach in $x$ and $Q^2$ increase with $\sqrt{s}$. The $x - Q^2$ coverage of CC DIS measurements at an EIC is shown on the left-hand panel of Fig.~\ref{fig:CCXSEC} along with the available kinematic
range for three center-of-mass energies. The available phase space for CC production is largest for the largest $\sqrt{s}$. These measurements will constrain flavor-dependent polarized PDFs over a wide range in $x$. In addition, the $x - Q^2$ coverage afforded by the largest $\sqrt{s}$ values provides significant overlap 
with the HERA region \cite{Aaron:2009aa}. Thus the EIC can improve on the statistically limited unpolarized HERA results while extending them to higher-$x$. Finally, the kinematic reach provided by the largest $\sqrt{s}$ available at an EIC means that for fixed $x\geq 0.08$, there will be a decade 
or more coverage in $Q^2$. This will allow the study of the $Q^2$ evolution 
of both the CC cross-section and single-spin asymmetry over a wider $x$ range than is possible at lower $\sqrt{s}$.

The dependence of  the cross-section on $\sqrt{s}$ can be seen in the right-hand panel of Fig.~\ref{fig:CCXSEC} which shows the CC DIS cross-section integrated over $Q^2 > Q^{2}_{\mathrm{min}}$ for several values of $Q^{2}_{\mathrm{min}}$. For $Q^{2}_{\mathrm{min}}$ = 100~GeV$^2$, the drop in
the cross-section as a function of $\sqrt{s}$ is relatively modest. However, as $Q^{2}_{\mathrm{min}}$
increases, the dip in the cross-section at lower $\sqrt{s}$ values becomes quite 
dramatic, making the collection of sufficient statistics at high $Q^2$ difficult even with large integrated luminosities. The higher $\sqrt{s}$ values eliminate the need to compensate for the falling cross-section with higher instantaneous luminosities.  

\end{multicols}

\FloatBarrier
			
	\subsection{Structure Functions in Nuclei}
			\label{section:nuclearStructureFunctions}
\begin{multicols}{2}
	
The $e^\pm$+$p$ DIS experiments at HERA~\cite{Abramowicz:2015mha} yielded very accurate 
information on unpolarized proton structure in a wide kinematic range down to $x=10^{-4}$ 
for $Q^2 \gtrsim 10 \, {\rm GeV}^2$, where perturbative computations are reliable. The HERA 
experiments performed numerous measurements with neutral-current, charged-current, as well 
as jet and heavy-quark tagged cross-sections. These measurements provide the main data set 
to unravel the proton's internal structure and form the backbone of all the present day 
global fits of Parton Distribution Functions (PDFs) \cite{Forte:2013wc,Rojo:2015acz} in a 
proton. These PDFs are crucial in searches for new physics at the LHC.

Similarly, $e^\pm$+A DIS scattering experiments at an EIC \cite{Arneodo:1992wf} will extract 
the nuclear PDFs (nPDFs) \cite{Eskola:2012rg,Paukkunen:2014nqa}, that are important for a 
deeper understanding of heavy-ion collisions. So far, the kinematic reach of the available 
fixed target $e^\pm$+A cross-section measurements is much more restricted than for protons, 
with very little data available for low $x \leq 10^{-2}$ at $Q^2\sim 10$ GeV$^2$. 
Furthermore, as noted previously, even at intermediate $x$ they do not provide the required 
lever arm to extract information on gluons through scaling violations. As a consequence, 
gluon and sea quark nuclear PDFs are widely unconstrained.

An important recent development is the release of the nPDF global fit, 
EPPS16~\cite{Eskola:2016oht}, that includes LHC data from the \pPb\ run 1. However, at 
low-$x$ and low-$Q^2$, the latest LHC \pPb\ data does little to constrain nPDFs. The reason 
for this moderate impact lies in the need to evolve the data ``backwards", i.e. from low to 
high $Q^2$, which requires much higher precision data. In sharp contrast to PDFs in the 
proton, the role of gluons in nuclei is still {\it terra incognita}.

\begin{figure*}[t!]
	\centering \includegraphics[width=\linewidth]{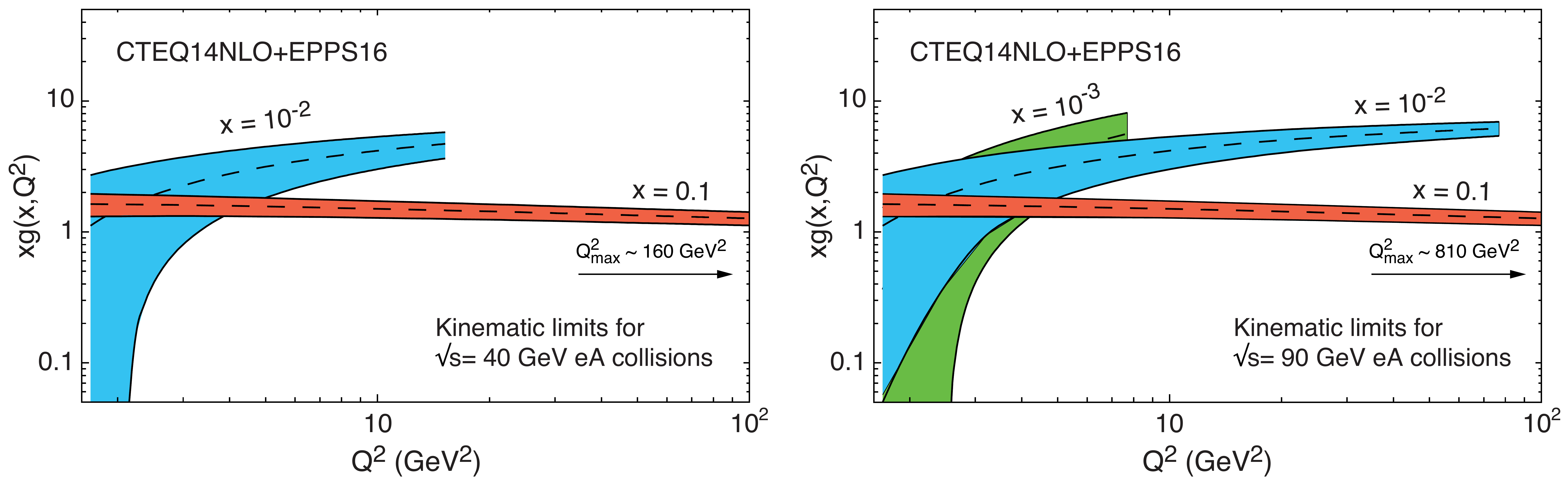}
	\vspace{-6mm}
	\caption{Nuclear Parton Distribution Functions of gluons as functions of $Q^2$ for various $x$ values obtained by multiplying the gluon distribution in the proton extracted by the CTEQ  collaboration to NLO  \cite{Dulat:2015mca} with the nuclear modification ratio for gluons extracted by EPPS16~\cite{Eskola:2016oht}. The PDFs are cut off at the kinematic limits imposed by the indicated  energies of $\sqrt{s}=40$ GeV ({\it left}) and 90 GeV ({\it right}), proposed for \eA\ collisions at an EIC. We will show later in Fig.~\ref{fig:epps16ratios} that these uncertainties will be greatly constrained by EIC data. For instance, at $x=10^{-3}$ they are reduced by a factor of $\sim 5$. }
	\label{fig:CTEQ14-Q2dep_eA_combo}
\end{figure*} 

As an example, the nuclear gluon distribution for \eA\ collisions at an EIC as a function of 
$Q^{2}$ and for different values of $x$ is shown in Fig.~\ref{fig:CTEQ14-Q2dep_eA_combo}. 
The gluon distribution was obtained using the NLO parameters from the CTEQ collaboration 
\cite{Dulat:2015mca} multiplied by the corresponding nuclear correction factor from EPPS16. 
The uncertainty bands reflect the combined uncertainties from both distributions. In 
contrast to the \ep\ case shown in Fig.~\ref{fig:CTEQ14-Q2dep_ep_combo} of Section 
\ref{sec:lessonsDIS}, the DGLAP evolution generates gluon distributions to good accuracy 
only for high values of $x$. At $x \sim 10^{-2}$ the limited lever arm in $Q^{2}$ 
complicates the precise extraction of the gluon density at $\sqrt{s}=40$ GeV. It is only 
when increasing the energy by a factor of 2 or more that one can access the higher $Q^{2}$ 
where the gluon density can be reliably determined. Furthermore, the reach of the 
insufficiently explored low-$x$ domain is feasible only at $\sqrt{s}=90$ GeV center-of-mass 
energy. We will show later that the uncertainties from current world data on nuclear gluon 
distributions will be significantly reduced by EIC data.

\begin{figurehere}
	\vspace{0mm}
	\begin{center}
		\includegraphics[width=0.55\columnwidth]{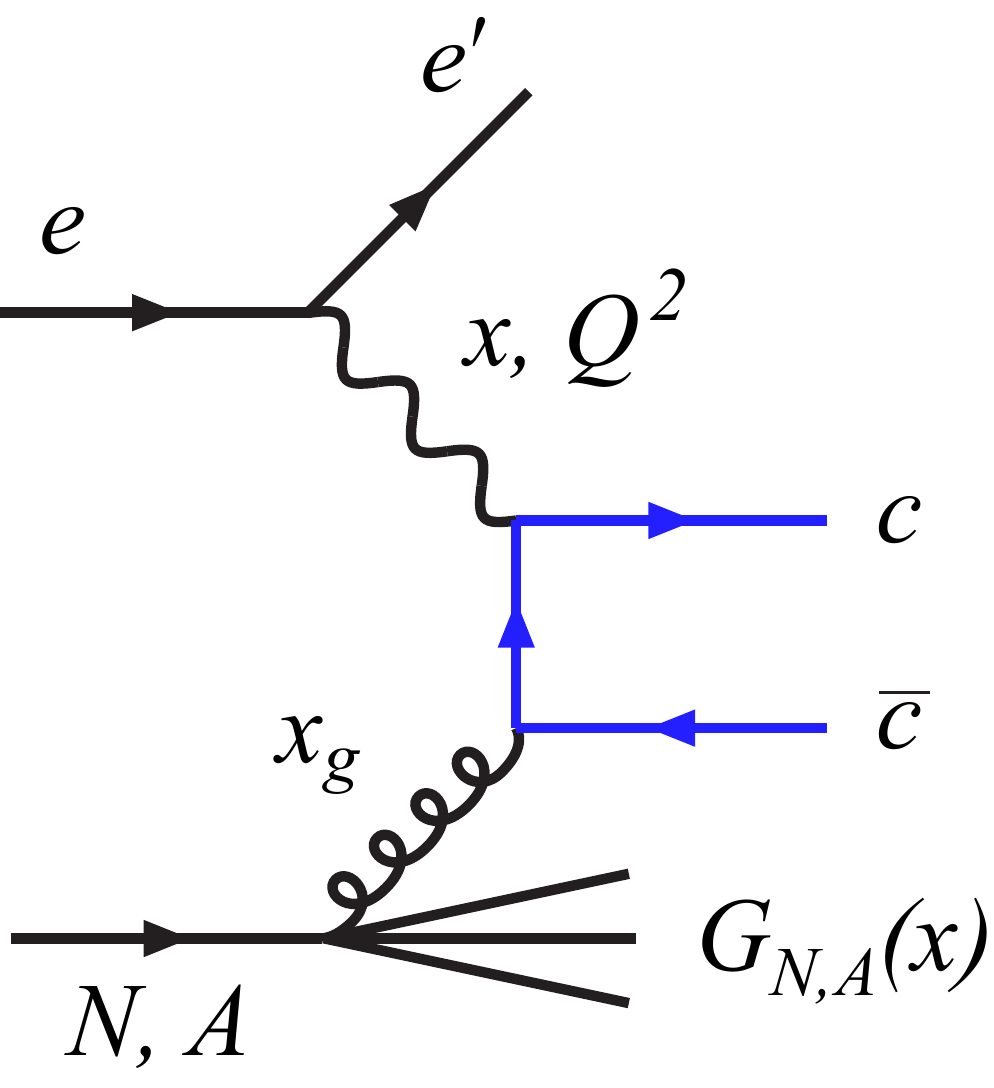} 
	\end{center}
	\vspace{0mm}
	\caption{Charm pair production via photon-gluon fusion.}
	\label{fig:CharmProd} 
\end{figurehere}
\vspace{5mm}

Therefore, an EIC with a wide lever arm in $x$ and $Q^2$ is critical for unambiguous 
determination of the parton structure of nuclei. Such a determination is an important first 
step towards a deeper examination of outstanding questions regarding i) how color is 
confined in a nucleus as opposed to a proton, ii) the nature of the residual color forces 
that bind nucleons together at short distances, and iii) the response of the nuclear medium 
to colored probes.

In DIS processes, the fully inclusive reduced cross-section can be written in terms of the 
structure functions $F_2$ and $F_L$ as
\begin{equation}
\sigma_\text{reduced} =
F_2(x,Q^2) - \frac{y^2}{1+(1-y)^2}F_L(x,Q^2),
\label{Eq:StructFunc}
\end{equation}

\noindent 
where $F_2$ is sensitive to the sum of the quark and anti-quark momentum distributions and 
$F_L$ is sensitive to the gluon distribution. For EIC kinematics, up to $10-15\%$ of the 
inclusive cross-section is from production of charm quarks--the charm structure function can 
be measured in nuclei for the first time. Since the dominant process is the production of 
charm-anticharm pairs through photon-gluon fusion (illustrated in Fig.~\ref{fig:CharmProd}) 
the measurement of this cross-section allows for an independent extraction of the gluon 
distribution in nuclei.

Simultaneous measurements of the $F_2$, $F_L$, and $F_L^{c\bar c}$ structure functions are 
key to uniquely constrain PDFs. The current theoretical description, even in the case of 
proton PDFs, has an ambiguity when the heavy quark production thresholds are 
crossed~\cite{Thorne:2008xf}. This ambiguity, usually called a ``mass scheme", has a 
significant impact on the PDFs extracted and can be resolved by determining the heavy quark 
structure function $F_L^{c\bar c}$. Such measurements provide precise values of heavy quark 
masses.

\begin{figurehere}
	\begin{center}
		\includegraphics[width=\columnwidth]{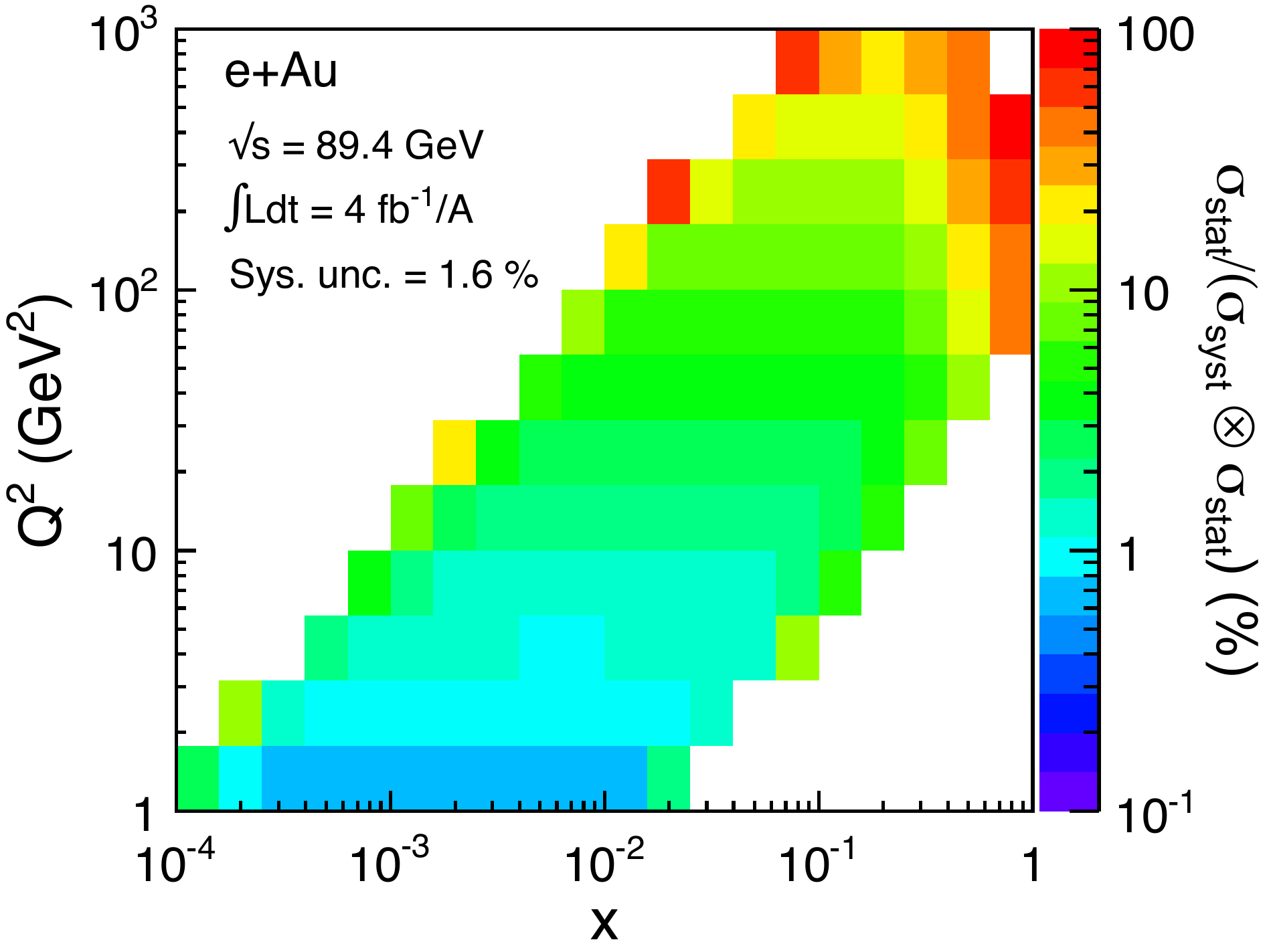} 
	\end{center}
	\vspace{-5mm}
	\caption{Fraction of statistical uncertainty over total uncertainty in measuring the reduced cross-section at $\sqrt{s}=89.4$~GeV in bins of $x$ and $Q^2$. The assumed systematic uncertainty is 1.6\%.}
	\label{Fig:sys} 
\end{figurehere}
\vspace{5mm}

\begin{figure}[h!]
	\centering
	\includegraphics[width=0.95\textwidth]{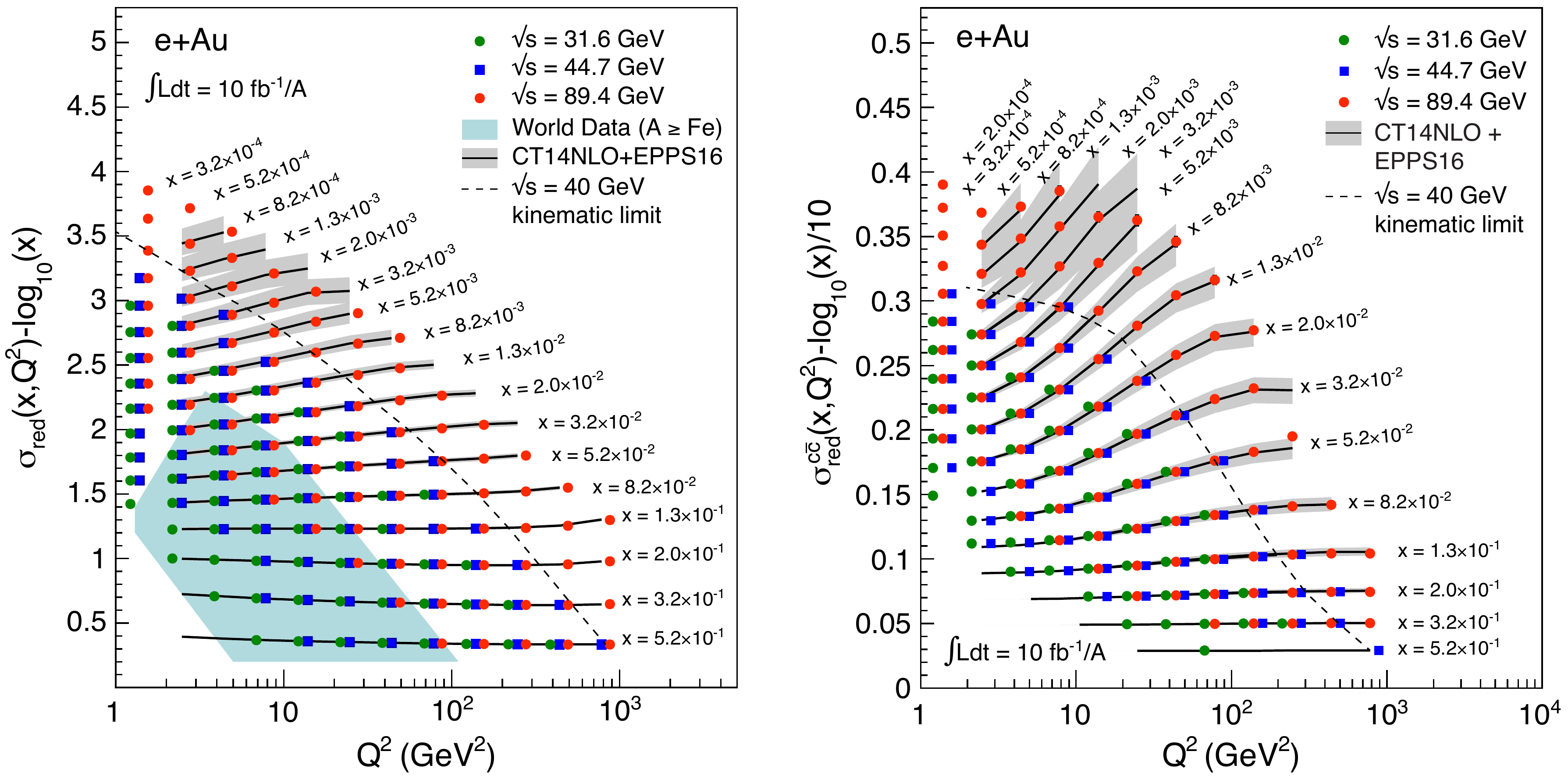}   
	\vspace{-2mm}
	\caption{Inclusive ({\it left}) and charm ({\it right}) reduced cross-sections plotted as functions of $Q^2$ and $x$ for both EIC pseudo-data and the EPPS16 model 
		(gray-shaded curves)~\cite{Eskola:2016oht,Aschenauer:2017oxs}.  The uncertainties represent statistical and systematics added in quadrature.  Also shown on the left plot is the region covered by currently available data.}
	\label{Fig:xred}
\end{figure}

For a quantitative estimate of what can be achieved, 
collisions at three different $\sqrt{s}$ were simulated with PYTHIA 6.4 
generator~\cite{Sjostrand:2006za} including nuclear modifications~\cite{Eskola:2009uj}. A 
collection of all the simulated measurements of DIS reduced cross-sections at an EIC, 
together with the EPPS16 uncertainties, are shown in Fig.~\ref{Fig:xred} for inclusive 
(left) and charm (right) production. In both cases, the points are shifted by 
$-\text{log}_{10}(x)$ for visibility. The statistical and systematic uncertainties are added 
in quadrature. This study corresponds to a combined integrated luminosity of 
$10~\text{fb}^{-1}$. For the corresponding statistics, the experimental uncertainties are 
dominated by systematic errors, as shown in Fig.~\ref{Fig:sys}.

Figure~\ref{Fig:xred} (left) depicts in the shaded region, for comparision, the current 
world data from DIS off heavy-ions. There are no charm measurements in \eAu\ collisions. The 
dashed line in both plots corresponds to the kinematic limit at the lower 40 GeV 
center-of-mass energy. We observe that the current extrapolated uncertainties, depicted by the grey 
bands~\cite{Eskola:2016oht,Aschenauer:2017oxs}, become substantially larger beyond this 
dashed line. Thus data from the higher center-of-mass energy will significantly constrain these uncertainties, and thereby, QCD  evolution of PDFs to smaller $x$.

\begin{figure}[h!] 
	\centering
	\includegraphics[width=0.95\textwidth]{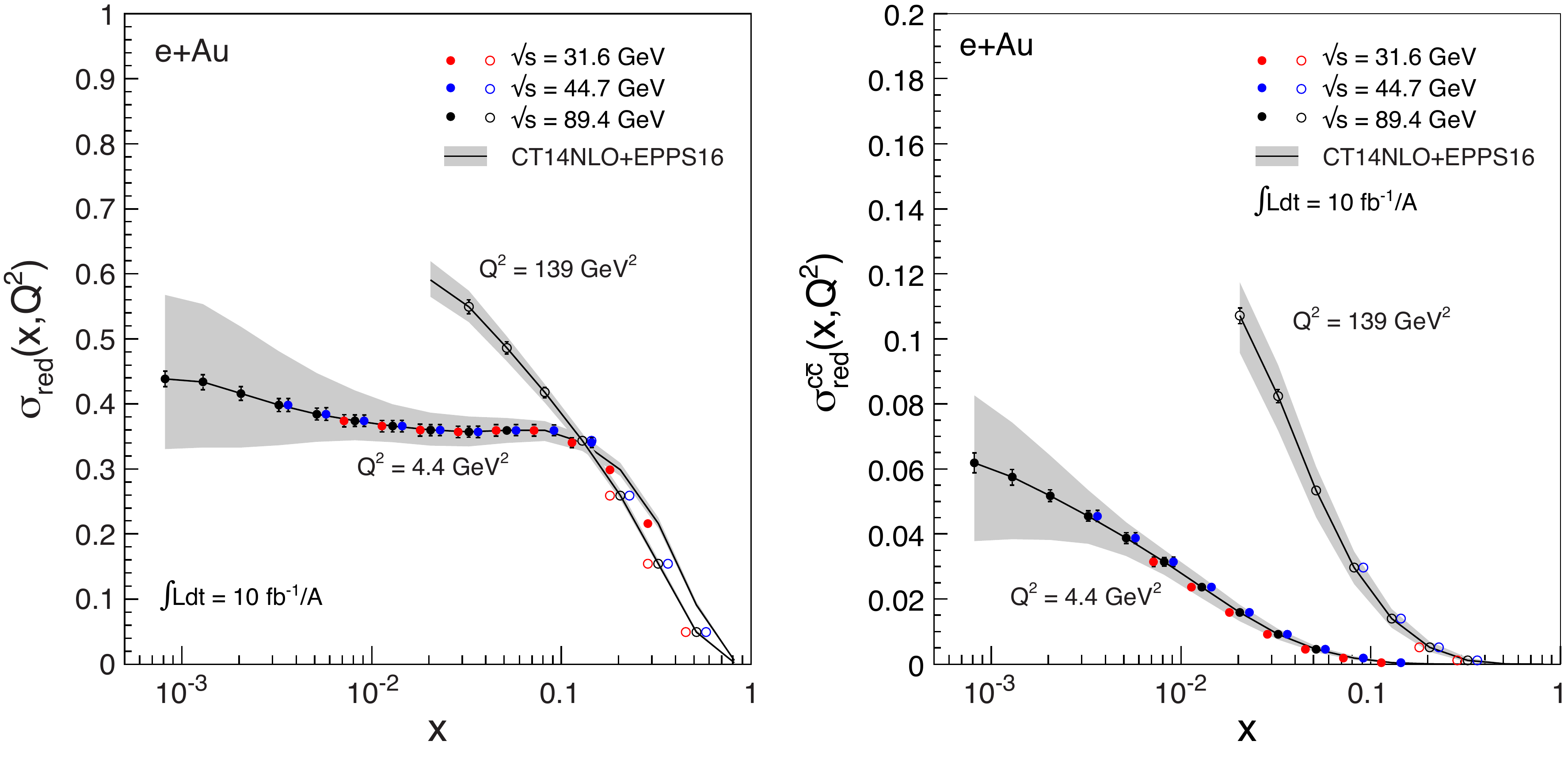}   
	\vspace{-3mm}
	\caption{Inclusive ({\it left}) and charm ({\it right}) reduced cross-sections as a function of $x$ at the $Q^{2}$ values of 4.4~GeV$^{2}$ (solid circles) and 139~GeV$^{2}$ (open circles) at three different center-of-mass energies. See text for details.}
	\label{Fig:xred2}
\end{figure}

To emphasize the precision achievable at an EIC, two examples of the reduced cross-section 
as a function of $x$ at the $Q^{2}$ values of 4.4~GeV$^{2}$ and 139~GeV$^{2}$ are shown in 
Fig.~\ref{Fig:xred2} for inclusive (left) and charm (right) production.  It is clear from 
Fig.~\ref{Fig:xred2} that at large values of $x$, the uncertainties are very small. It is 
only at $x < 10^{-2}$ and small $Q^2$ that the expected experimental errors on the EIC 
measurements become much smaller than the uncertainties from the EPPS16 parametrization that 
are largest at the smallest $x$ values; these will clearly be significantly constrained by 
data at these $x$ values.

Measuring the longitudinal structure function $F_{\rm L}$, poses additional experimental 
challenges. This observable is typically very small at high values of $x$ and $Q^2$ but 
increases with smaller $x$ and, at a fixed value of $x$, also rises with $Q^2$. It can be 
extracted through a Rosenbluth separation analysis and requires measurements from collisions 
at a minimum of three different center-of-mass energies. Using Eq. \ref{Eq:StructFunc}, a 
fit to the data leads to the negative of the gradient, giving $F_L$. Experimental data with 
a wide range in center-of-mass energy gives the lever arm necessary for precise extraction 
of $F_{L}$. 

\begin{figure}[t!]
	\centering 
	\includegraphics[width=0.96\textwidth]{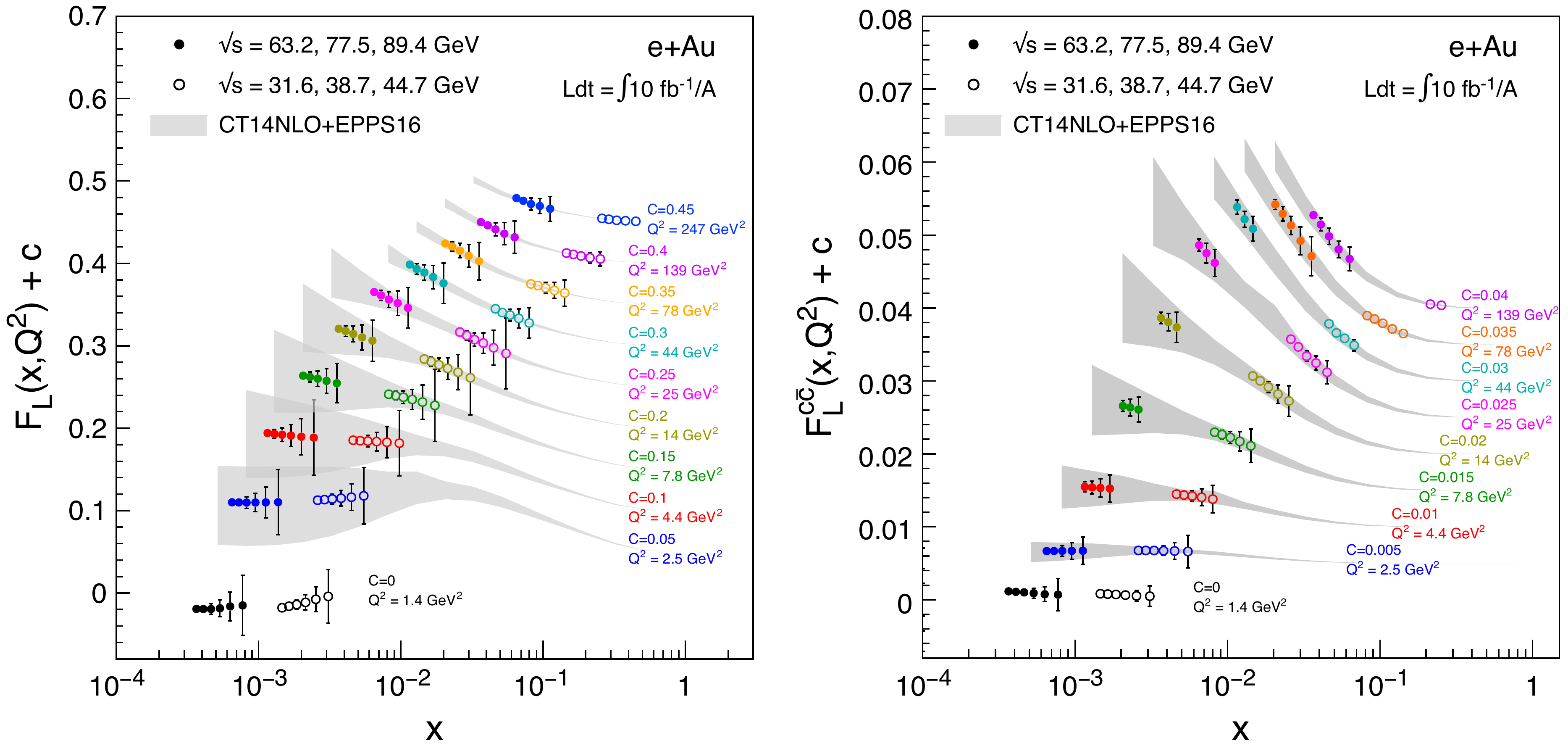}  
	\caption{Inclusive ({\it left}) and charm ({\it right}) $F_L$ structure function, plotted as a function of $Q^2$ and $x$. The uncertainties represent statistical and systematics added in quadrature. The gray-shaded bands depict the uncertainties in our current knowledge  of $F_L$ derived from the EPPS16 nuclear PDF~\cite{Eskola:2016oht,Aschenauer:2017oxs}. See text for further details.}
	\label{Fig:FL}
	\vspace{10mm}
\end{figure}

Figure~\ref{Fig:FL} shows a collection of the possible $F_{L}$ measurements in \eAu\ collisions at the EIC for both inclusive (left) and charm (right) production, plotted versus $Q^2$ for a number of $x$ values. The two sets of three different center-of-mass energies used in each extraction of $F_{L}$ are also indicated on the plot by full and open circles respectively.  For each set, the simulation assumes an integrated luminosity of 10~fb$^{-1}$. The gray-shaded bands indicate the uncertainties in our current knowledge  of $F_L$ derived from the EPPS16 nuclear PDF~\cite{Eskola:2016oht,Aschenauer:2017oxs}.
With this luminosity, an EIC can perform very precise measurements of $F_{L}$ in several $x, Q^{2}$ bins. This accuracy is crucial for a significant measurement of a quantity that is expected to be very small. The comparison of EIC pseudo-data and errors with the current depicted uncertainties (gray band) demonstrate dramatically the need for higher energies allowing one to  reach lower $x$ values where uncertainties are large. For $F_L$ at $Q^2 > 10$ GeV$^2$ and for charm $F_L$ the lower energy range does not provide any substantial improvement.  It is also important to note that EIC can achieve a comparable precision in measuring $F_{L}$ for the proton, improving even on the existing measurements from HERA~\cite{Andreev:2013vha} where kinematics overlap. 
\end{multicols}

\FloatBarrier
	
	\subsection{Nuclear Modifications of Parton Distribution Functions}
            \label{section:nuclearPDFs}
	\begin{multicols}{2}
		
			\begin{figure*}[t!]
			\centering
			\includegraphics[width=\textwidth]{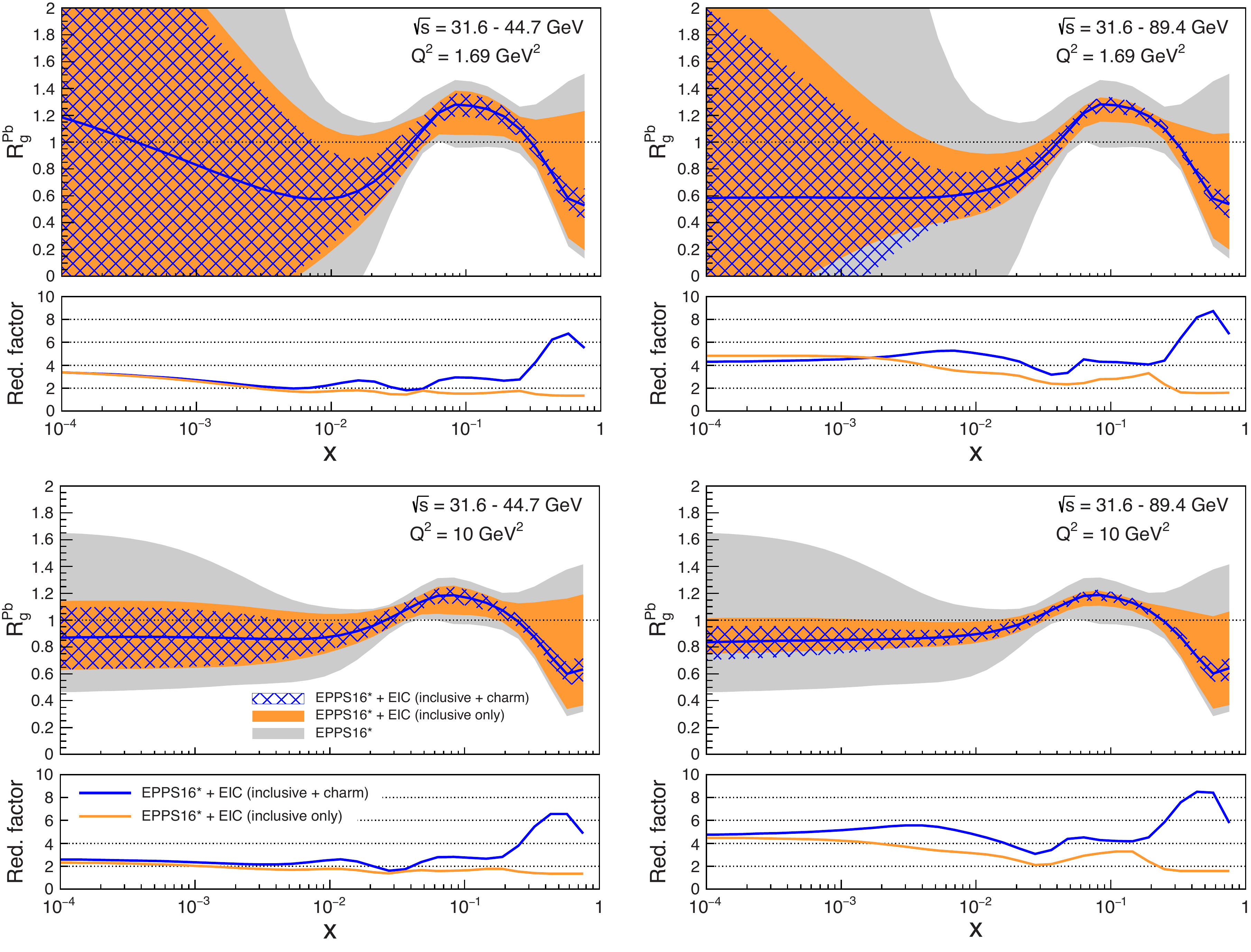} 
			\caption{The ratio $R_g^{\textrm{Pb}}$, from EPPS16*, of gluon distributions in a lead nucleus relative to the proton, for the low ({\it left}) and high ({\it right}) $\sqrt{s}$, at $Q^{2}=1.69~\text{GeV}^{2}$ and $Q^{2}=10~\text{GeV}^{2}$ (upper and lower plots, respectively). The grey band represents the EPPS16* theoretical uncertainty. The orange (blue hatched) band includes the EIC simulated inclusive (charm quark) reduced cross-section data. The lower panel in each plot shows the reduction factor in the uncertainty with respect to the baseline fit.}
			\label{fig:epps16ratios}
		\end{figure*}
		
The simulations discussed in the previous section suggest that an EIC will have an enormous 
impact on the global extractions of nuclear PDFs, particularly for the gluons. While the LHC 
data have achieved a substantial broadening of coverage in the kinematical space, the newly 
explored regions scan a $Q^{2}$ range where the DGLAP RGE significantly wash away the 
nuclear effects, leaving the low $x$ gluon nearly unconstrained.

The modification introduced by the nuclear environment can be quantified in terms of the 
ratio between the nucleus $A$ and the free proton PDF ($R_f^A$, $f=q,g$) for quarks and 
gluons, with deviations from unity being manifestations of nuclear effects. A depletion of 
this ratio relative to unity is often called shadowing. The impact study of EIC simulated 
data shown in Fig.~\ref{Fig:xred} was done by incorporating these data into the EPPS16 
fit~\cite{Eskola:2016oht}. 
However, as the parameterization is too stiff in the as yet unexplored low $x$ region, 
additional free parameters for the gluons have been added to the functional 
form~(EPPS16*~\cite{Hannu:DIS2017,Aschenauer:2017oxs}). The corresponding 
$R_g^{\textrm{Pb}}$ from EPPS16* is shown in Fig.~\ref{fig:epps16ratios}.

The grey band represents the EPPS16* theoretical uncertainty. The orange band is the result 
of including the EIC simulated inclusive reduced cross-section data in the fit. The lower 
panel of each plot shows the reduction factor in the uncertainty (orange curve) with respect 
to the baseline fit (gray band).  It is clear that the higher center-of-mass energy has a 
significantly larger impact in the whole kinematical range with the relative uncertainty 
roughly a factor of 2 smaller than for the lower center-of-mass energy.

We also examined the simulated charm quark reduced cross-section (blue hatched band), for 
which no data currently exist. The impact of its measurement for nuclear gluon distributions 
is shown in Fig.~\ref{fig:epps16ratios}. While it brings no additional constraint on the low 
-$x$ region, its impact at high-$x$ is remarkable providing up to a factor 8 reduction in 
uncertainty (blue curve).

\end{multicols}

\FloatBarrier

\label{subsubsection:heavyIons}
\subsubsection*{Impact on Heavy-Ion Physics}

\begin{multicols}{2}

Measurements over the last two decades, first at RHIC and later at the LHC, have provided 
strong evidence for the formation of a strongly coupled plasma of quarks and gluons (sQGP) 
in high energy collisions of heavy nuclei. This sQGP appears to behave like a nearly perfect 
liquid and is well described by hydrodynamics at around 1 fm/$c$ after the initial impact of 
the two nuclei \cite{Kolb:2003dz,Teaney:2000cw,Luzum:2008cw,Schenke:2011bn}. For reviews, 
see \cite{Heinz:2013th,Gale:2013da,deSouza:2015ena,Song:2017wtw}.

Despite the significant insight accumulated in the past 17 years, little is understood about 
how the initial non-equilibrium state, whose properties are little known, evolves towards a 
system in thermal equilibrium. A conjectured picture of the initial phase, based on the CGC 
framework, suggests that at leading order the collision can be approximated by the collision 
of ``shock waves" of classical gluon fields (Glasma fields), 
~\cite{Kovner:1995ja,Krasnitz:1998ns,Schenke:2012wb} resulting in the production of 
non-equilibrium gluonic matter.

\begin{figure*}[t!]
	\begin{center}
		\includegraphics[width=\textwidth]{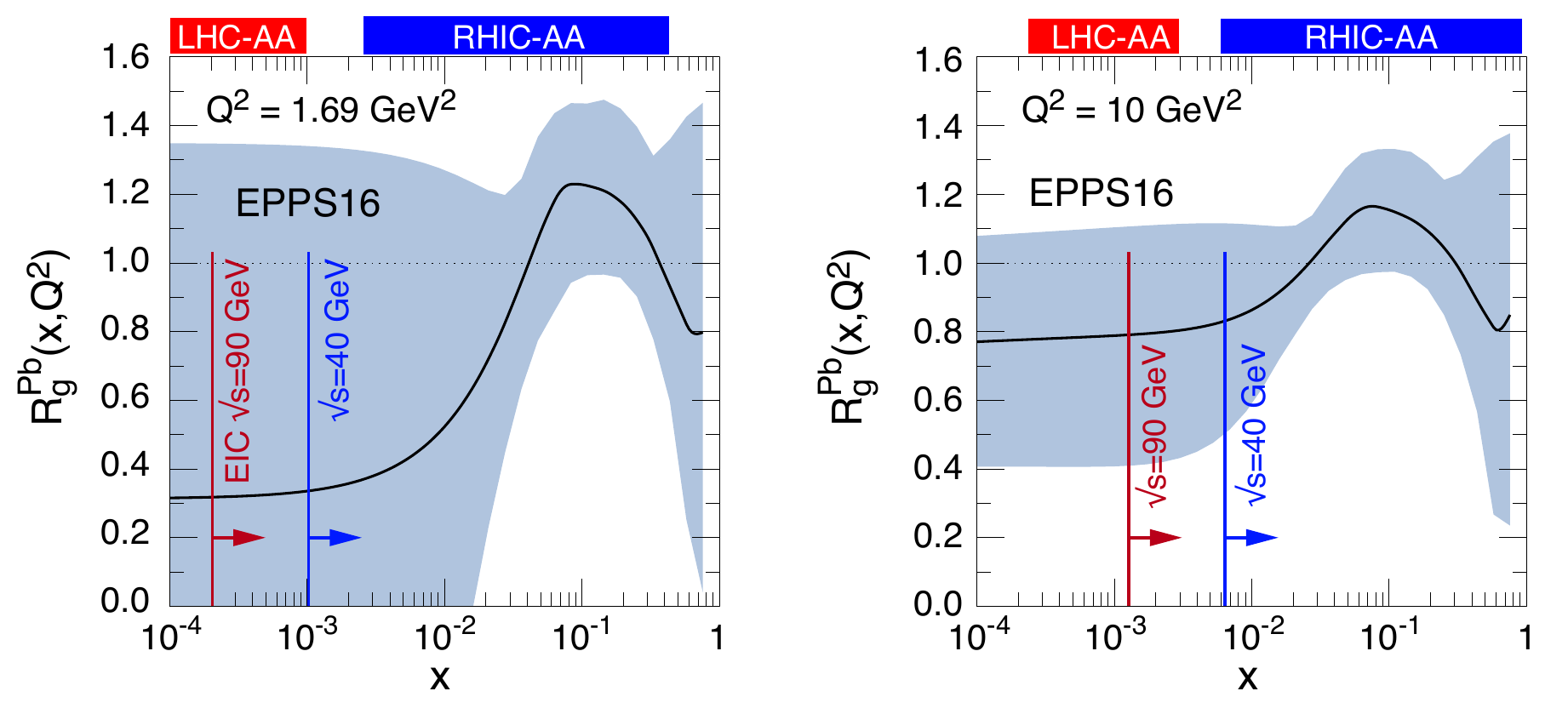}
	\end{center} 
	\vspace{-6mm}
	\caption{ \label{fig:eicHImapping} EPPS16 ratio of gluon PDF in a Pb nucleus relative to that of the proton ($R_g^{\textrm{Pb}}$), and its uncertainty band at $Q^2 = 1.69$ and 10 GeV$^2$ 
		\cite{Eskola:2016oht}. The plot for $Q^2 = 1.69$ GeV$^2$ is 
		indicative for processes that produce more than 90\% of all final 
		state particles in a heavy-ion collision at mid-rapidity. The 
		bands on the top of each panel reflect the referring kinematic 
		acceptance of the typical RHIC and LHC experiment. For details, 
		see text. The vertical red and blue lines indicate the 
		kinematic limits for different EIC center-of-mass energies.}
\end{figure*}

Unfortunately, heavy-ion collisions themselves cannot teach us much about the initial state 
because most of the details are wiped out during the evolution of the plasma. The final 
observables are sensitive to both, the initial state and the final state, whose transport 
parameters one ultimately seeks to extract. Therefore, information on the initial state 
needs to be extracted from experiments on \pA\ and ultimately \eA\ with small and well 
understood final state effects.

\label{page:incoherentdiff} It was demonstrated in \cite{Mantysaari:2016ykx} how \ep\ data 
can be successfully used to understand shape fluctuations of the proton. Here, the authors 
studied measurements of coherent and incoherent diffractive vector meson production at HERA 
to constrain the density profile of the proton and the magnitude of event-by-event 
fluctuations. Working within the CGC picture, they found that the gluon density of the 
proton has large geometric fluctuations. No such data for \eA\ collisions exists. 
Assumptions on initial state fluctuations and anisotropies that govern many aspects of the 
observed collective flow phenomena are rather speculative at present.

 Data from an EIC can therefore have a profound impact on our understanding of the 
properties of the \emph{initial state} in heavy-ion collisions, such as the momentum and 
spatial distributions of gluons and sea quarks. Nuclear effects, such as shadowing and 
saturation, can be studied. By varying the scale and energy of the collision the interplay 
between the soft non-perturbative and the hard perturbative regimes can be addressed.

In order to illustrate how the EIC energy maps onto the kinematic range in \AA\ collisions 
we focus on the longitudinal momentum distributions in the nucleus, the nPDFs described 
earlier in this section. Figure \ref{fig:eicHImapping} shows the EPPS16 
\cite{Eskola:2016oht} nuclear PDF and it's uncertainty band at $Q^2 = 1.69$ and 10 GeV$^2$. 
The plot for $Q^2 = 1.69$ GeV$^2$ is indicative for processes that produce more than 90\% of 
all final state particles. The bands on the top of each panel reflect the referring 
kinematic acceptance of the typical RHIC and LHC experiments. We used $x \approx 
p_T/\sqrt{s}\ \exp({\pm \eta})$ where $p_T \approx Q$; we chose for the pseudo-rapidity 
window $\eta = \pm 1$, typical for the central barrel acceptance of heavy-ion experiments. 
The horizontal red and blue lines indicate the EIC kinematic limits for two different 
center-of-mass energies \sqrts= 40 and 90 GeV, respectively. While data from \sqrts=40 GeV 
will provide an important constraint on the RHIC \AA\ data, it will not reach into the 
regime where the bulk of LHC \AA\ data comes from. On the other hand, for \sqrts=90 GeV, the 
nPDFs cover the kinematics of semi-hard processes at both RHIC and the LHC. This expanded 
coverage is of great importance for a common quantitative framework of \AA\ collisions at 
both colliders in the quest for a deeper understanding of the intial conditions and the 
transport properties of the sQGP. 

\end{multicols} 
\vfill

	\subsection{Gluon Saturation in Nuclei}
			\label{section:saturation}


A key goal of the EIC is to access the high parton density regime of the QCD landscape depicted in Fig.~\ref{fig:bigpicture}. The saturation scale $Q_s^2$, characterizing the QCD dynamics in this regime, is expected to scale as $A^{1/3}$. Hence, DIS off heavy nuclei at large per nucleon center-of-mass energies \sqrtsNN\ makes it possible to cleanly access this novel intrinsically nonlinear regime of the theory. When $Q_s^2$ is larger than the intrinsic QCD scale, weak coupling methods may be applicable and enable one to relate the nonlinear dynamics of gluons and quarks to experimental measurements. 

A strong hint in the theory  that such a novel regime must exist follows from the unitarity bound on QCD cross-sections. This fundamental bound would be violated if the observed rapid rise of gluon distributions with decreasing $x$ persists at even lower $x$. Remarkably, there exist weakly coupled albeit strongly interacting many-body interactions in the theory that cause gluons at small $x$ to recombine  into harder gluons at the same rate at which they like to shed softer gluons. A deeper understanding of this emergent  effect, and the wider framework in which such phenomena are embedded, has the potential to radically transform the study of intrinsically nonlinear dynamics in QCD. 

Although there is a significant body of data at small $x$ from HERA, RHIC and the LHC that can be described in saturation models, there are important caveats that stand in the way of a discovery claim. While saturation models do an excellent job describing a wide variety of HERA data~\cite{Rezaeian:2012ji}, the corresponding saturation scales, as shown in Fig.~\ref{fig:QsReach_eRHIC_JLEIC}, are very small. Larger (nuclear) saturation scales are  accessed  in proton-nucleus and nucleus-nucleus collisions at RHIC and the LHC, but interpretation of data in terms of the evolving parton dynamics of the nuclear wavefunction is complicated by strong final state effects occuring after the collision. Both of these concerns are mitigated in \eA\ collisions at the EIC. 
Figure~\ref{fig:QsReach_eRHIC_JLEIC} clearly shows the significant increase in $Q_s^2$ relative to HERA, and final state effects can be controlled fully. 

In the following, we shall review the center-of-mass energy requirements for two observables that are especially sensitive to saturation effects. We will first discuss in Sec. \ref{subsection:dihadrons} the evolution of the back-to-back correlation of the two produced hadrons in double inclusive scattering. Here, the available \sqrtsNN\ range directly determines the magnitude of the saturation scale that is accessed. 
In Sec.~\ref{subsection:diffraction},  we will return to diffractive scattering that was briefly mentioned in the context of imaging in Sec.~\ref{section:imaging}. Diffractive measurements have fundamentally impacted physics over the centuries. 21st century diffraction measurements at the EIC hold similar potential for discovery. We will demonstrate in a simple saturation model, the likelihood that novel physics will first manifest itself in these processes. 
			\subsubsection{Dihadron Suppression}

\label{subsection:dihadrons}

\begin{multicols}{2}
	
\begin{figure*}[t!]
		\begin{center}
			\includegraphics[width=\textwidth]{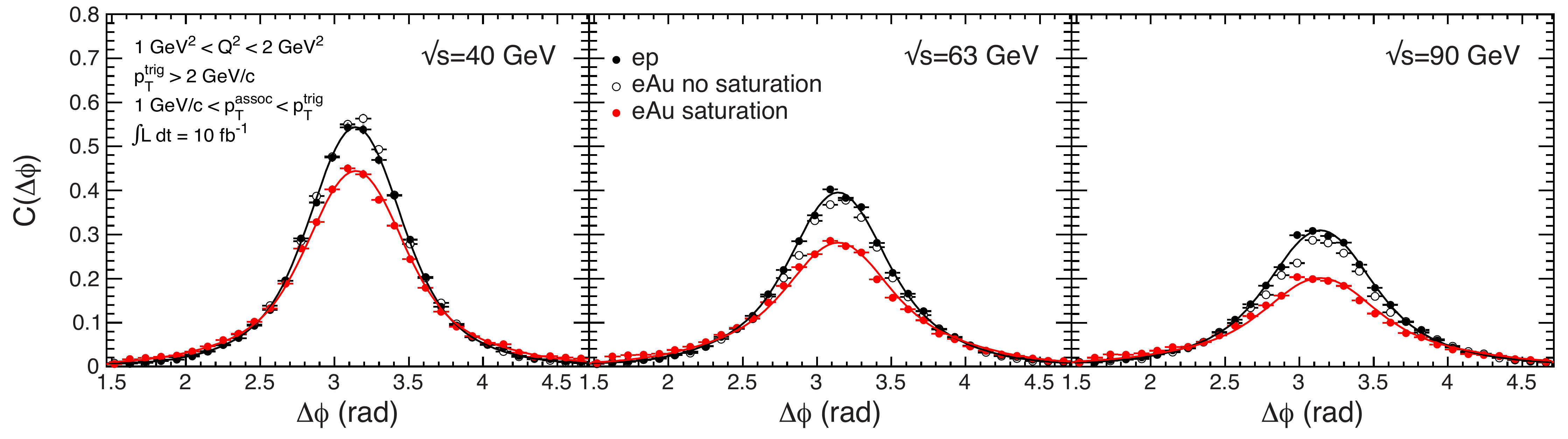}
		\end{center} 
		\vspace{-6mm}
		\caption{\label{fig:dihadronCurves} Comparison of dihadron correlation functions from 
			a saturation model prediction for \eAu\ collisions (red curve) with \ep\ collisions (black 
			curve) and calculations from a conventional non-saturated model (hollow data points) for 
			three different center-of-mass energies ranging from \sqrts=40 to 90 GeV. For details see 
			text.}
\end{figure*} 
Multiparton correlations allow us to reconstruct the internal structure of protons far more than single parton distributions alone permit.  A key measurement of multiparton correlations 
in \eA\ is the distribution of the azimuthal angle between two hadrons $h_1$ and $h_2$ 
in the process $e + \mathrm{A} \rightarrow e^\prime + h_1 + h_2 + X$. This process was discussed previously in the  EIC White Paper~\cite{Accardi:2012qut}. These correlations are  sensitive to the transverse momentum dependence of the gluon distribution as well as gluon correlations for which first principles 
computations are now becoming available \cite{Lappi:2012nh, Lappi:2016fmu}. The precise 
measurement of these dihadron correlations at an EIC would allow one not only to determine 
whether the saturation regime has been reached, but study as well the nonlinear evolution of spatial multi-gluon correlations. The $A$ dependence of this measurement provides another handle to study the nonlinear evolution of such correlations, and to ascertain their universal features. 

The saturation scale $Q_s$ for a given nucleus depends on the gluon 
momentum fraction $x_g$. Even though $x_g$ is not directly accessible experimentally, one can effectively constrain the underlying $x_g$ distribution in controlling the experimentally measured $x$ by  
varying beam energies. Dihadron correlations are relatively simpler to study at a collider. They 
are measured in the plane transverse to the beam axis, and are plotted as a function of 
the azimuthal angle $\Delta \phi$ between the momenta of the produced hadrons in that 
plane. The near-side peak ($\Delta\phi$=0) of this $\Delta \phi$ distribution is dominated 
by the fragmentation from the leading jet, while the away-side peak ($\Delta\phi$=$\pi$) 
is expected to be dominated by back-to-back jets produced in the hard 2$\rightarrow$2 
scattering. 

Saturation effects in this channel correspond to a progressive disappearance 
of the back-to-back correlations of hadrons with increasing atomic number. A comparison 
of the heights and widths of the dihadron azimuthal distributions in \ep\ and \eA\ 
collisions respectively would then be a clear experimental signature of such an effect.
The highest transverse momentum hadron in the dihadron correlation function is called the 
``trigger" hadron, while the other hadron is referred to as the ``associated" hadron with 
$p_T^{assoc} < p_T^{trig}$. The selected $p_T$ ranges affect the effective $Q^2$, that 
together with $x_g$, are the key parameters that govern the process.

In order to elucidate the importance of the center-of-mass dependence of this measurement 
we generated dihadron correlations for three different energies, \sqrts = 40, 63, and 90 
GeV in \ep\ and \eAu\ collisions following the procedures described in 
\cite{Zheng:2014vka}. Only charged pions $\pi^{\pm}$s were used. The calculations were performed  
for $1 < Q^2 < 2\ \mathrm{GeV}^2$ and include a Sudakov form factor to account for the radiation generated by parton showers. The hadrons were selected to have $p_T^{trig} 
> 2$ GeV/$c$ and $1\ \mathrm{GeV}/c< p_T^{assoc} < p_T^{trig}$. Statistical error bars 
correspond to 10 fb$^{-1}$/A integrated luminosity.

The away-side correlation peak for the three different energies is shown in 
Fig.~\ref{fig:dihadronCurves}.  Each panel depicts the \ep\ reference curve in black, as 
well as the predictions from saturation models in \eAu\ in red. It is important to verify 
how precisely the suppression of the away-side peak can be studied at an EIC and how the saturation model  predictions can be clearly distinguished from a conventional leading twist 
shadowing (LTS) scenario~\cite{Frankfurt:2003gx, Frankfurt:2011cs}. Such scenarios include nonlinear interactions only in the initial conditions but not in the QCD evolution of the distributions. 

To obtain results for the LTS scenario, we use a hybrid Monte Carlo generator, consisting of PYTHIA-6~\cite{Sjostrand:2006za} for parton generation, showering and fragmentation, DPMJet-III \cite{Roesler:2000he} for the nuclear geometry, and a cold matter energy-loss afterburner~\cite{Armesto:2004hz}. We employ the nPDFs from EPS09 \cite{Eskola:2009uj} to describe the  shadowed parton distributions. 

The resulting LTS correlation function is indicated in Fig.~\ref{fig:dihadronCurves} by black hollow points. As expected, the difference between \ep\ and the nonsaturated \eAu\ is minuscule. Final state nuclear effects hardly alter the correlation peak, an observation that is in agreement with findings at RHIC when 
comparing \pp\ with \pA\ collisions at mid-rapidity. The difference between the saturated
\eAu\ case and the \ep\ reference is already visible at lower energies but becomes 
striking at the highest energies, namely, at the lowest-$x_g$ range accessible.

\begin{figure}[h!]
	\centering
	\begin{minipage}[b]{0.49\textwidth}
		\centering
		\includegraphics[width=\columnwidth]{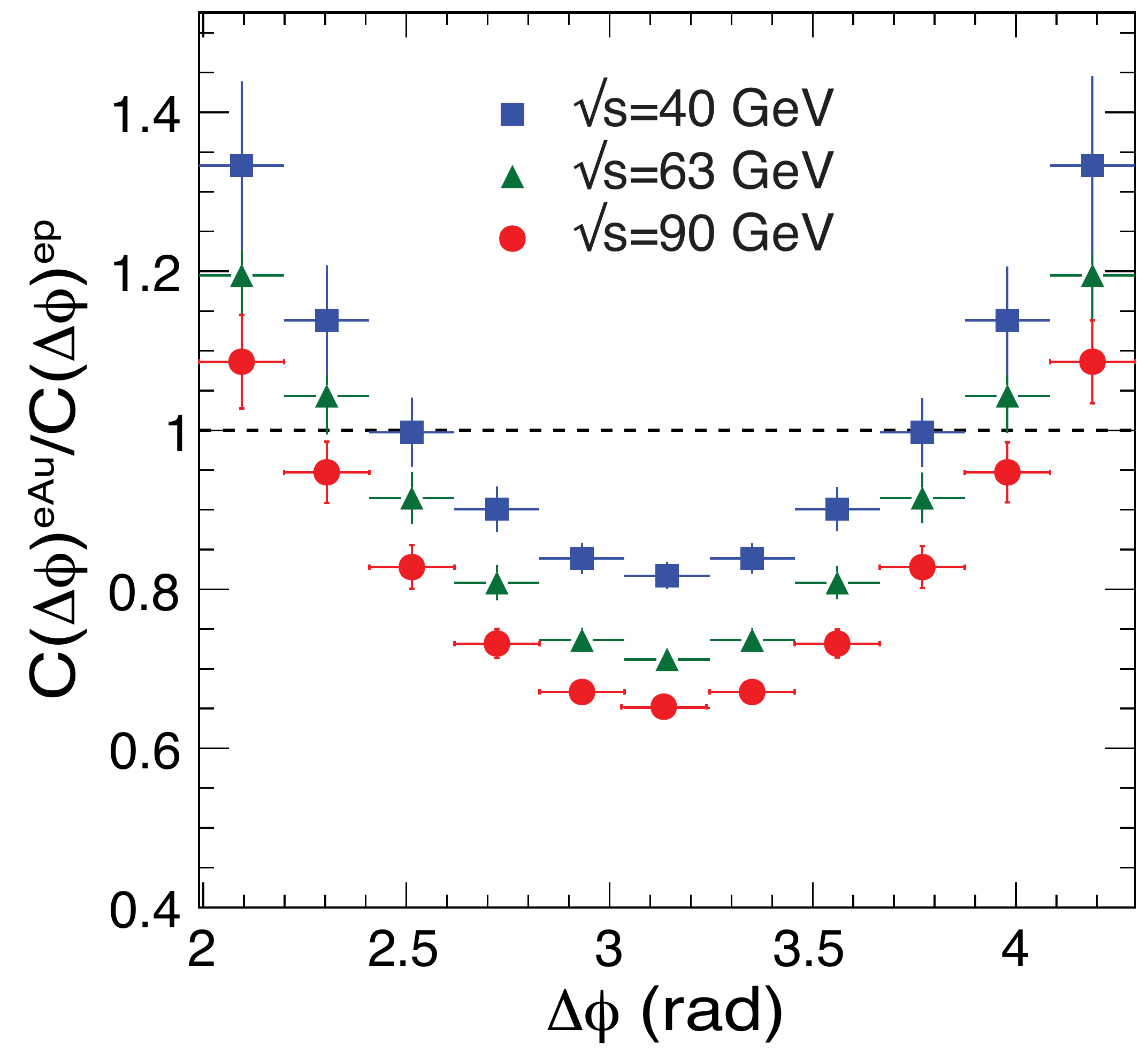}
		\caption{\label{fig:dihadronRatio} Ratio of the dihadron correlation functions in \eAu\ 
			collisions over those in \ep\ for the three center-of-mass energies.}
		\label{fig:file1}
		\vspace{2.6\baselineskip}
	\end{minipage}
	\hfill
	\begin{minipage}[b]{0.49\textwidth}
		\centering
		\includegraphics[width=\columnwidth]{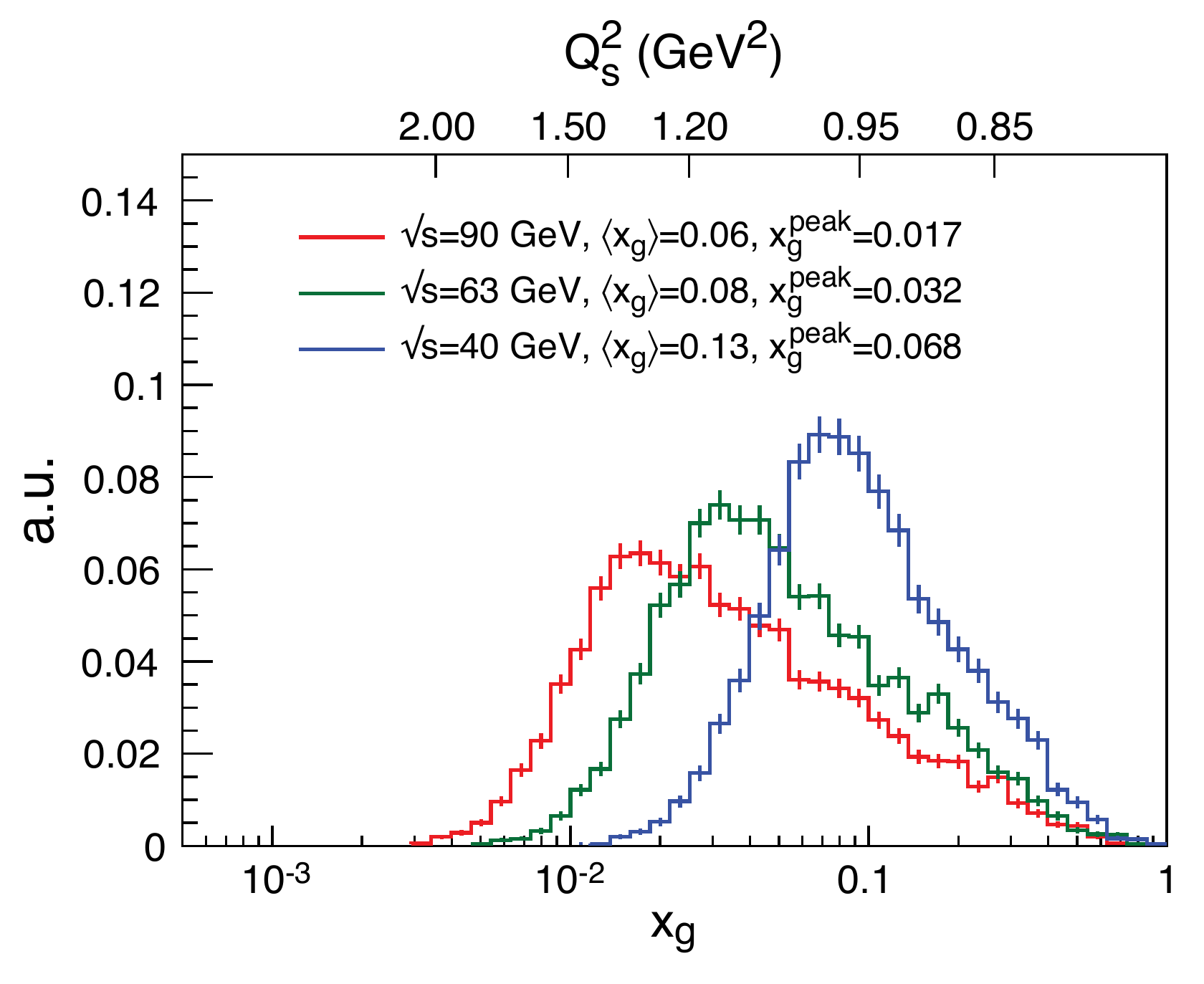}
		\caption{\label{fig:xgCurves} $x_g$ distributions probed by the correlated hadron pairs for 
			different center-of-mass energies, \sqrts = 40, 63, and 90 GeV in \eAu\ collisions. The 
			average and peak values for the distributions are shown. The gluon saturation scales 
			$Q^2_s$ corresponding to $x_g$ values are displayed on top of the plot. }
		\label{fig:file2}
	\end{minipage}
\end{figure}

To better illustrate the energy dependence, we plot in Fig.~\ref{fig:dihadronRatio} the 
ratio of the correlation functions in \eAu\ over those in \ep\ for all three energies. 
Note that the suppression is stronger by a factor of $\sim 2$ at \sqrts = 90 GeV when 
compared to the lowest simulated energy of 40 GeV. Measuring a suppression greater than 20\% relative to \ep\ will be crucial in the comparison of data with saturation model calculations that typically  carry uncertainties of at least in this order \cite{Zheng:2014vka}. The ability to study dihadron suppression over a wide range of $x_g$ is of the utmost importance for this observable. 

Fig.~\ref{fig:xgCurves} shows the corresponding $x_g$ distributions for dihadrons produced 
at the three different center-of-mass energies discussed. The larger the energy, the 
smaller the $x_g$ values one can access, and the further we reach into the saturation 
regime. Since the saturation scale is a function of $x_g$ alone, we also show the reference 
$Q_s^2$ values on the top of the plot. Only a sufficiently wide lever arm will allow one
to study the non-linear evolution in $x_g$ and $Q^2$ and extract the saturation scale 
with high precision. As we have shown, this requires center-of-mass energies in the 
range of \sqrts=90 GeV.
\end{multicols}

\vfill
\FloatBarrier
			\subsubsection{Diffraction}
					\newcommand{\xpom}{{x_\mathbb{P}}}
\label{subsection:diffraction}
\begin{multicols}{2}
In diffractive DIS, the incoming electron interacts with the target proton or nucleus without exchanging net color. Experimentally, such a scattering manifests itself as a rapidity gap in the detector between the target remnants and the diffractively produced system. This is an indication of a colorless exchange with ``vacuum" quantum numbers between the projectile fragments and that of the target. In contrast, if the exchange carried color, confinement dictates that the QCD string corresponding to this exchange would fragment into a shower of hadrons that fill up this gap in rapidity. Diffractive experiments are therefore in principle outstanding probes of how confinement operates. However, as we will discuss, diffraction is also a sensitive probe of saturation. Novel diffractive measurements that are feasible at the EIC thence offer an opportunity to study phenomena that are sensitive to the strong color fields generated by both weak and strong coupling dynamics. 

\vspace{4mm}
\begin{figurehere}
	\centering
	\includegraphics[width=\columnwidth]{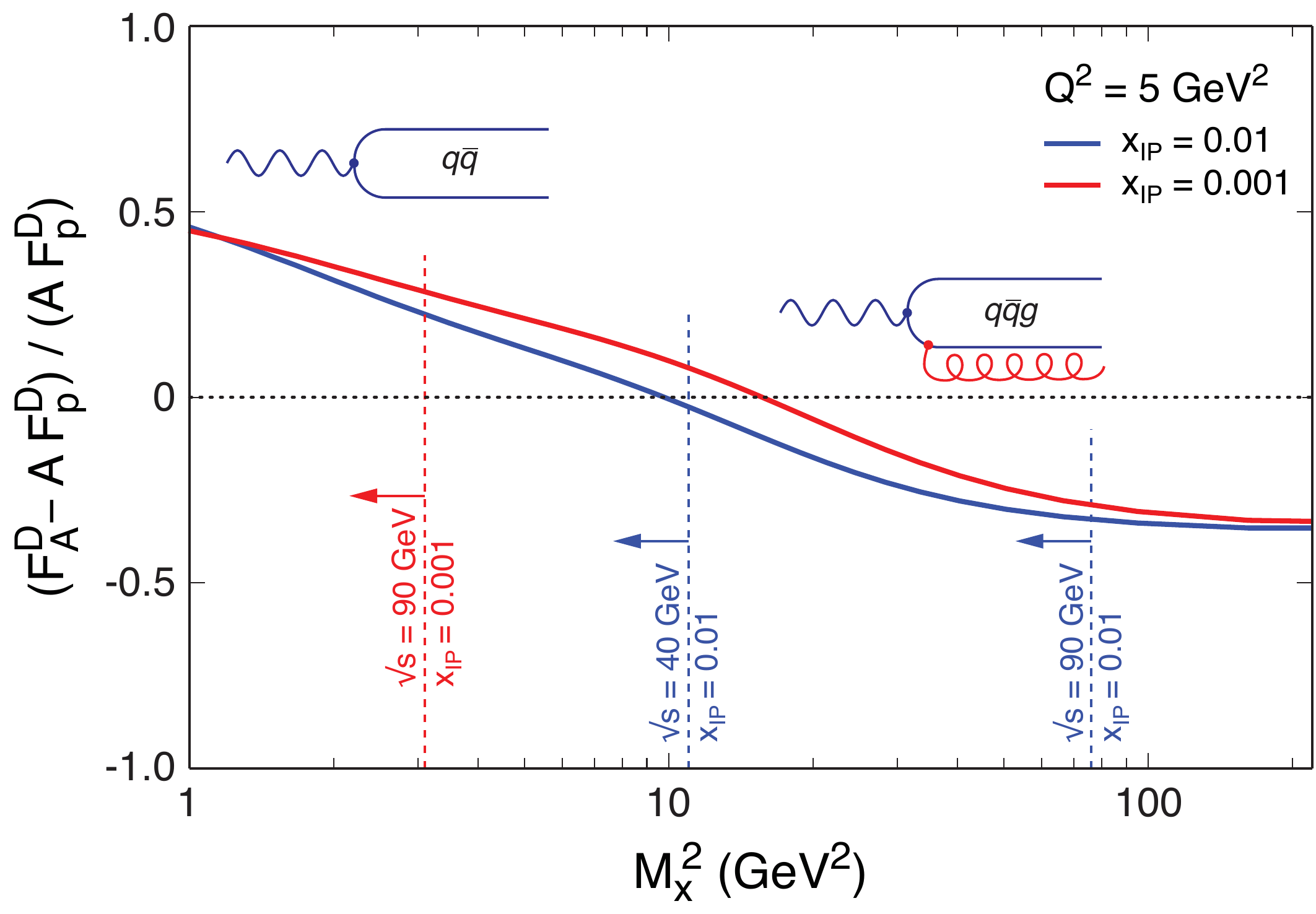} 
	\vspace{-5mm}
	\caption{Relative modification of the diffractive structure function of the nucleus as a function of the invariant mass of the diffractive system for two different values of $x$ and $Q^2 = 5$ GeV$^2$.}
	\label{fig:diffractive_structure_function_ratio}
\end{figurehere}
\vspace{4mm}

At leading order in perturbative QCD, diffractive scattering can be understood as the color singlet exchange of two gluons. The  cross-section is therefore proportional to the \emph{square} of the target's gluon distribution function. Diffractive events are also sensitive to the geometric structure of hadrons. As noted in Sec.~\ref{section:imaging}, the Fourier transform of the $t$ dependence of diffractive cross-sections gives the spatial distribution of gluon configurations inside hadrons. 

In the dipole picture we discussed previously in Sec.~\ref{sec:dipole}, the strong sensitivity of diffractive DIS to gluon saturation manifests itself as a strong dependence of the cross-section on the ratio $Q_s^2/Q^2$. Specifically, it is proportional to the squared dipole amplitude, and has the functional form:
\begin{equation}
\left[1 - e^{-Q_s^2/Q^2}\right]^2.
\end{equation}

Since the  cross-section depends on {\it both $Q_s^2$ and $Q^2$}, a large lever arm in $Q^2$, in addition to that in $x$, can dramatically reveal the onset of saturation. This was illustrated  in Fig.~\ref{fig:q2qs_shaded} of Sec.~\ref{sec:dipole}, where the kinematically allowed $Q^2$ range where the dipole amplitude is probed is shown for the two different EIC energies. It was also shown there that the impact of a wider lever arm is much greater for diffractive processes relative to that for inclusive DIS final states.

\begin{figure*}[t!]
	\center
	\includegraphics[width=0.495\linewidth]{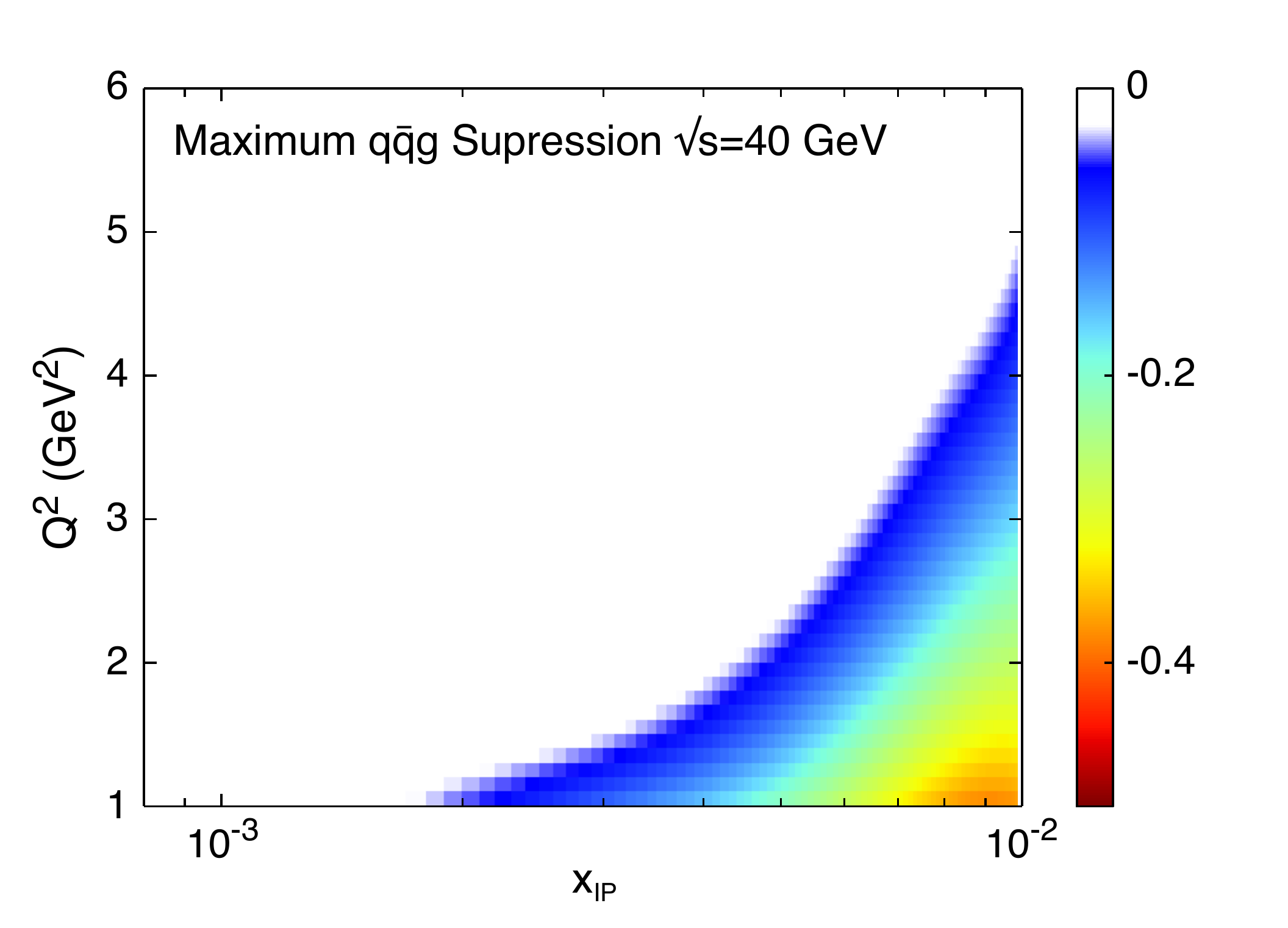}
	\includegraphics[width=0.495\linewidth]{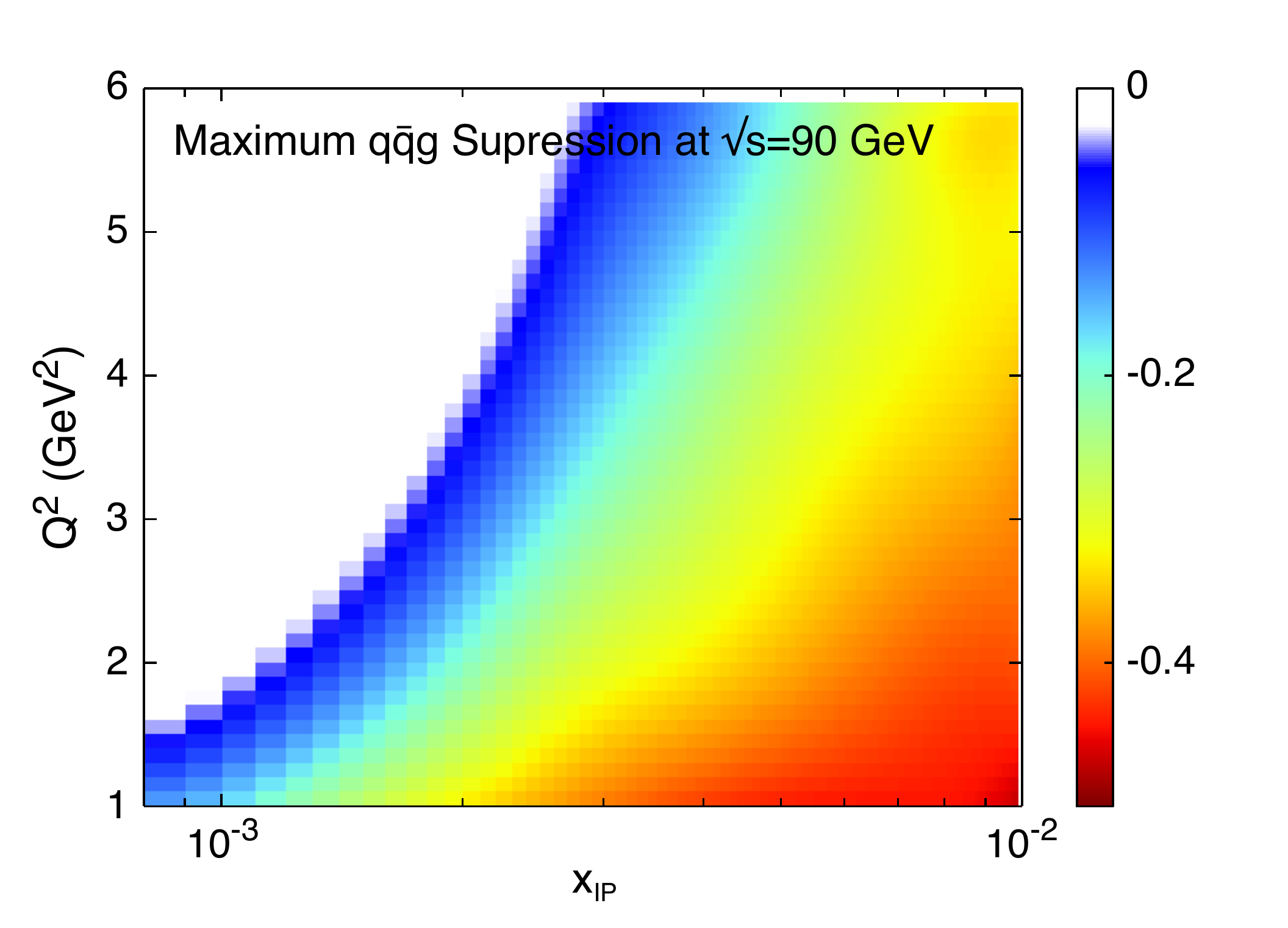}
	\vspace{-7mm}
	\caption{Maximum $q\bar q g$ dipole suppression observable at \sqrtsNN\ = 40 GeV (left) and at \sqrtsNN\ = 90 GeV (right).}
	\label{fig:qqg}
\end{figure*}

A striking result from HERA was that approximately $15\%$ of the \ep\ scattering events were diffractive~\cite{Abramowicz:1998ii}. As discussed in the White Paper, a strong prediction of the saturation picture is that the ratio of diffractive to total cross-section should be enhanced in a heavy nucleus compared to that of the proton. This is in stark contrast to leading twist shadowing calculations that predict a suppression.

A simple observable to study nuclear effects is the diffractive structure function, $F^D$, which is proportional to the total diffractive cross-section. At high energy, the dipole picture of diffraction predicts that the virtual photon emitted by the electron fluctuates into $|q\bar q\rangle$, $|q\bar q g\rangle$, or  higher ``Fock" states, which then scatter elastically off the target without exchanging net color. These states subsequently  hadronize to a sytem with invariant mass $M_x$. Smaller values of $M_x$ are primarily from the fragmentation of the $|q\bar q\rangle$ while large $M_x$ values correspond to significant contributions from 
the higher Fock states. In saturation model computations~\cite{Kowalski:2008sa}, the $q\bar q$ dipole contribution (small $M_x$) is enhanced in a nucleus relative to a proton. In contrast, because the strong color field of the nucleus absorbs a $q\bar q g$ dipole more strongly than the proton, the diffractive cross-section at large $M_x$ is suppressed in the nucleus relative to a proton. 

This is demonstrated in Fig.~\ref{fig:diffractive_structure_function_ratio} that depicts the  modification of the diffractive structure function in heavy nuclei, $F^D_A$, relative to that of the proton, $F^D_p$, as a function of the invariant mass of the diffractive system.  The corresponding kinematical coverages of the two proposed EIC energies, $\sqrt{s}=40\gev$ and $\sqrt{s}=90\gev$ are shown. At  $Q^2=5$ GeV$^2$, where one has perturbative control over the DIS probe, we observe for $\sqrt{s}=90$ GeV that one can scan the entire region in $M_x$ where one sees a sign change from enhancement to suppression. 

The variable $\xpom$ denotes the momentum fraction of the colorless exchange with respect to the hadron.
The logarithm of $\xpom$ is proportional to the size of the rapidity gap; for $\xpom\leq 0.01$, a clean separation in rapidity exists between projectile and target fragments. Observation of such a sign change, for the kinematics noted, would provide strong evidence that DIS is probing the gluon saturation regime. On the other hand, while the nuclear enhancement of the diffractive cross-section can be identified at $\sqrt{s}=40$ GeV, the sign change is out of reach. For the larger gaps of $\xpom=0.001$, the sign change is not accessible at either energies; however, for $\sqrt{s}=90$ GeV, studying the QCD evolution of the diffractive structure function with $\xpom$ is feasible.

The maximal suppression of diffractive cross-sections in the nucleus relative to the proton that can be measured at both design energies is shown in Fig.~\ref{fig:qqg}. One can investigate for how wide a region in  $\xpom$ -- $Q^2$  this suppression is seen. Towards this end, we compute the same ratio as the one plotted in  Fig.~\ref{fig:diffractive_structure_function_ratio}, at the highest kinematically allowed $M_x^2$. Recall that the greatest suppression of the ratio of cross-sections is at the largest $M_x^2$. This ratio is only shown in the region where the sign is negative; it is only at the highest EIC energy that a significant window in $\xpom$ -- $Q^2$ exists. This is especially the case if we demand $Q^2$ be large enough for a perturbative treatment of the DIS probe. 

 \begin{figurehere}
 	\vspace{4mm}
	\centering
	\includegraphics[width=\columnwidth]{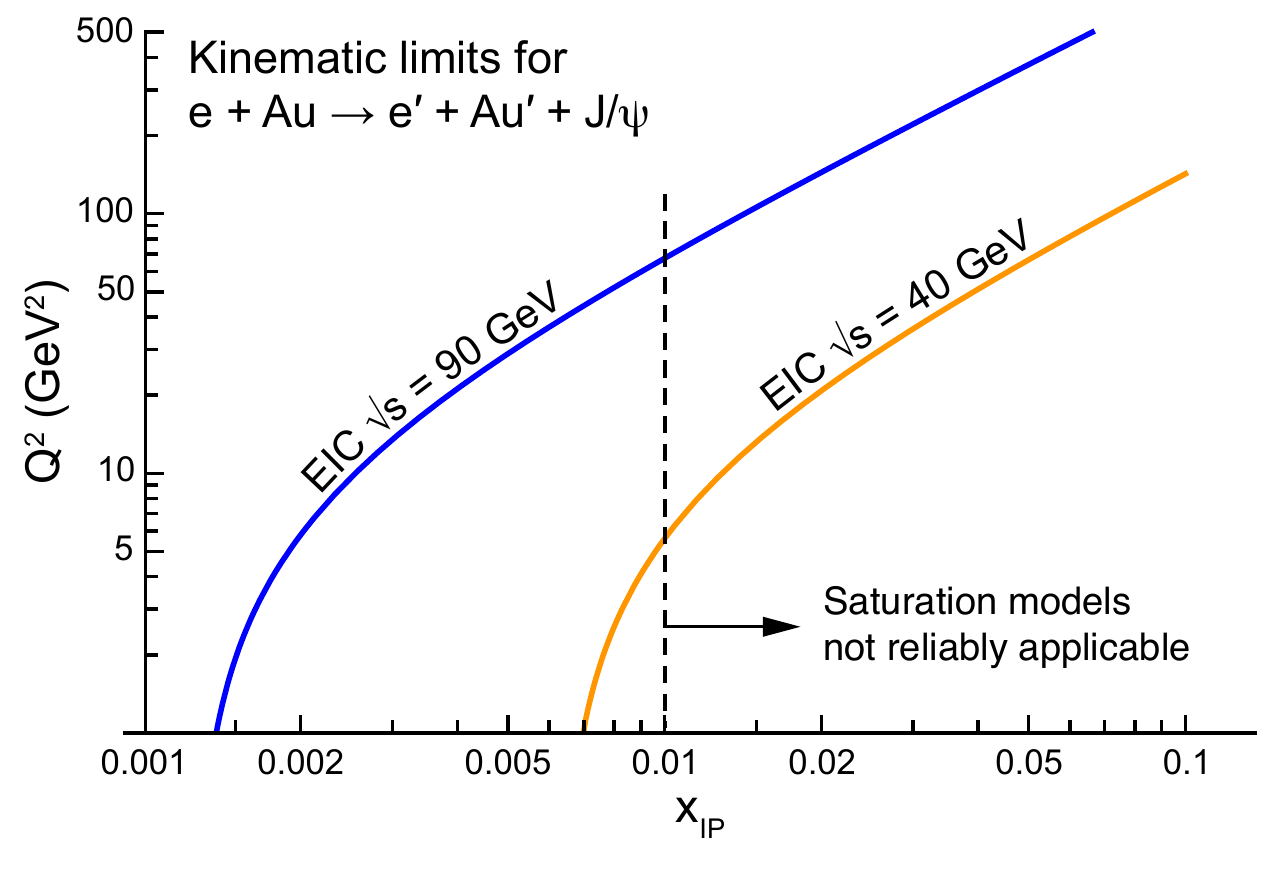} 
	\vspace{-8mm}
	\caption{Kinematical coverage for diffractive $J/\Psi$ production.}
	\label{fig:jpsi_diffraction_coverage}
 	\vspace{8mm}
\end{figurehere}

In addition to inclusive diffraction, exclusive vector meson production provides an additional handle to study small $x$ gluon distributions. The advantage is that such a process is experimentally clean, and it is  easy to measure the squared transverse momentum transfer $t$ to the target. As noted previously, the momentum transfer is Fourier conjugate to the impact parameter. Therefore, as discussed in Sec.~\ref{section:imaging},  measuring exclusive vector meson production differentially in $t$ makes it possible to study the impact parameter profile of the small $x$ structure. Recent phenomenological studies have shown that exclusive $J/\Psi$ production can be used to construct the average density profile, and its event-by-event fluctuations, for both protons and nuclei~\cite{Toll:2012mb,Mantysaari:2016ykx,Mantysaari:2016jaz}. 
As we show here in Fig.~\ref{fig:jpsi_diffraction_coverage},  \sqrtsNN\ = 90 GeV is needed in order to have access to $J/\Psi$ production in the region where saturation model computations are valid.

\end{multicols}
\FloatBarrier	
	 \subsection{Jet Physics}
	       \label{section:jets}

Since the earliest days of collider physics, jets have been an important tool 
in the exploration of QCD and have provided important discoveries and insights in all 
colliding systems, including $e^{+}$+$e^{-}$, \ep\, hadron+hadron, and nucleus+nucleus. (See for example \cite{Ali:2010tw}.) With the advances in experimental techniques, and corresponding advances in 
theoretical understanding over time, jets have become precision tools for studying the parton structure of matter.  Jets are guaranteed to contribute at the EIC to a variety of key electron-nucleus and electron-hadron physics topics, such as:

\begin{itemize}
	\item The study of hadronization, to shed light on the nature of color neutralization and confinement (see \cite{Boer:2011fh})
	\item Parton shower evolution in strong color fields to measure cold nuclear matter transport coefficients (see \cite{Boer:2011fh})
	\item The study of diffractive dijet production, which can possibly provide  direct access to the gluon Wigner function (see \cite{Hatta:2016dxp,Hagiwara:2016kam})
	\item Constraints on high-$x$ quark and gluon PDFs
	\item Precision measurements of  (un)polarized hadronic photon structure (see \cite{Chu:2017mnm})
	\item Measurement of the gluon helicity distribution in the proton, $\Delta g$, and its evolution via the photon-gluon fusion process. 
\end{itemize}

\begin{figure}[h!]
	\centering
	\begin{minipage}[b]{0.49\textwidth}
		\centering
		\includegraphics[width=\textwidth]{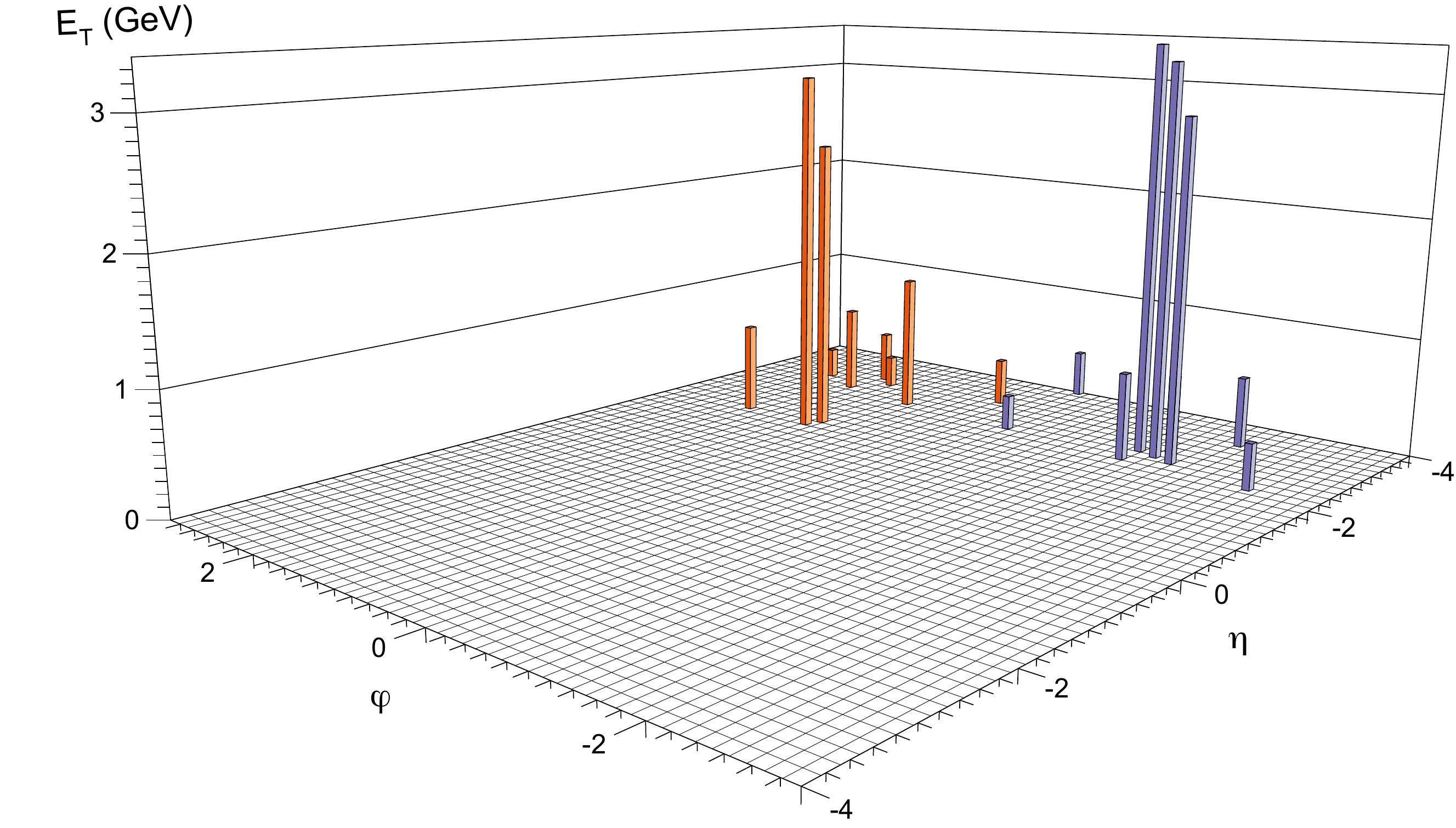}
		\caption{Visualization of a typical \ep\ dijet event showing the transverse energy and distribution of particles in rapidity-phi space in the Breit frame.}
		\label{fig:DIJETEVENT}
	\end{minipage}
	\hfill
	\begin{minipage}[b]{0.49\textwidth}
		\centering
		\includegraphics[width=\textwidth]{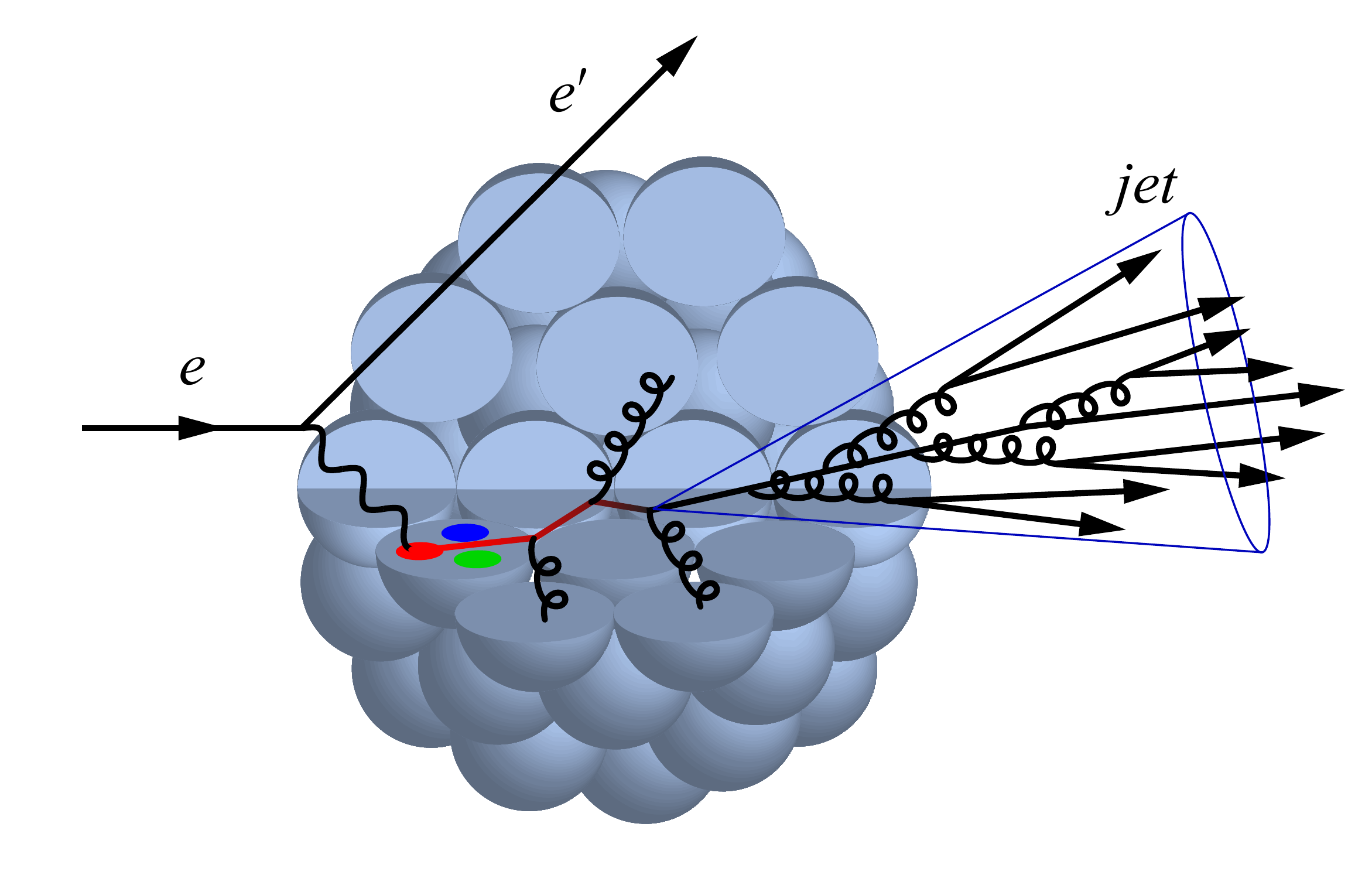}
		\caption{Schematic depiction of a struck parton propagating through cold nuclear matter resulting in the formation of a single jet.}
		\label{fig:EAJETEVENT}
	\end{minipage}
\end{figure}

While jets are familiar objects in high-energy physics analyses, 
thanks to the modern colliders such as the LHC \cite{Ellis:2007ib}, those produced at an EIC 
will have important differences to the ones produced at hadron-hadron 
machines. Figure 
\ref{fig:DIJETEVENT} presents a typical \ep\ dijet event arising from the 
photon-gluon fusion process. In contrast to events in hadron-hadron scattering, \ep\ events 
are very clean, with little energy present that is not associated with the 
jets. Additionally, the jets themselves contain relatively few particles, and 
the particles have moderate energies. This will make precision tracking essential 
for control of jet energy scale systematics.

There are at least two features inherent to jets which make them attractive 
alternatives to single-hadron observables. The first such feature is that 
they are better surrogates for the scattered partons. As jets contain more of the final state particles which 
arise as a parton fragments, they represent more accurately the energy and 
momentum of the initial parton. This is important in cases where it is 
necessary to reconstruct the partonic kinematics.  Examples of these are 
determining $x_\gamma$, 
the momentum fraction of the parton with respect to the exchanged photon, to isolate resolved photon events for studies of the 
partonic structure of the photon or when studying the azimuthal correlations 
discussed in Sec. \ref{subsection:dijets}. The second advantage lies 
with the fact that jets have substructure which can be characterized and 
studied systematically. A comparison of substructure in \eA\ to baseline \ep\ collisions,
as well as a comparison of jets which form outside and inside the nuclear medium (illustrated in Fig.~\ref{fig:EAJETEVENT}), 
will provide a wealth of information about the propagation of 
partons through nuclear matter and the dynamics underlying the emergence of 
hadrons from colored partons \cite{Boer:2011fh}.

\FloatBarrier

	       \subsubsection{Effect of Collision Energy on Jet Observables}	       
			\label{section:jetsKin}

\begin{multicols}{2}

To facilitate the studies presented in this section, DIS events were 
generated using PYTHIA 6.4 \cite{Sjostrand:2006za} and stable particles with 
transverse momenta above 250~MeV/$c$ were clustered into jets in the Breit 
frame using the anti-k$_T$ algorithm \cite{Cacciari:2008gp} as implemented in the 
FastJet package \cite{Cacciari:2011ma}. 
When selecting a dijet event, it was required that one of the jets had $p_{T}$ 
greater than 5~GeV/$c$ and the other greater than 4~GeV/$c$. It was also 
required that the jets be more than 120 degrees apart in azimuthal angle.

\begin{figure*}[t!]
	\centering \includegraphics[width=0.86\linewidth]{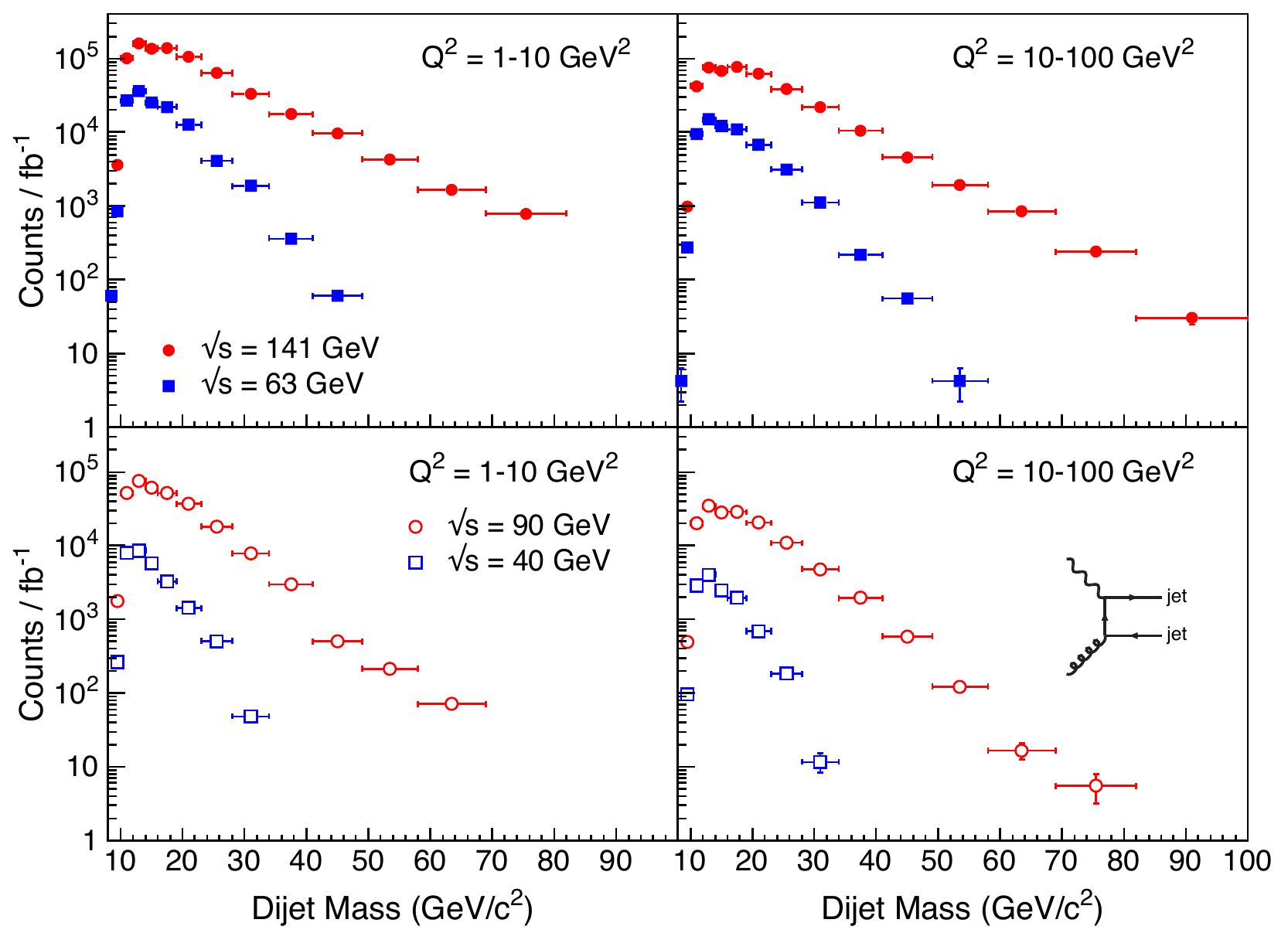}
	\caption[Dijet Cross Section]{Dijet yields as a function of invariant mass  scaled to a luminosity of 1 fb$^{-1}$ for $Q^2 = 1-10$~GeV$^2$ (left column) and $Q^2 = 10-100$~GeV$^2$ (right column). The top row compares proposed \ep\ center-of-mass  energies while the bottom row compares \eA\ energies.}
	\label{fig:DIJETMASS}
\end{figure*}

Because jets are expected to be important observables for many physics topics, 
it is crucial that an EIC be in position to utilize them to their fullest 
potential. There are several aspects of jet measurements which benefit from higher center-of-mass 
energies. Some of these benefits can be seen in Fig.~\ref{fig:DIJETMASS},
which shows expected dijet yields as a function of invariant mass assuming 
1 fb$^{-1}$ of integrated \ep\ 
luminosity for two $Q^2$ ranges and several $\sqrt{s}$ values representing 
proposed \ep\ and \eA\ energies. We see that the production cross-section increases with $\sqrt{s}$, compensating for the high luminosities that would otherwise be needed to collect sufficient statistics. We see this to be especially true for high invariant mass, where the dijet cross-section drops much more rapidly for lower energies. 

High energies will be needed to take advantage of the largest jet $p_{T}$ and dijet mass ranges, which will be important for characterizing observables such as 
cross-sections and spin asymmetries (among others) that depend on these variables \cite{Chu:2017mnm,Klasen:2017kwb}.
It is worth noting that the correction of raw jet spectra by means of the standard unfolding procedures
is affected by the steepness of the spectra. Higher energies imply harder spectra, as illustrated 
in Fig.~\ref{fig:DIJETMASS}, and subsequently require smaller unfolding factors thus reducing the overall
uncertainties in the final spectra.

As seen above, access to higher $p_{T}$'s enabled by larger collision energies is 
important for extending the kinematic reach of jet measurements. In 
addition, the significant yield of jets at high $p_{T}$ may be important for 
certain analyses 
which utilize the substructure of these objects.
Figure \ref{fig:DIJETEVENT} shows that a typical jet at an EIC will contain 
relatively few particles. This is quantified in Fig.~\ref{fig:NUMPARTS},
which shows (for inclusive jets) the average number of particles inside a jet 
as a function of the jet 
transverse momentum for two $Q^2$ ranges and $\sqrt{s}$ values. It is seen that the particle content of a jet grows as a function of $p_{T}$ and is largely 
insensitive to $Q^2$ or $\sqrt{s}$. 

Analyses which utilize jet substructure have become quite important in both hadron-hadron \cite{Altheimer:2012mn,Aad:2011kq,Chatrchyan:2013kwa} and heavy ion 
\cite{Connors:2017ptx} collisions and will certainly be important in both \ep\ and \eA\ at an EIC. Substructure will be invaluable to the study of the nuclear 
medium, where observables such as the jet fragmentation function, jet profile, and first 
splitting (among many others) will quantify jet modifications and thus shed light on how partons lose energy while traversing the medium. 

Observables sensitive to 
jet substructure can also be used to identify and separate jets that arise  

\begin{figurehere}
	\centering \includegraphics[keepaspectratio=true,
	width=\columnwidth]{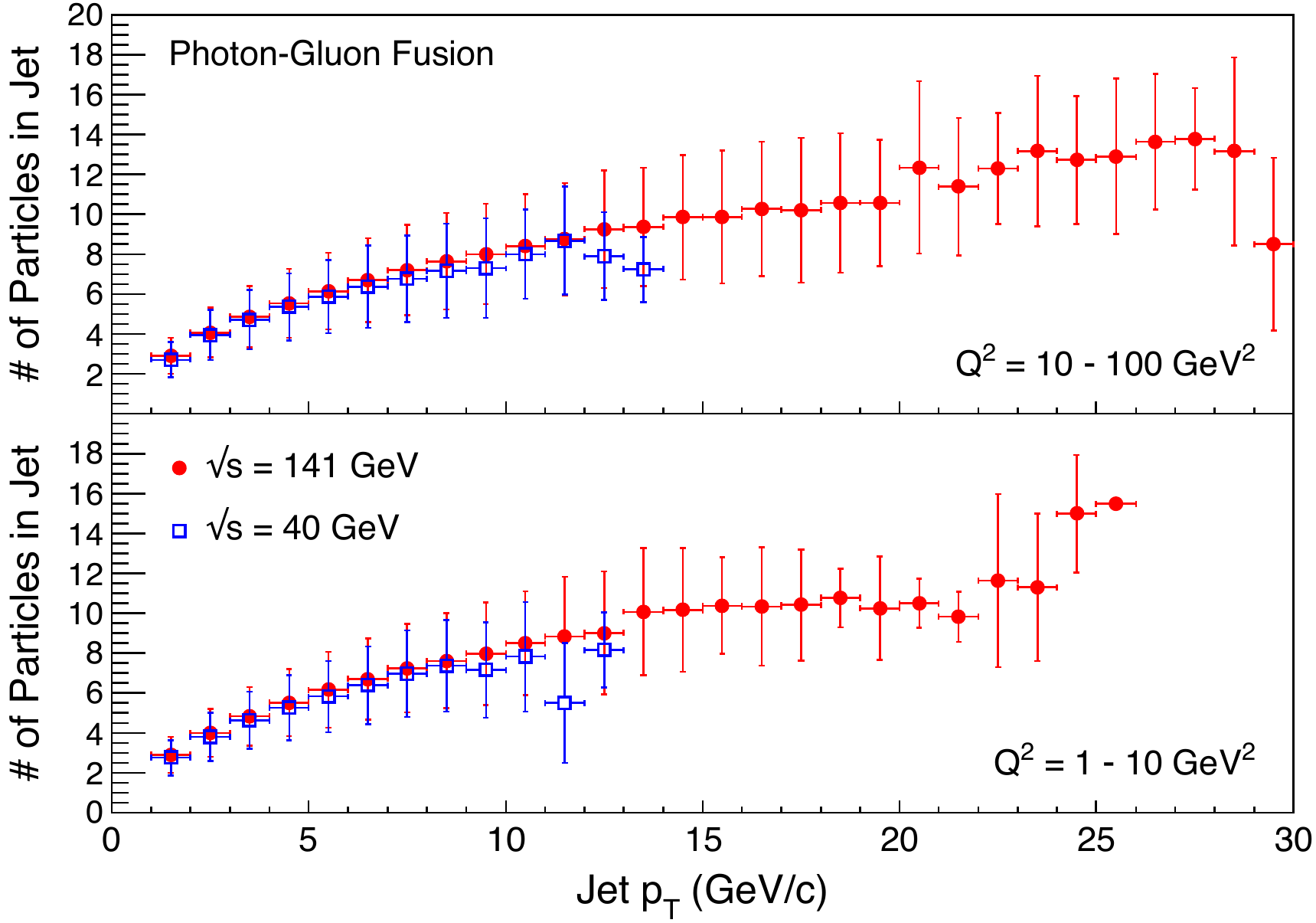}
	\caption[Number of Particles]{Average number of particles with transverse momentum greater than 250 MeV/$c$ within a jet as a function of jet transverse momentum (with the root-mean-square of the distribution represented by the error bars) for $Q^2 = 1-10$~GeV$^2$ (bottom) and $Q^2 = 10-100$~GeV$^2$ (top) and two center-of-mass energies.}
	\label{fig:NUMPARTS}
\end{figurehere} 
\noindent
from quarks versus those from gluons as the different fragmentations of the two parton types will 
lead to different energy distributions within a jet. For measurements which take into account 
jet substructure, it will be important that the jet contains 
enough particles to construct meaningful observables. The larger yield of high 
$p_{T}$ jets available at higher center-of-mass energies will ensure that 
precision measurements utilizing jet substructure can be carried out.

\end{multicols}

\FloatBarrier

		    \subsubsection{Azimuthal Asymmetry in Dijets}
\label{subsection:dijets}

	\begin{figure}[b!]
	\begin{center}
		\includegraphics[width=\textwidth]{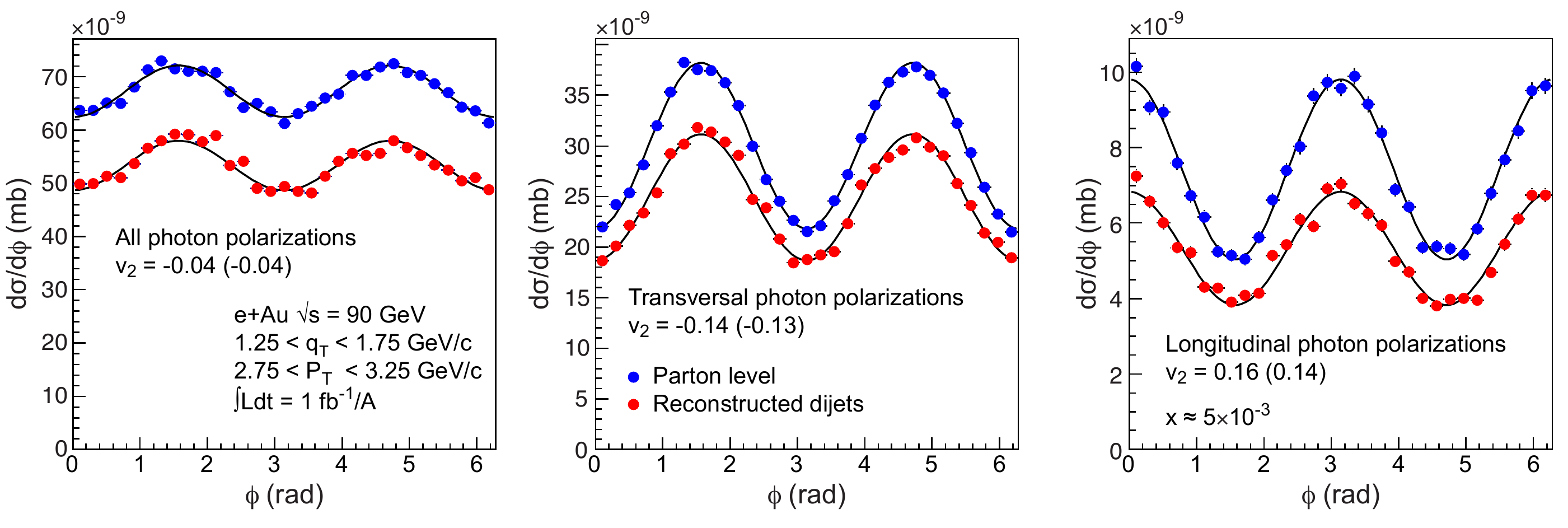}
	\end{center} 
   \vspace{-6mm}
	\caption{\label{fig:dijetCombo} $\protect\mathrm{d}\sigma/\protect\mathrm{d}\phi$ 
		distributions for parton pairs (blue points) generated with the MC-Dijet 
		\protect\cite{mcdijet2017} generator and corresponding reconstructed dijets (red points) 
		in $\sqrt{s}$=90 GeV \eAu\ collisions for $1.25 < q_T < 1.75$ GeV/$c$ and $2.75 < P_T <
		3.25$ GeV/$c$. The error bars reflect an integrated luminosity of 1 fb$^{-1}$/A. The left 
		plot shows the azimuthal anisotropy for all virtual photon polarizations, and the middle 
		and right plot for transverse and longitudinal polarized photons, respectively. For 
		details, see text.
	}
\end{figure} 

\begin{multicols}{2}
	
Progress has been achieved in developing frameworks to extend our understanding of parton structure beyond the one dimensional PDFs. One such example are the GPDs that provide information on the  transverse spatial structure of hadrons. Transverse momentum dependent parton distributions (TMDs) provide another powerful example~\cite{Collins:1981uw,Mulders:2000sh,Meissner:2007rx}. The TMDs depend not only on the longitudinal momentum fraction $x$ of the parton but also on its transverse momentum $k_T$ and therefore contain much more detailed information on the internal structure of polarized and unpolarized protons relative to the PDFs.

\begin{figure*}[b!]
	\begin{center}
		\includegraphics[width=\textwidth]{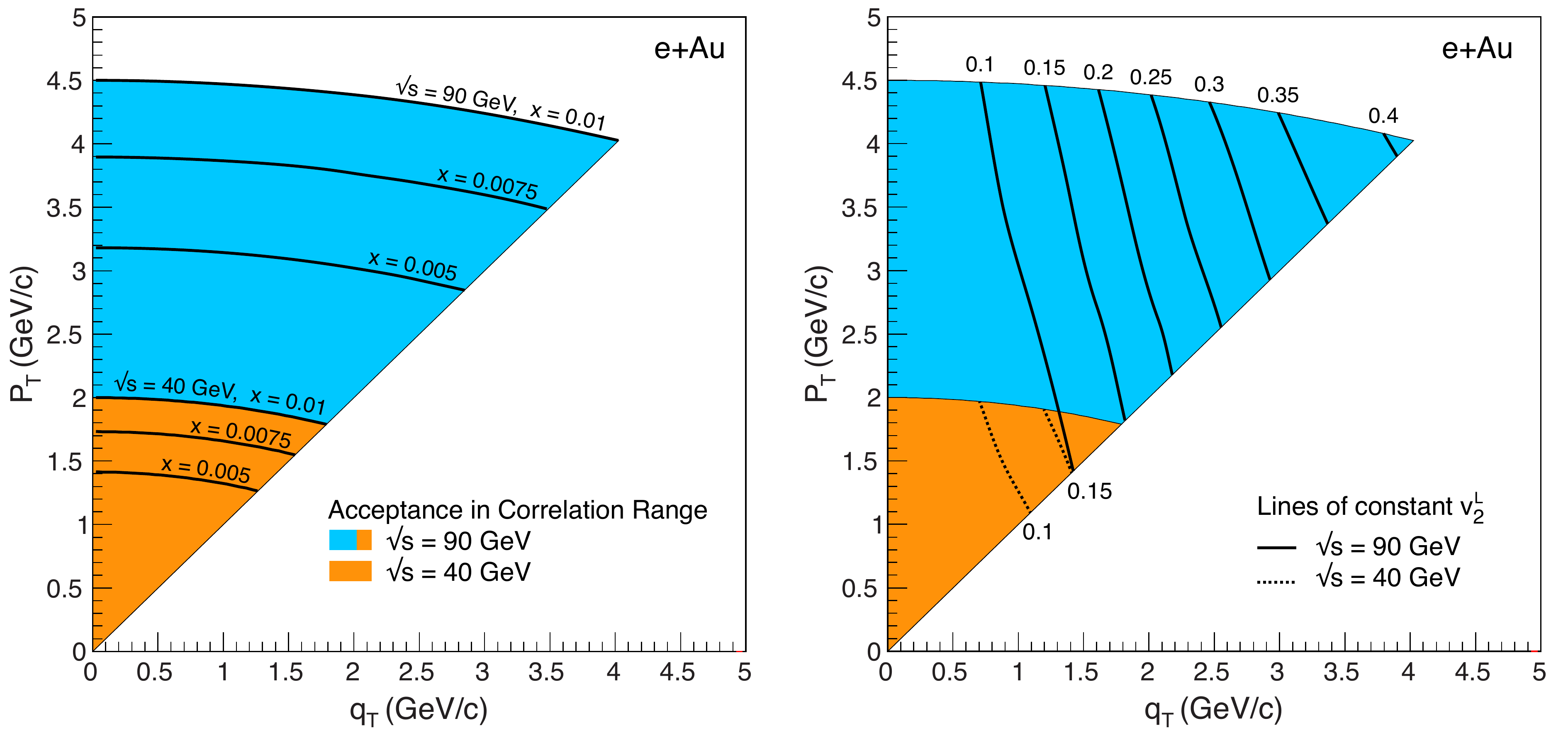}
	\end{center} 
\vspace{-4mm}
	\caption{\label{fig:KinePieCombo} Kinematic range in $q_T$ versus $P_T$ of the relevant 
		correlation region, ${q_T} < {P_T}$, for two EIC energies, \sqrts=40 and 90 GeV. On the 
		left plot we depict lines of constant $x$ for the referring energies and on the right we 
		show lines of constant azimuthal anisotropy for longitudinally polarized virtual photons.}
\end{figure*} 

Thus far, the main focus of studies has been on quark TMDs while the available studies of gluon TMDs are rather sparse. Of particular interest, is the distribution of linearly polarized gluons inside an unpolarized hadron, $h_\perp^{(1)}$ \cite{Boer:2009nc,Metz:2011wb}. It has been shown that this distribution can be accessed through measuring azimuthal
anisotropies in processes such as jet pair (dijet) production in \ep\ and \eA\ scattering 
\cite{Boer:2010zf,Dominguez:2011br,Dumitru:2014yza,Dumitru:2014vka,Dumitru:2015gaa,Dumitru:2016jku}. 
Furthermore, it is recognized that these gluon distributions play a central role in
small $x$ saturation phenomena \cite{Dumitru:2015gaa}. Given the important {\em dual} role 
of these measurements, we conducted first feasibility studies of dijet measurements at an 
EIC in \eA\ collisions and investigated their dependence on $\sqrt{s}$.

Recent studies in \pp\ \cite{Khachatryan:2010gv} and \pPb\ collisions
\cite{Khachatryan:2015waa} at the LHC have revealed long-range near-side 
azimuthal angular $\cos 2\phi$ correlations for particle production in high multiplicity 
events. Such correlations are commonly referred to as  ``ridge'' correlations. They can also be quantified by measurements of $v_2 = \langle \cos 2\phi \rangle$. Since the azimuthal angle correlation in dijet production in \eA\ at high energies originates  from long-ranged eikonal interactions 
\cite{Dumitru:2015gaa}, one can make this connection explicit by parameterizing the 
azimuthal structure arising from the linearly polarized gluon distribution in terms of 
$v_2$, defined as in the above. However, the azimuthal angle $\phi$ is here defined as the angle between 
the transverse momentum vector of the jets, $\vec{P_T}$, and the transverse momentum 
imbalance, $\vec{q_T}$.

We studied the production of a $q\bar{q}$ dijet at leading order in the high energy (small $x$) scattering
of an electron off a gold nucleus. For our simulations, we used the MC-Dijet event generator \cite{mcdijet2017} to generate the correlated partons. The generator determines the distribution of linearly 
polarized gluons of a dense target at small $x$ by solving the B-JIMWLK renormalization 
evolution equation \cite{Balitsky:1995ub,Weigert:2000gi,Kovner:2000pt,Ferreiro:2001qy}. 
We restrict ourselves to kinematic configurations where ${q_T} < {P_T}$, referred to as 
the ``correlation limit" or TMD regime \cite{Dominguez:2011wm,Dominguez:2011br}. Only in 
this limit can the underlying theory be expressed in terms of a specific gluon TMD, part 
of which is the distribution of linearly polarized gluons, $h_\perp^{(1)}$ that one seeks to determine. 
The partons are then passed to fragmentation algorithms 
from the PYTHIA8 event generator \cite{Sjostrand:2007gs} for showering and hadronization 
into jets. After experimental acceptance and kinematic cuts, all remaining final state 
particles are used as input to a jet finder algorithm (FastJet \cite{Cacciari:2011ma}) and 
the relevant kinematic variables are calculated.

Figure \ref{fig:dijetCombo} shows the resulting $\mathrm{d}\sigma/\mathrm{d}\phi$
distributions for the original parton pairs (blue points) and the reconstructed dijets 
(red points) in \sqrts=90 GeV \eAu\ collisions for $1.25 < q_T < 1.75$ GeV/$c$ and $2.75 < 
P_T < 3.25$ GeV/$c$. The error bars reflect an integrated luminosity of 1 fb$^{-1}$/A. The 
left plot shows the azimuthal anisotropy for all virtual photon polarizations, and the 
middle and right plot for transversal and longitudinal polarized photons, respectively. 
The quantitative measure of the anisotropy ($v_2$) is listed in the figures. The values shown are those for  parton pairs; the accompanying numbers in parenthesis denote the values derived from the reconstructed dijets.

Note the characteristic phase shift of $\pi/2$ between the anisotropy in longitudinal 
versus transversally polarized photons. Despite this shift, the sum of both polarizations 
still adds up to nonzero net $v_2$ due to the dominance of transversly polarized 
photons, as shown in the leftmost plot in Fig.~\ref{fig:dijetCombo}. 
While the  polarization of the virtual photon cannot be measured directly in this process, one will be able to 
disentangle both contributions by either analyzing their dependence on $Q^2$, $q_T$, and $P_T$,
or by two-component fits constraining $v_2^L$ and $v_2^T$ using the relation $v_2 = (R v_2^L + v_2^T)/(1+R)$, where $R$ is a kinematic factor depending only on known kinematic quantities such as $Q^2$ and $P_T$.

The reconstructed dijets reflect the original anisotropy at the parton level remarkably 
well despite the dijet spectra not being corrected for efficiency in this study. The loss 
in dijet yield, mostly due to loss of low $p_T$ particles, is on the order of $\sim20$\%. This is seen from their cross-sections, when compared to that of parton pairs. We also 
studied background effects by applying the same kinematic cuts to \ep\ events generated by 
the PYTHIA event generator that contains no azimuthal anisotropies effects. We find, 
with the appropriate cuts, a signal to background level of around 4:1. The background
contributions show no modulations and can be easily subtracted via two-component fits.

To illustrate the energy dependence of this measurement, we plot in 
Fig.~\ref{fig:KinePieCombo} the kinematic range in $q_T$ versus $P_T$ of the relevant 
``correlation'' region  ${q_T} < {P_T}$, for two EIC energies, \sqrts=40 and 90 GeV. In 
the left plot, we depict lines of constant $x$ for these energies, and in the right, 
we show lines of constant azimuthal anisotropy for longitudinally polarized virtual 
photons ($v_2^L$). It becomes immediately clear that substantial anisotropies, $v_2^L \geq 
0.15$, can only be observed at the larger energy. 

From an experimental point of view, even more important is the magnitude of the average transverse momentum $P_T$. This is because  jet reconstruction requires sufficiently large jet energies to be feasible. The lower the jet  energy, the more particles in the jet cone fall below the typical tracking thresholds 
($p_T\sim 250$ MeV/$c$ in collider detectors), making jet reconstruction de facto 
impossible.  This point was also emphasized in Sec.~\ref{section:jetsKin}. Ultimately, the extraction of the gluon distribution $h_\perp^{(1)}$, requires a wide range of coverage in $q_T$, $P_T$ and 
thereby $x$ and $v_2$ \cite{Dumitru:2015gaa}, that only will be feasible at the higher EIC 
energy.

\end{multicols}

\FloatBarrier

%
%
\section{Summary}
			\label{sec:summary}

\begin{multicols}{2}
	
In this report, we assessed the case for an EIC with the highest energies discussed 
within the scope of the EIC White Paper. We began by first taking the big picture view
of understanding of the parton structure of 
protons and complex nuclei. We observed that while there are corners of the QCD 
landscape where a great deal is understood, there are fundamental questions across the 
wide swath of this landscape that demand answers. Chief among those is the puzzling 
confining many-body dynamics of quarks and gluons. Little is understood about how the 
parton structure of protons and nuclei changes under boosts, how quarks and gluons 
arrange themselves spatially, and what their distributions are in transverse momentum. 
How is the spin of the proton divided up amongst the helicities of quarks and 
gluons and their confined orbital motion? Are the quantum numbers of protons 
and nuclei carried only by valence partons with large momentum fractions $x\sim 1$ of 
the hadron's momenta, or are part of these also carried by the ``wee" small $x$ 
partons?

We know little of the correlations amongst partons, and how these change with energy 
and resolution across the landscape. We discussed the conjectured phenomenon of gluon 
saturation, whereby matter inside protons and nuclei is weakly coupled even though the 
corresponding color electric and magnetic fields are amongst the strongest in all 
nature. If this conjecture is confirmed by experiment, the physics of saturated gluons 
will be a remarkable example of fully nonlinear dynamical phenomena whose properties 
can nevertheless be computed. An understanding of this corner of the landscape may 
provide a window to other regimes of confining dynamics where equally strong color 
fields exist but where the absence of small parameters impair the comparison of theory 
to experiment. An example of such a phenomenon is the physics of how struck quarks and 
gluons transform into different species of hadrons.

The EIC brings penetrating and varied tools to obtain answers to the above questions. As 
the world's first DIS collider with polarized proton beams, it will have a much wider 
lever arm relative to previous experiments, to extract the helicity distributions of 
quarks and gluons and gain insight into their confined orbital motion. The high 
luminosities will enable the extraction of transverse spatial and momentum 
distributions of quarks and gluons. Charged-current probes of polarized protons will 
allow for clean extraction of polarized flavor distributions. Furthermore, the wide 
lever arm in $x$ will reveal the extent to which the proton's spin is transmitted to 
small $x$. In general, a significant contribution from small $x$ to the proton's 
quantum numbers may lead to novel empirical insight into the role of the QCD 
vacuum in hadron structure.

As the world's first electron-nucleus collider, the EIC will enable first measurements 
of nuclear gluon distributions in DIS. These will reveal the extent of nuclear shadowing, and 
how these distributions change with $Q^2$. Measurements of nuclear fragments, in 
coincidence with hadron final states, will provide novel information on the parton 
structure of the composite objects that transmit short-range nuclear forces. Nuclei are also a laboratory to understand 
multiple scattering, energy loss and hadronization of quarks and gluons in QCD media. 
How jet showers develop in the nuclear medium, and the propagation of heavy quarks in 
this medium, will be studied for the first time. It is anticipated that gluon 
saturation sets in at much larger values of $x$ (lower energies) in large nuclei than 
in the proton. The EIC should be able to unambiguously extract how the nuclear 
saturation scale evolves with $x$, and reveal its dependence on the atomic number. Such 
studies can also look for evidence that the physics of gluon saturation is 
universal--that phenomena in this regime do not depend on the quantum numbers that are 
specific to each nucleus.

To fully exploit the novel capabilities of this machine with its versatile polarized 
proton and nuclear beams, and in the high luminosities that will be available, it is 
important to assess the center-of-mass energy requirements of the key measurements that 
will fulfill the promise of a deep and varied exploration of the QCD landscape. To illustrate the energy dependence of these measurements, we chose two center-of-mass energy scenarios, $\sqrt{s} 
\approx 60$ GeV and 140 GeV for \ep\ and $\sqrtsNN \approx 40$ and 90 GeV for \eA\ 
collisions. We assumed further that the higher energy machine can also run at lower 
energies.

In assessing the energy case, we first examined the relevant lessons provided by past 
DIS experiments and those provided by heavy-ion experiments. We found that 
large lever arms in energy and resolution are crucial in maximizing the discovery 
potential in both of the cases studied. In the DIS case, scaling violation effects from 
the $Q^2$ evolution of quark and gluon distributions are more pronounced at small $x$. 
Because of this feature of the theory, further reach towards small $x$ has a bigger 
role in constraining uncertainties than simple kinematic considerations alone might 
suggest. As a specific case study, we explored in a simple dipole model, the impact of 
a larger lever arm in $Q^2$ for a possible discovery of gluon saturation. While the 
saturation scale $Q_s^2$ does not vary strongly with $x$ from low to high energies, the 
additional factor of 5 in $Q^2$ can have a dramatic effect. This is because inclusive, 
and especially, exclusive and diffractive measurements have a strong nonlinear 
dependence on $Q_s^2/Q^2$, not on $Q_s^2$ alone.

A striking effect of the dependence of a measurement on center-of-mass energy was 
observed in simulations of the helicity distributions of the proton.  Our studies 
showed a rapid shrinkage of the uncertainties in the parton helicities at the higher 
collision energy relative to the lower energy. This is because the uncertainties in the 
polarized $g_1$ structure functions grow very rapidly with decreasing $x$. Therefore, 
even a factor 5 enhanced reach relative to available data will strongly constrain these 
uncertainties.

The EIC White Paper laid out how high luminosities at an EIC, combined with a large 
kinematic reach, open up a unique opportunity to go far beyond our present largely 
one-dimensional ``images" of the proton. Our studies in this context illustrate that a 
high energy EIC will be the ideal machine for detailed quantitative studies of hard 
exclusive reactions and the unexplored sea quark and gluon GPDs. 
We showed that a large lever arm in $Q^2$  and a wide $x$ coverage are essential in determining the transverse spatial distribution of partons obtained from the measurement of the DVCS cross-sections.

The measurements of structure functions in DIS provide the main data set to unravel the 
internal structure of hadrons and nuclei and form the basis of all present-day PDFs. 
While highly precise measurements for the proton were performed at HERA, our knowledge 
of structure functions of nuclei is poor.  We presented in this report the results of 
simulations of the
 $F_2$, $F_L$, as well as $F^{c\bar{c}}_2$ and $F^{c\bar{c}}_L$ 
structure  functions of nuclei for the two center-of-mass energies.  We showed that the higher center-of-mass energy has a significantly larger impact on the extraction of the nuclear gluon distribution over the whole kinematic range. The relative uncertainties for the higher energy are approximately a factor 2 smaller compared to the lower energy range. While this lower energy range does improve our current knowledge of nuclear PDFs 
substantially at larger $x$, it provides only moderate constraining power at 
lower $x$, a range that is relevant for our understanding of \pA\ and \AA\ 
collisions at the LHC. We also studied the luminosity requirements of these 
measurements. We find that the experimental errors are dominated by systematic 
uncertainties over 
most of the available $x$ -- $Q^2$ phase space. 
In contrast with other measurements, these studies require only modest luminosities; statistics beyond a few fb$^{-1}$ do not improve the precision achieved.

The ability to reach low $x$ in \eA\ collisions is mandatory to explore the realm where 
gluon saturation can be measured. To study the energy dependence of probes sensitive to 
saturation we selected the suppression of dihadron production as an example of a key 
measurement of saturation. We find that only the larger energy provides enough lever 
arm to study the nonlinear evolution in $x_g$ and $Q^2$. At lower energies, the 
suppression is rather small and competes with systematic uncertainties in the 
measurements and in theoretical models.

As discussed in the EIC White Paper, a genuine prediction from the saturation picture 
is the enhancement of the ratio of diffractive to total cross-section in the nucleus 
compared to that of the proton. This is in contrast to conventional leading twist 
shadowing calculations that predict a suppression in this ratio. We expanded our 
studies to higher $Q^2$ values than those explored in the EIC White Paper. We find a 
transition from an enhancement of the diffractive structure function in nuclei to a 
suppression as a function of  increasing invariant-mass of produced particles. This 
``sign flip" signals the strong absorption of quark-antiquark-gluon and higher Fock 
states by the saturated gluon medium.  Our simulations also showed that this sign flip 
signature of saturation is not accessible at the lower center-of-mass energy.
 
We also added in this report the important topic of jet measurements that was omitted 
in the EIC White Paper.  Jets have become a precision tool in the exploration of QCD 
and have provided important discoveries and insights in many colliding systems. In 
addition to performing a general study of the effect of collision energy on jet 
observables, we looked at the measurement of azimuthal anisotropies in dijets. These 
conjectured asymmetries allow one to extract the transverse momentum dependent 
distribution of linearly polarized gluons in nuclei. The results of our studies are not 
surprising for energy hungry hard probes such as jets. We showed that a high energy EIC 
will be indeed essential to take advantage of the full potential of jet measurements. 
We also noted that the larger yield of high $p_T$ jets and the harder jet spectra 
available at the higher collider energies cannot be compensated by higher luminosities. 
The potential of an EIC for measuring dijet asymmetries at the lower energy is minimal.

In Table~\ref{Table:Summary}, we list the key measurements at an EIC for which the energy dependence
was assessed, either in this document or already in the White Paper~\cite{Accardi:2012qut}.
The last column ranks the role of the chosen higher energy, relative to that of the lower energy, for each measurement. 
Based on these results, we conclude that the greater reach provided by the higher energy chosen for our study greatly enhances the physics potential of an EIC and amplifies the discovery potential of these measurements. 


\end{multicols}
\newpage
\newcolumntype{M}{>{\centering\arraybackslash\hsize=.8\hsize}X}%
\newcolumntype{S}{>{\centering\arraybackslash\hsize=.6\hsize}X}
\newcolumntype{L}{>{\arraybackslash\hsize=2.\hsize}X}
\renewcommand\thefootnote{{\fnsymbol{footnote}}}
\setcounter{footnote}{1} 

{\small
\begin{tabularx}{\linewidth}{ | S  |  M  | L |  S |}%
\hline

	{\bf Process} & {\bf Observables} &  {\bf What we learn}  & {\bf Impact of high energy on physics}\\
	\hline \hline
	\multirow{11}{*}{\centering Inclusive DIS}   &     \multirow{8}{*}{Unpolarized $\frac{\text{d}^{2}\sigma}{\text{d}x\text{d}Q^{2}}$}  & Gluon momentum distributions $g_{A}(x,Q^2)$, nuclear wave function   &  \multirow{2}{*}{\centering Significant} \\  \cline{3-4}
	&  & Collective nuclear effects at intermediate-$x$  &  Significant \\   \cline{3-4}
	&  & $Q^{2}$ evolution: onset of DGLAP violation, saturation  & \multirow{2}{*}{\centering Indispensable} \\     \cline{3-4}
	&  & Beyond DGLAP A-dependence of shadowing and antishadowing &  \multirow{2}{*}{\centering Indispensable} \\  \cline{3-4}
	&  & Parton distribution functions in nuclei & Moderate \\  
	\cline{2-4}
	& \multirow{3}{*}{\parbox{3.cm}{\centering Polarized structure function $g_1$} } & Unravel the different parton contributions to the spin of the proton (polarized gluon distribution $\Delta G$, $\Delta \Sigma$, $L_q$, $g$) & \multirow{3}{*}{\centering Indispensable} \\
	\hline
	\multirow{19}{*}{\parbox{2.5cm}{\centering Semi-inclusive DIS}}    &  Production cross-section for identified hadrons~\footnotemark[\value{footnote}] &  Polarized quark and antiquark densities, quark contribution to proton spin; asymmetries like $\Delta \bar{u} - \Delta \bar{d}$; $\Delta s$ &  \multirow{3}{*}{\centering Indispensable} \\ \cline{2-4}
	& (Un)polarized cross-section in $W$ production~\footnotemark[\value{footnote}]  & Flavor separation at medium $x$ and contribution of quarks to the proton spin & \multirow{3}{*}{\centering Indispensable} \\  \cline{2-4}
	& Production of light and heavy identified hadrons~\footnotemark[\value{footnote}]  & Transport coefficients in nuclear matter, color neutralization: mass dependence of hadronization, multiple scattering and mass dependence of energy loss medium effect of heavy quarkonium production  &  \multirow{4}{*}{\parbox{2.4cm}{\centering Moderate}} \\  \cline{2-4}
	& Dihadron correlations  &  $k_T$-dependent gluons $f(x, k_{T})$; gluon correlations, non-linear QCD evolution/universality; saturation scale $Q_{s}$   &  \multirow{3}{*}{\centering Significant} \\ \cline{2-4}
	& Differential cross-sections and spin asymmetries with longitudinal and transverse polarization~\footnotemark[\value{footnote}]  & Sivers function \& unpolarized quark and gluon TMDs, quantum interference \& spin-orbital correlations, 3D imaging of quark and gluon's motion, QCD dynamics in an unprecedented $Q^2(P_{hT})$ range  & \multirow{6}{*}{\parbox{2.4cm}{\centering Moderate for quarks; significant for gluons}} \\ 
	\hline 
         \multirow{4}{*}{\parbox{2.5cm}{\centering Exclusive and diffractive DIS in $e+p$ \& $e+A$ collisions}}    &  Spin asymmetries, $\text{d}\sigma/\text{d}t$  &  GPDs, transverse spatial distributions of quarks and gluons; total angular momentum & \multirow{2}{*}{\centering Significant} \\  \cline{2-4}
	&  $\sigma_\text{diff}/\sigma_\text{tot}$, $\text{d}\sigma/\text{d}t$, $\text{d}\sigma/\text{d}Q^{2}$  &  Spatial distribution of gluons in nuclei; non-linear small-$x$ evolution; saturation dynamics  &  Indispensable \\ 
	\hline
	 \multirow{4}{*}{\parbox{2.5cm}{\centering Inclusive jets, dijets, photon-jet}}  &  $\text{d}\sigma/\text{d}t$ for diffractive dijets   &  Direct access to the gluon Wigner function   &   \multirow{2}{*}{\centering Indispensable} \\   \cline{2-4}
	&  (Un)polarized dijet cross-sections  &  Constraints for high-$x$ quark and gluon PDFs, and for (un)polarized photon PDF  & \multirow{2}{*}{\centering Indispensable} \\ 
	\hline
	\caption{\label{Table:Summary}Summary of key measurements at the EIC along with our assessment of 
		the  impact of the EIC White Paper's higher center-of-mass energy range on these measurements.}	
\end{tabularx}
}
\vfill
\footnotetext{Energy dependence already discussed in the EIC White Paper~\cite{Accardi:2012qut}.}
\newpage

\begin{appendices}
\section{Kinematic Variables}
			\label{app:variables}
\small
\begin{tabularx}{\linewidth}{ l   X }
	{\bf Variable} & {\bf Description} \\
	\hline
	A & Atomic Number  \\
    $b_T$ & Transverse position of parton inside a nucleon/nucleus. Often referred to as impact parameter.\\
    $\Delta g$ & Gluon helicity contribution to the total spin of the proton. $\Delta g$ is a function of $x$ and $Q^2$.\\
        $\Delta G$ & Integrated gluon helicity contribution to the total spin of the proton, i.e. $\int_0^1 \text{d}x \Delta g(x,Q^2)$. $\Delta G$ is  a function of $Q^2$ only. \\
    $\Delta \Sigma$ & Quark helicity contribution to the total spin of the proton. $\Delta \Sigma$ is a function of $x$ and $Q^2$. \\   
    $\eta$ & Pseudo-rapidity of particle or jet.   \\       
    $F_2$ & Structure function sensitive to the sum of quark and anti-quark momentum distributions in the nucleon/nucleus.  $F_2$ is a function of $x$ and $Q^2$.  \\ 
    $F_L$ & Longitudinal structure function dominated by the gluon momentum distribution in the
    nucleon/nucleus at low $x$.  $F_L$ is a function of $x$ and $Q^2$.\\   
    $F_2^D$ & Diffractive structure function. $F^D$ is a function of $x$ and $Q^2$.  \\ 
    $g_1$ & Polarized structure function. $g_1$ is a function of $x$ and $Q^2$.\\  
     $h_T^{(1)}$ & Distribution of linearly polarized gluons inside an unpolarized hadron. \\
    $k_{T}$	& Intrinsic transverse momentum of partons in the nucleon/nucleus. \\
    ${\cal{L}}$ & Orbital angular momenta of quarks and gluons. ${\cal{L}}$ is a function of $Q^2$ only.\\
    $M_X^2$ &  In diffraction, is the squared mass of the diffractive final state.  \\    
    $p_T$ &  Transverse momentum of a hadron or  jet.\\
    $P_T$ & Average transverse momentum of a dijet.\\
    $q_T$ & Difference in momenta of a dijet, or transverse momentum imbalance.\\ 
    $Q^2$ & Squared momentum transfer to the lepton, equal to the virtuality of the exchanged photon.
    $Q^2$ can be interpreted as the resolution power of the scatter. Note the relation $Q^2 \approx x y s$.\\
    $Q^2_s$ & Saturation scale, indicating the $Q^2$ value at a given $x$ were gluon saturation
starts to dominate. $Q^2_s$ is a function of $x$ only.\\
    $R_f^\text{A}$ & Ratio of a the parton distribution function in a nucleus, A,  over that in the proton.\\
    $s$ & Squared center-of-mass energy. In \ep\ collisions $s \approx 4 E_e  E_p$.  \\   
    $\sqrtsNN$	& Nucleon-nucleon center-of-mass energy in heavy-ion collisions. \\
    $\sigma_\text{reduced} $ & Inclusive DIS cross-section, simplified (reduced) by dividing out the Mott cross-section.\\  
    $t$	& Square of the momentum transfer at the hadronic vertex, $({p}_\text{in} - {p}_\text{out})^2$. \\
    $v_2$ &  The second harmonic  coefficient of the azimuthal Fourier decomposition of a given momentum distribution. Measure of azimuthal momentum space anisotropy of particle emission. \\
	$x$ & In the parton model, is the fraction of the nucleon or nucleus momentum carried by the struck parton ($0 < x < 1$ in \ep). \\
     $x_{g}$	& Longitudinal momentum fraction of a gluon involved in hard interactions. \\
     $\xpom$ & In diffraction, is the momentum fraction of the colorless exchange (Pomeron) with respect to the hadron.\\
    $x_\gamma$ & The momentum fraction carried by the parton from a resolved virtual photon, where resolved refers to the case when the photon fluctuates into a virtual hadronic state and can contribute a quark or gluon to a hard-scatter with a parton from the target.    \\
     $\xi$ &  In Deeply Virtual Compton Scattering (DVCS), reflects the asymmetry in the longitudinal momentum fraction of the struck parton in the initial and final state.\\
    $y$ & Inelasticity defines as the fraction of the lepton’s energy lost in the nucleon rest frame. It it thus also the fraction of the incoming electron energy carried by the exchange boson in the rest frame of the nucleon.  Note that ($0 < y < 1$). 
\end{tabularx}

\end{appendices}

%
%
\bibliographystyle{h-physrev5}
\clearpage
\phantomsection
\addcontentsline{toc}{section}{References}

{
\small
\bibliography{./tex/bibliography}{}
}

\newpage
\thispagestyle{empty}
\mbox{}
\clearpage
\thispagestyle{empty}
\includepdf[width=8.51in]{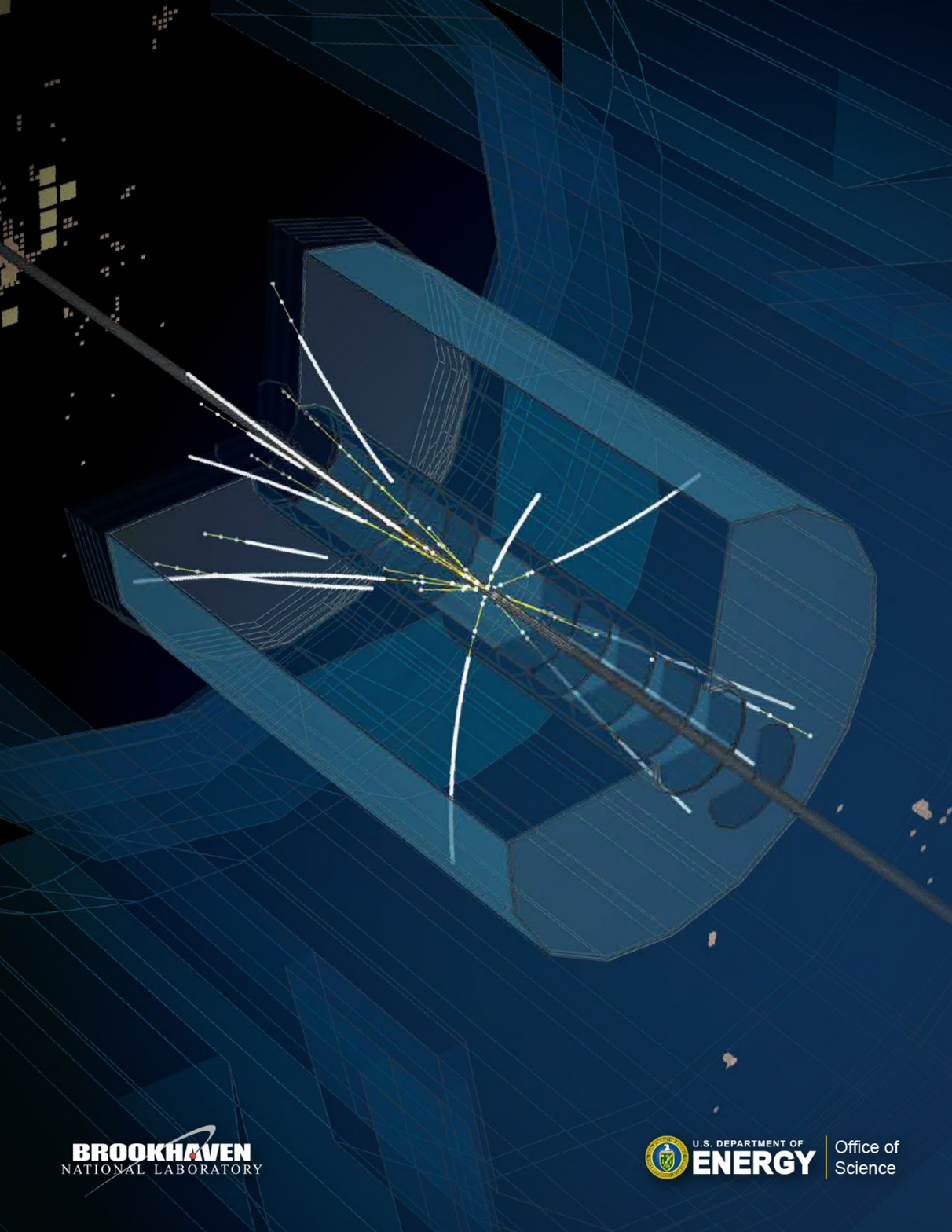}

\end{document}